\documentclass[useAMS,usenatbib]{mn2e}

\pdfoutput=1
\usepackage{longtable}
\usepackage{amssymb}
\usepackage{graphicx}
\usepackage[usenames,dvipsnames]{color}
\usepackage{verbatim}
\usepackage{morefloats}
\usepackage{pdflscape}

\usepackage{times}
\usepackage{mathptmx}

\usepackage{ulem} 

\newcommand{\um}[1]{\,\mathrm{#1}}
\newcommand{\mic}[1]{\,\umu\mathrm{#1}}
\newcommand{\HAng}[3]{#1\!:\!#2\!:\!#3}
\newcommand{\DAng}[3]{#1^\circ\,#2'\,#3''}

\title[A Radio Characterization of Galactic compact Bubbles]{A Radio Characterization of Galactic compact Bubbles}
\author[A. Ingallinera]{{\large A. Ingallinera$^1$\thanks{E-mail:ingallinera@oact.inaf.it}, C. Trigilio$^2$, G. Umana$^2$, P. Leto$^2$, A. Noriega-Crespo$^3$, N. Flagey$^4$, R. Paladini$^5$, C. Agliozzo$^1$, C. Buemi$^2$}\\
$^1$Universit\`a di Catania, Dipartimento di Fisica e Astronomia, via Santa Sofia, 64, 95123 Catania, Italy\\
$^2$INAF-Osservatorio Astrofisico di Catania, Via S. Sofia 78, 95123 Catania, Italy\\
$^3$Spitzer Science Center, California Institute of Technology, Mail Code 314-6, Pasadena, CA 91125, USA\\
$^4$Jet Propulsion Laboratory, California Institute of Technology, Pasadena, CA, USA\\
$^5$NASA Herschel Science Center, California Institute of Technology, Pasadena, CA, USA
}

\begin{document}

\date{}

\pagerange{\pageref{firstpage}--\pageref{lastpage}} \pubyear{2013}

\maketitle

\label{firstpage}

\begin{abstract}
We report the radio observations of a sub-sample of the 428 galactic compact bubbles discovered at $24\mic{m}$ with the MIPSGAL survey. Pervasive through the entire Galactic plane, these objects are thought to be different kinds of evolved stars. The very large majority of the bubbles ($\sim\!70\%$) are however not yet classified. We conducted radio observations with the EVLA at $6\um{cm}$ and $20\um{cm}$ in order to obtain the spectral index of 55 bubbles. We found that at least 70 per cent of the 31 bubbles for which we were effectively able to compute the spectral index (or its lower limit) are likely to be thermal emitters. We were also able to resolve some bubbles, obtaining that the size of the radio nebula is usually similar to the IR size, although our low resolution (with respect to IR images) did not allow further morphological studies. Comparisons between radio flux densities and IR archive data from Spitzer and IRAS suggest that at least 3 unclassified bubbles can be treated as planetary nebula candidates.
\end{abstract}


\section{Introduction}
Over 400 compact roundish objects, presenting diffuse emission, were identified at $24\mic{m}$ from visual inspection of the MIPSGAL Legacy Survey (\citealt{Carey2009}; \citealt{Mizuno2010}) mosaic images, obtained with MIPS\footnote{The Multiband Imaging Photometer for Spitzer.} \citep{Rieke2004} on board of the \textit{Spitzer Space Telescope}. These small ($\leq1'$) rings, disks or shells (hereafter denoted as `bubbles'') are pervasive throughout the entire Galactic plane in the mid-infrared (IR). Their distribution is approximately uniform in Galactic latitude and longitude, and the average density is found to be around 1.5 bubbles per square degree. A further analysis of the GLIMPSE\footnote{The Galactic Legacy IR Mid-Plane Survey Extraordinaire, conducted with the InfraRed Array Camera (IRAC) on board of the Spitzer Space Telescope.} ($3.6\mic{m}$ to $8.0\mic{m}$) and MIPSGAL ($70\mic{m}$) images indicates that the bubbles are mostly detected at $24\mic{m}$ only. The absence, for most of these objects, of a counterpart at wavelengths shorter than $24\mic{m}$ could be interpreted either as a sign of extreme extinction, which would explain the non-detection of these objects in previous visible or near-IR surveys, or as intrinsic property of the objects. The main hypothesis about the nature of the bubbles is that they are different type of evolved stars (planetary nebulae, supernova remnants, Wolf--Rayet stars, asymptotic giant branch stars, etc.).

Some bubbles present a central source in the middle of the nebula in the MIPSGAL images. Studies by \citet{Wachter2010} show how this central source is usually well detected at shorter wavelengths (down to 2MASS \textit{J} band or even optical for exceptional cases). In particular the authors spectroscopically examined 62 bright source surrounded by a 24-$\umu$m shell, being able to characterize the nature of 45 central sources. They found that 19 of them are compatible with Oe/WN, Wolf--Rayet (WR) and luminous blue variable (LBV) stars. Furthermore, they also pointed out that it is possible to explain that many bubbles emit only at $24\mic{m}$ assuming that this emission is not a continuum from warm dust but it rises from an intense \mbox{[O\,{\sc iv}]} line emission at $25.89\mic{m}$, as found by \citet{Morris2006}, resulting in an almost pure gas nebula.

The presence of very massive stars can also be inferred by the morphology of the nebula. \citet{Gvaramadze2010} found that many bubbles, showing central sources, resemble known nebulae surrounding blue supergiant (BSG), LBV, or WR stars. They confirmed the nature of some bubbles, inferred by a morphological analysis, by means of spectroscopic identification of their central sources, showing that the mere presence and shape of the nebula can suggest the possibility of these massive stars.

Mid-IR spectroscopic observations with IRS\footnote{The IR Spectrograph on board of the Spitzer Space Telescope.} were carried out for 14 bubbles, 4 in high-resolution mode \citep{Flagey2011} and 10 in low-resolution (Nowak et al., \textit{in prep.}). Among the 4 bubbles observed in high-resolution mode, two show a dust-poor spectrum dominated by highly ionized gas lines of [O \textsc{iv}], [Ne \textsc{iii}], [Ne \textsc{v}], [S \textsc{iii}], and [S \textsc{iv}], typical of planetary nebulae with a very hot central white dwarf ($\gtrsim 200\ 000\um{K}$). The other two spectra are dominated by a dust continuum and lower-excitation lines. These two bubbles also show a central source and are, respectively, a nebula surrounding a WR star \citep{Stringfellow2012} and a LBV candidate \citep{Wachter2010}.

An extensive search of available catalogues had allowed us to identify less than 15 per cent of these objects. The majority of the already known bubbles were found to be planetary nebulae (PNe). Three supernova remnants (SNRs) and one post-asymptotic giant branch (AGB) star were also identified. Therefore, about 90 per cent of the objects within the MIPSGAL bubbles were new discoveries. Further studies on the bubble catalogue allowed to extend the number of classified bubbles to, presently, about 30 per cent.

Massive stars play a pivotal role in the evolution of their host galaxies. They are among the major contributors to the interstellar ultraviolet radiation and, via their strong stellar winds and final explosion, provide enrichment of processed material (gas and dust) and mechanical energy to the interstellar medium, strongly influencing subsequent local star formation. Still, the details of post-main sequence (MS) evolution of massive stars are poorly understood. On one side, theoretical modelling depends on mass-loss from the stars, which in turn is function of poorly constrained parameters such as metallicity and rotation (e.g. \citealt{Leitherer1991}; \citealt{Chieffi2013}). On the other side, empirical studies had relied on a low number of objects at different stages of post-MS evolution \citep{Clark2005}, and only recently IR observations have permitted the discovery of hundreds of new WR and LBV stars (e.g. \citealt{Shara2009}; \citealt{Shara2012}; \citealt{Wachter2010}; \citealt{Wachter2011}; \citealt{Mauerhan2011}; \citealt{Stringfellow2012}; \citealt{Stringfellow2012b}). Besides being a powerful `game reserve' for evolved massive stars, the mid-IR bubbles catalogue is the right place where to look for the missing Galactic population of embedded PNe. The small number of known PNe (i.e. $\sim2000$) compared to those expected to populate the Galaxy disk ($\sim23\,000$, \citealt{Zijlstra1991}) is usually explained in terms of strong interstellar extinction in the Galactic plane. Part of the missing PNe population are thought to consist of more rare objects embedded in thick circumstellar envelopes. Such heavily obscured PNe may descend from the most massive AGB stars and they might be the key objects for understanding the late evolution of the most massive ($M\geq2M_\odot$) PNe progenitors.

In this paper we present results of a radio continuum study of a sub-sample of the bubbles aimed at understanding their nature. Spectral information, as derived from multi-frequency radio observations, are an unique tool for a first assessment of the content of non-thermal and thermal radio emitters in our sample, sorting out SNRs or, more generally, shocked nebulae (synchrotron emission), from nebulae associated to evolved massive stars (LBV and WR) and PNe (thermal free-free emission). It is eventually showed how radio and IR observations can be combined to establish more exhaustive classification schemes.

\section{Observations and data reduction}
\subsection{Sample selection}
From the original sample of the MIPSGAL bubbles only sources with $\delta\geq-40^\circ$ (to be visible with the EVLA\footnote{The Expanded Very Large Array.}) were selected, resulting in a total `northern sample' of 367 sources. We then checked a $1'\times1'$ field centred on each of the MIPSGAL positions in both the NVSS\footnote{http://www.cv.nrao.edu/nvss/NVSSlist.shtml} catalogue and in MAGPIS\footnote{http://third.ucllnl.org/gps/catalogs.html} for radio emission, ending up with a total of 55 sources possibly detected at $20\um{cm}$. Despite the fact that, for our targeted sources, either NVSS or MAGPIS (or both) data already exist, these cannot be used for the purpose of identifying the $24\mic{m}$ MIPSGAL bubbles. In fact, the existing NVSS/MAGPIS data suffer from three main issues: the available data were obtained with a typical rms of $0.3-0.45\um{mJy/beam}$, on average one order of magnitude worst than what is achievable with EVLA; the existing observations were taken at a different time with respect to ours and time variability effects could potentially affect the spectral index analysis; the combination of VLA and EVLA data can be, in principle, very problematic from a technical point of view.

The available NVSS and MAGPIS data yet provided very useful indications regarding the size and flux of our selected sample of sources, and this information was used to guide our observing strategy in terms of configuration and time request. Remarkably, 11 of the objects selected for EVLA observations were already classified, according to the SIMBAD\footnote{http://simbad.u-strasbg.fr} database.

\subsection{Observing strategy}
Observations of the bubbles sample were made with the EVLA at $6\um{cm}$ (central frequency $4.959\um{GHz}$ -- $C$ band) in configuration D during March 2010 and at $20\um{cm}$ ($1.4675\um{GHz}$ -- $L$ band) in configuration C and CnB during, respectively, March and May 2012.

For $C$-band observations the sample was split in four subset, observed in four different days. Each bubble was observed for slightly less than 10 minutes and in two 128-MHz wide spectral windows (resulting therefore in a total bandwidth of $256\um{MHz}$) allowing us to achieve a theoretical noise level of $\sim\!10\mic{Jy/beam}$. We note that calibration errors and required flagging introduce further sources of noise that eventually dominate over the theoretical thermal noise.

For observations in $L$ band, the previous 6-cm observations were used to select a sub-sample on which focusing our attention. In particular we selected a sub-sample of 34 bubbles detected or possibly detected at $6\um{cm}$, excluding some bubbles that appeared too extended at $6\um{cm}$ or whose classification was certain. The larger field-of-view at $20\um{cm}$ allowed to include other 6 bubbles as field sources, resulting in a total sample of 40 bubbles. Though the total bandwidth was as wide as $1\um{GHz}$, a lot of radio-frequency interferences (RFI) contaminated our data and the signal-to-noise ratio was much lower than we expected, sometimes one order of magnitude or more.

Since observations were made toward the galactic plane, it was also necessary to check the confusion limit. At $6\um{cm}$ we expected a value around $7\mic{Jy/beam}$, while at $20\um{cm}$ slightly less than $20\mic{Jy/beam}$. Both limits were well below our expected noise levels.

In Table \ref{tab:obsVLA} all observed objects are reported along with their coordinates and the date and duration of each observation. Beside the official designation (MGE$l\!\pm\!b$), for each bubble we list a shorter identification name (second column), derived from shorthands used during the identification phase, that will be used in this work as a compact notation.

\begin{table*}
\caption{Observations summary. Source dimensions at $24\mic{m}$ are from the bubbles catalogue \citep{Mizuno2010}.}
\begin{minipage}{\textwidth}
\begin{center}
\begin{tabular}{cccccccccc}
\hline
Designation & Bubble & RA & DEC & Obs. day & Obs. time & Obs. day & Obs. time & Dimension & Classified\\
{[MGE]} & ID & (J2000) & (J2000) & (2010) & (min) & (2012) & (min) & at $24\mic{m}$ & in SIMBAD?\\
& & & & $C$ band & $C$ band & $L$ band & $L$ band\\
\hline
010.5569+00.0188 & 3153 & $\HAng{18}{08}{50.5}$ & $-\DAng{19}{47}{39}$ & 13--Mar & 12 & -- & -- & $\phantom{0}25''$\\
013.5944+00.2139 & 3173 & $\HAng{18}{14}{17.1}$ & $-\DAng{17}{02}{16}$ & 14--Mar & 12 & -- & -- & $\phantom{0}24''$\\
014.1176+00.0816 & 3177 & $\HAng{18}{15}{48.9}$ & $-\DAng{16}{38}{27}$ & 14--Mar & 12 & 06--Mar\footnote{Observed as field source.} & 10 & $\phantom{0}15''$\\
016.1871+00.1202 & 3188 & $\HAng{18}{19}{45.1}$ & $-\DAng{14}{48}{02}$ & 14--Mar & 10 & 06--Mar & 10 & $\phantom{0}16''$\\
015.9774+00.2955 & 3192 & $\HAng{18}{18}{42.2}$ & $-\DAng{14}{54}{09}$ & 14--Mar & 12 & 06--Mar$^{\textstyle a}$ & 10 & $\phantom{0}18''$\\
016.1274+00.3327 & 3193 & $\HAng{18}{18}{51.7}$ & $-\DAng{14}{45}{10}$ & 14--Mar & 12 & 06--Mar & 10 & $\phantom{0}18''$\\
019.6492+00.7740 & 3214 & $\HAng{18}{24}{04.0}$ & $-\DAng{11}{26}{16}$ & 14--Mar & 10 & -- & -- & $\phantom{0}18''$ & PN\footnote{[1] \citealt{Miszalski2008}; [2] \citealt{Green2009}; [3] \citealt{Kerber2003}; [4] \citealt{Chevalier2005}; [5] \citealt{Parker2006}; [6] \citealt{Kohoutek2001}.}$^{\textstyle [1]}$\\
030.1503+00.1237 & 3222 & $\HAng{18}{45}{55.2}$ & $-\DAng{02}{25}{08}$ & 14--Mar & 10 & 06--Mar & 10 & $\phantom{0}26''$\\
023.4499+00.0820 & 3259 & $\HAng{18}{33}{43.3}$ & $-\DAng{08}{23}{35}$ & 14--Mar & 10 & -- & -- & $\phantom{0}25''$\\
023.6857+00.2226 & 3269 & $\HAng{18}{33}{39.5}$ & $-\DAng{08}{07}{08}$ & 13--Mar & 10 & -- & -- & $\phantom{0}44''$\\

026.4700+00.0209 & 3282 & $\HAng{18}{39}{32.2}$ & $-\DAng{05}{44}{20}$ & 14--Mar & 10 & -- & -- & $\phantom{0}80''$\\
027.5373+00.5473 & 3309 & $\HAng{18}{39}{37.4}$ & $-\DAng{04}{32}{56}$ & 14--Mar & 10 & 06--Mar & \phantom{0}9 & $\phantom{0}15''$\\
027.3891--00.0079 & 3310 & $\HAng{18}{41}{19.9}$ & $-\DAng{04}{56}{06}$ & 13--Mar & \phantom{1}5 & 06--Mar$^{\textstyle a}$ & \phantom{0}9 & $250''$ & SNR$^{\textstyle b[2]}$\\
028.4451+00.3094 & 3313 & $\HAng{18}{42}{08.2}$ & $-\DAng{03}{51}{03}$ & 14--Mar & 10 & 06--Mar & \phantom{0}9 & $\phantom{0}80''$\\
029.0784+00.4545 & 3328 & $\HAng{18}{42}{46.8}$ & $-\DAng{03}{13}{17}$ & 14--Mar & 10 & 06--Mar$^{\textstyle a}$ & \phantom{0}9 & $\phantom{0}28''$ & PN$^{\textstyle b[3]}$\\
028.7440+00.7076 & 3333 & $\HAng{18}{41}{16.0}$ & $-\DAng{03}{24}{11}$ & 14--Mar & 10 & 06--Mar & 10 & $\phantom{0}23''$\\
030.8780+00.6993 & 3347 & $\HAng{18}{45}{12.0}$ & $-\DAng{01}{30}{32}$ & 14--Mar & 10 & 06--Mar & 10 & $\phantom{0}18''$\\
031.7290+00.6993 & 3354 & $\HAng{18}{46}{45.2}$ & $-\DAng{00}{45}{06}$ & 13--Mar & 10 & 06--Mar\footnote{Observed also on 13--May.} & 18 & $\phantom{0}44''$\\
032.8593+00.2806 & 3362 & $\HAng{18}{50}{18.3}$ & $\phantom{-}\DAng{00}{03}{48}$ & 13--Mar & 10 & 06--Mar & 10 & $\phantom{0}15''$\\
032.4982+00.1615 & 3367 & $\HAng{18}{50}{04.3}$ & $-\DAng{00}{18}{45}$ & 13--Mar & 10 & 06--Mar$^{\textstyle c}$ & 15 & $\phantom{0}16''$\\
034.8961+00.3018 & 3384 & $\HAng{18}{53}{56.8}$ & $\phantom{-}\DAng{01}{53}{08}$ & 13--Mar & 10 & 06--Mar$^{\textstyle c}$ & 18 & $\phantom{0}21''$\\
042.0787+00.5084 & 3438 & $\HAng{19}{06}{24.6}$ & $\phantom{-}\DAng{08}{22}{02}$ & 13--Mar & 10 & 06--Mar$^{\textstyle c}$ & 21 & $\phantom{0}21''$\\
042.7665+00.8222 & 3448 & $\HAng{19}{06}{33.6}$ & $\phantom{-}\DAng{09}{07}{20}$ & 13--Mar & 10 & 13--May & \phantom{0}9 & $\phantom{0}33''$\\
065.9141+00.5966 & 3558 & $\HAng{19}{55}{02.4}$ & $\phantom{-}\DAng{29}{17}{20}$ & 13--Mar & 10 & -- & -- & $\phantom{0}33''$ & PN$^{\textstyle b[3]}$\\
040.3704--00.4750 & 3654 & $\HAng{19}{06}{45.8}$ & $\phantom{-}\DAng{06}{23}{53}$ & 13--Mar & 10 & 13--May & \phantom{0}4 & $\phantom{0}27''$ & PN$^{\textstyle b[3]}$\\
031.9075--00.3087 & 3706 & $\HAng{18}{50}{40.1}$ & $-\DAng{01}{03}{09}$ & 13--Mar & \phantom{1}6 & 13--May & \phantom{0}4 & $\phantom{0}25''$ & PN$^{\textstyle b[3]}$\\
029.4034--00.4496 & 3724 & $\HAng{18}{46}{35.9}$ & $-\DAng{03}{20}{43}$ & 14--Mar & 10 & 06--Mar & 10 & $\phantom{0}16''$\\
027.3839--00.3031 & 3736 & $\HAng{18}{42}{22.5}$ & $-\DAng{05}{04}{29}$ & 14--Mar & 10 & 06--Mar & 10 & $\phantom{0}37''$\\
016.2280--00.3680 & 3866 & $\HAng{18}{21}{36.9}$ & $-\DAng{14}{59}{41}$ & 14--Mar & 10 & 06--Mar & \phantom{0}5 & $\phantom{0}18''$\\
011.1805--00.3471 & 3910 & $\HAng{18}{11}{28.9}$ & $-\DAng{19}{25}{29}$ & 13--Mar & \phantom{1}6 & -- & -- & $260''$ & SNR$^{\textstyle b[4]}$\\
010.6846--00.6280 & 3915 & $\HAng{18}{11}{30.8}$ & $-\DAng{19}{59}{41}$ & 13--Mar & 12 & -- & -- & $\phantom{0}15''$\\
001.0178--01.9642 & 4409 & $\HAng{17}{55}{43.1}$ & $-\DAng{29}{04}{04}$ & 27--Mar & 10 & 08--May$^{\textstyle a}$ & 10 & $\phantom{0}28''$ & PN$^{\textstyle b[5]}$\\
003.5533--02.4421 & 4422 & $\HAng{18}{03}{18.4}$ & $-\DAng{27}{06}{22}$ & 13--Mar & 12 & -- & -- & $\phantom{0}30''$ & PN$^{\textstyle b[3]}$\\
356.7168--01.7246 & 4436 & $\HAng{17}{44}{29.6}$ & $-\DAng{32}{38}{11}$ & 27--Mar & 10 & 08--May & 10 & $\phantom{0}18''$\\
359.5381--01.0838 & 4443 & $\HAng{17}{48}{46.6}$ & $-\DAng{29}{53}{34}$ & 21--Mar & 10 & -- & -- & $\phantom{0}23''$\\
001.2920--01.4680 & 4452 & $\HAng{17}{54}{23.6}$ & $-\DAng{28}{34}{51}$ & 21--Mar & 10 & 08--May & 10 & $\phantom{0}15''$\\
002.0599--01.0642 & 4463 & $\HAng{17}{54}{34.4}$ & $-\DAng{27}{42}{51}$ & 27--Mar & 10 & 08--May & 10 & $\phantom{0}27''$\\
002.2128--01.6131 & 4465 & $\HAng{17}{57}{03.9}$ & $-\DAng{27}{51}{30}$ & 27--Mar & 10 & 08--May & 10 & $\phantom{0}18''$\\
003.4305--01.0738 & 4467 & $\HAng{17}{57}{42.5}$ & $-\DAng{26}{32}{05}$ & 13--Mar & 12 & -- & -- & $\phantom{0}19''$\\
005.6102--01.1516 & 4473 & $\HAng{18}{02}{48.4}$ & $-\DAng{24}{40}{54}$ & 13--Mar & 12 & 08--May & 10 & $\phantom{0}20''$\\
009.4257--01.2294 & 4479 & $\HAng{18}{11}{10.6}$ & $-\DAng{21}{23}{15}$ & 13--Mar & 12 & -- & -- & $\phantom{0}20''$ & PN$^{\textstyle b[5]}$\\
351.2381--00.0145 & 4485 & $\HAng{17}{23}{04.4}$ & $-\DAng{36}{18}{20}$ & 21--Mar & 10 & 08--May & 10 & $\phantom{0}40''$\\
352.3117--00.9711 & 4486 & $\HAng{17}{29}{58.3}$ & $-\DAng{35}{56}{56}$ & 21--Mar & 10 & 08--May & 10 & $\phantom{0}18''$\\
356.8155--00.3843 & 4497 & $\HAng{17}{39}{21.3}$ & $-\DAng{31}{50}{44}$ & 27--Mar & 10 & 08--May & 10 & $\phantom{0}15''$\\
006.5850--00.0135 & 4530 & $\HAng{18}{00}{35.2}$ & $-\DAng{23}{16}{18}$ & 21--Mar & \phantom{1}5 & 08--May$^{\textstyle a}$ & 10 & $\phantom{0}12''$\\
349.7294+00.1747 & 4534 & $\HAng{17}{17}{59.3}$ & $-\DAng{37}{26}{09}$ & 21--Mar & \phantom{1}5 & -- & -- & $\phantom{0}68''$\\
356.1447+00.0550 & 4552 & $\HAng{17}{35}{54.5}$ & $-\DAng{32}{10}{35}$ & 27--Mar & 10 & 08--May & 10 & $\phantom{0}25''$\\
355.7638+00.1424 & 4555 & $\HAng{17}{34}{35.2}$ & $-\DAng{32}{26}{58}$ & 27--Mar & 10 & -- & -- & $\phantom{0}16''$\\
001.5280+00.9171 & 4580 & $\HAng{17}{45}{40.7}$ & $-\DAng{27}{09}{15}$ & 21--Mar & 10 & 08--May & 10 & $\phantom{0}18''$\\
001.9965+00.1976 & 4583 & $\HAng{17}{49}{32.1}$ & $-\DAng{27}{07}{32}$ & 21--Mar & 10 & 08--May & 10 & $\phantom{0}12''$\\
001.6982+00.1362 & 4584 & $\HAng{17}{49}{04.9}$ & $-\DAng{27}{24}{47}$ & 21--Mar & 10 & 08--May & 10 & $\phantom{0}18''$\\
005.2641+00.3775 & 4589 & $\HAng{17}{56}{13.4}$ & $-\DAng{24}{13}{13}$ & 27--Mar & 10 & 08--May & 10 & $\phantom{0}16''$\\
006.9367+00.0497 & 4595 & $\HAng{18}{01}{06.4}$ & $-\DAng{22}{56}{05}$ & 21--Mar & 10 & 08--May & 10 & $\phantom{0}20''$\\
009.3523+00.4733 & 4602 & $\HAng{18}{04}{38.9}$ & $-\DAng{20}{37}{27}$ & 13--Mar & 12 & 08--May & 10 & $\phantom{0}28''$ & PN?$^{\textstyle b[6]}$\\
008.9409+00.2532 & 4607 & $\HAng{18}{04}{36.3}$ & $-\DAng{21}{05}{26}$ & 13--Mar & 12 & 08--May & 10 & $\phantom{0}18''$\\
\hline
\end{tabular}
\label{tab:obsVLA}
\end{center}
\end{minipage}
\end{table*}

\subsection{Data reduction}
The entire data reduction process was performed using the package \textsc{casa}. As a first step, the data were edited and flagged in order to identify and delete not properly working antennas, bad baselines and border (and usually noisy) channels. For $C$-band observations the editing process revealed no great corruptions in our data, while for $L$-band observations a large amount of flagging was needed in order to filter out the conspicuous RFI, leading to less 33 per cent of useful data remaining.

For all the observations, the bandpass and flux calibrations were done using 3C286 as calibrator. In order to improve the quality of our gain calibration, depending on the distance from the source (typically within $10^\circ$), we used a variety of standard calibrators spanning a range of flux densities.

\subsection{Imaging}
\label{sec:ima}
Data imaging was made using the Clark implementation \citep{Clark1980} of the CLEAN algorithm \citep{Hogbom1974}, convolving the resulting `clean components' with a Gaussian PSF.

For $C$ band, since all observations were carried out with the EVLA in the same configuration (D), no significant differences were found in synthesis beam sizes. Therefore all the images were built using a $4''$ pixel and a total size of $256\times256$ pixels, in such a way that each map covers approximately a $17'\times17'$ area (the primary beam is about $9'$ FWHM). In some maps we were able to clean down to a $\mathrm{rms}\sim30\mic{Jy/beam}$, with an average beam size around $25''\times15''$. The typical noise was one order of magnitude greater than the confusion limit.

For $L$ band instead, since multiple configurations were used, for each image a best choice between a pixel size of $3''$ or $4''$ was adopted. Also, the size of images was allowed to vary to best accommodate for field sources cleaning. The typical rms was about $0.5-1\um{mJy/beam}$ with an average beam size of $18''\times12''$. The typical noise was two orders of magnitude greater than the confusion limit.

In $C$ band we expected that only sources with dimensions significantly less than $2'$ (EVLA largest angular scale) could be reasonably well imaged, and this would also permit a total flux density recovery. The more a source is extended the less reliable is its flux density measurement. Therefore, at the end of the imaging process, we cautiously excluded eight bubbles (namely 3259, 3282, 3310, 3328, 3558, 3910, 4485 and 4595) from the remainder of this work since they were suspected to be resolved out by the EVLA. A single-dish analysis for these bubbles is in progress.

Radio maps and 24-$\umu$m images of some Bubbles are presented in appendix A (online only).

\section{Spectral index analysis}
\label{sec:detsrc}

\subsection{Detections and flux densities calculation}
\label{sec:det}
The majority of the bubbles observed were detected in both bands. In particular for $C$ band we detected 44 bubbles out of 55, with 3 uncertain detections and 8 non-detections. For $L$ band we detected 23 bubbles out of 40, with 3 uncertain detections and 14 non-detections.

Since one of the main goals of this work was to characterise the radio emission of the bubbles as an important aid to their classification, a very accurate flux density determination was needed. To avoid introducing methodological errors or biases, a unique procedure in this calculation was adopted. First of all the sources were divided into two classes depending on whether they were resolved or not.

For point sources (not resolved) the flux density was determined using the \textsc{casa} task \texttt{imfit}, which fits an elliptical Gaussian component to an image. Given that the maps units are jansky/beam, the total flux density for a point source is equal to the peak value of the fitted Gaussian, i.e. $S=S_p$. The error was computed as the quadratic sum of the error derived from the fit, the map rms and the calibration error (this one, negligible in both bands): 
\begin{equation}
\Delta S=\sqrt{\sigma_\mathrm{fit}^2+\sigma_\mathrm{rms}^2+\sigma_\mathrm{cal}^2}.
\end{equation}

The flux density calculation for extended sources proved much more difficult. For extended sources detected or resolved in one band only, the strategy was to localise the source boundary as the lowest brightness level at which we were confident to encompass only our object. Theoretically, one should go down to $\sigma_\mathrm{rms}$, below which the source becomes indistinct with respect to the background. However, the artefacts in interferometric images usually do not allow to look so deep and, for many bubbles, we were forced to stop at higher levels. Selected then an appropriate region for each object, the flux density was calculated by means of an integration over this area, performed directly with the \textsc{casa} \texttt{viewer}. The total error was estimated as the map rms multiplied by the square root of the integration area expressed in beams.

For sources resolved in both bands we proceeded as follows. First the map with the higher angular resolution was degraded by convolving the clean components with the lower resolution beam and adding back the residual map. Then, for each bubble, we selected a region large enough to cover the source in both bands, and used this to estimate the flux and corresponding error as in the previous case.

Furthermore an approximate size for resolved bubbles was calculated as follows: the observed size of the source, $\Omega_o$, is expressed as
\begin{equation}
\Omega_o=\Omega_s+\Omega_b
\end{equation}
where $\Omega_s$ is the `real' angular size of the source and $\Omega_b$ is the beam solid angle. The quantity $\Omega_o$ can also been expressed in term of number of beams, $N_b$, a quantity already computed for the determination of flux densities
\begin{equation}
\Omega_o=N_b\Omega_b,
\end{equation}
hence
\begin{equation}
\Omega_s=(N_b-1)\Omega_b
\end{equation}
and we calculated the corresponding mean size as
\begin{equation}
\langle\theta_s\rangle=\sqrt{b_\mathrm{maj}b_\mathrm{min}(N_b-1)},
\end{equation}
where $b_\mathrm{maj}$ and $b_\mathrm{min}$ are, respectively, the beam major and minor axis. The results obtained are listed, along with some useful characteristics of each map, in Table \ref{tab:fluxC} for $C$ band and in Table \ref{tab:fluxL} for $L$ band.

\begin{table*}
\caption{Flux densities at $6\um{cm}$. Among the 44 bubbles detected at this frequency 8 are, likely, resolved out (see section \ref{sec:ima}) and for one (Bubble 3173) the flux density measurement is not reliable. Therefore only 35 bubbles are listed.}
\begin{tabular}{cccccccc}
\hline
Bubble & Map rms & Beam & PA & Flux density & Resolved? & $\langle\theta_s\rangle$ & Notes\\
& (mJy/beam) & & & (mJy)\\\hline
3188 & 0.24 & $23.6''\times15.8''$ & $\phantom{0}{-5}^\circ$ & $\phantom{0}1.0\pm0.3$ & no?\\
3192 & 0.53 & $26.7''\times15.5''$ & $-34^\circ$ & $\phantom{0}1.2\pm0.6$ & no?\\
3193 & 0.61 & $22.5''\times15.0''$ & $-18^\circ$ & $\phantom{0}1.4\pm0.6$ & no\\
3214 & 0.11 & $22.4''\times16.0''$ & $\phantom{-0}0^\circ$ & $\phantom{0}4.0\pm0.2$ & no\\
3222 & 0.82 & $21.3''\times14.4''$ & $\phantom{-}41^\circ$ & $22.9\pm1.5$ & no\\
3309 & 0.21 & $19.8''\times15.5''$ & $\phantom{-}14^\circ$ & $\phantom{0}3.4\pm0.4$ & yes & $26''$\\
3313 & 0.30 & $19.6''\times15.1''$ & $\phantom{-}24^\circ$ & $\phantom{0}5.1\pm0.7$ & yes & $34''$\\
3333 & 0.15 & $19.3''\times15.1''$ & $\phantom{-}22^\circ$ & $\phantom{0}6.4\pm0.3$ & yes & $24''$\\
3347 & 0.14 & $20.2''\times14.1''$ & $\phantom{-}40^\circ$ & $\phantom{0}1.5\pm0.3$ & no\\
3354 & 0.04 & $21.1''\times18.5''$ & $\phantom{-}51^\circ$ & $12.3\pm0.1$ & yes & $23''$\\
3362 & 2.35 & $21.8''\times15.6''$ & $\phantom{-}34^\circ$ & $12.1\pm2.5$ & no\\
3367 & 0.79 & $21.3''\times15.4''$ & $\phantom{-}31^\circ$ & $\phantom{0}4.7\pm0.9$ & no\\
3384 & 0.16 & $27.2''\times21.4''$ & $\phantom{0}{-2}^\circ$ & $17.8\pm0.4$ & yes & $54''$ & \textit{Self-calibrated}\\
3438 & 0.09 & $19.8''\times16.1''$ & $\phantom{-}46^\circ$ & $10.5\pm0.1$ & no?\\
3448 & 0.13 & $20.6''\times16.1''$ & $\phantom{-}51^\circ$ & $12.7\pm0.4$ & no\\
3654 & 0.18 & $22.8''\times16.2''$ & $\phantom{-}52^\circ$ & $59.7\pm0.5$ & yes & $38''$\\
3706 & 0.40 & $23.1''\times15.7''$ & $\phantom{-}37^\circ$ & $19.6\pm0.8$ & yes & $22''$\\
3724 & 0.18 & $19.9''\times14.0''$ & $\phantom{-}36^\circ$ & $\phantom{0}3.2\pm0.4$ & yes & $31''$\\
3736 & 0.14 & $20.2''\times14.9''$ & $\phantom{-}27^\circ$ & $18.1\pm0.5$ & yes & $56''$\\
3866 & 0.30 & $24.0''\times16.0''$ & $\phantom{0}{-1}^\circ$ & $10.3\pm0.6$ & no\\
4409 & 0.03 & $68.2''\times12.7''$ & $-25^\circ$ & $\phantom{0}7.3\pm0.1$ & no\\
4422 & 0.06 & $35.1''\times13.5''$ & $\phantom{-}15^\circ$ & $40.3\pm0.2$ & no\\
4436 & 0.06 & $74.4''\times11.3''$ & $-15^\circ$ & $\phantom{0}5.8\pm0.1$ & no\\
4452 & 0.18 & $39.6''\times13.9''$ & $-30^\circ$ & $\phantom{0}2.1\pm0.4$ & yes? & $43''$\\

4465 & 0.04 & $52.3''\times12.2''$ & $-13^\circ$ & $\phantom{0}1.6\pm0.1$ & no\\
4473 & 0.52 & $31.8''\times13.4''$ & $-12^\circ$ & $38.1\pm0.8$ & no\\
4479 & 0.08 & $31.2''\times13.5''$ & $\phantom{-}23^\circ$ & $16.0\pm0.2$ & no\\
4486 & 0.58 & $43.9''\times14.2''$ & $-16^\circ$ & $15.5\pm0.9$ & no\\
4497 & 0.18 & $63.8''\times11.7''$ & $-16^\circ$ & $15.1\pm0.4$ & no\\
4552 & 0.48 & $66.0''\times11.7''$ & $-18^\circ$ & $15.0\pm0.7$ & no\\
4580 & 0.35 & $46.5''\times14.2''$ & $-37^\circ$ & $\phantom{0}2.0\pm0.4$ & no\\
4584 & 0.32 & $41.0''\times14.5''$ & $-33^\circ$ & $\phantom{0}2.1\pm0.5$ & yes? & $28''$\\
4589 & 0.08 & $47.4''\times12.3''$ & $-15^\circ$ & $\phantom{0}9.0\pm0.2$ & no\\
4602 & 0.25 & $30.1''\times13.8''$ & $\phantom{-}18^\circ$ & $17.7\pm0.7$ & no\\
4607 & 0.26 & $29.6''\times13.7''$ & $\phantom{-}16^\circ$ & $\phantom{0}8.2\pm0.4$ & no &  & \textit{Self-calibrated}\\
\hline
\end{tabular}
\label{tab:fluxC}
\end{table*}

\begin{table*}
\caption{Flux densities at $20\um{cm}$. The flux densities (or their upper limits) for 7 bubbles were not reliable and are not listed.}
\begin{tabular}{cccccccc}
\hline
Bubble & Map rms & Beam & PA & Flux density & Resolved? & $\langle\theta_s\rangle$ & Notes\\
& (mJy/beam) & & & (mJy)\\\hline
3188 & 1.46 & $25.1''\times14.4''$ & $-23^\circ$ & $<4.5$ & -- & & \textit{Upper limit only}\\
3192 & 0.74 & $29.1''\times14.1''$ & $-21^\circ$ & $<2.1$ & -- & & \textit{Upper limit only}\\
3193 & 0.74 & $29.1''\times14.1''$ & $-21^\circ$ & $<2.1$ & -- & & \textit{Upper limit only}\\
3222 & 2.26 & $18.4''\times14.4''$ & $-17^\circ$ & $21.5\pm3.1$ & no\\
3309 & 0.88 & $19.0''\times14.2''$ & $-12^\circ$ & $<5.2$ & -- & & \textit{Upper limit only}\\
3313 & 0.98 & $18.7''\times14.2''$ & $-10^\circ$ & $<6.9$ & -- & & \textit{Upper limit only}\\
3328 & 0.87 & $20.1''\times14.5''$ & $-27^\circ$ & $11.3\pm4.4$ & yes & $85''$ & \textit{Resolved-out at $6\um{cm}$}\\
3333 & 0.69 & $20.1''\times14.5''$ & $-27^\circ$ & $\phantom{0}4.0\pm1.2$ & yes & $24''$\\
3347 & 1.88 & $18.7''\times14.4''$ & $-21^\circ$ & $<5.7$ & -- & & \textit{Upper limit only}\\
3354 & 0.58 & $15.5''\times12.0''$ & $-36^\circ$ & $12.0\pm1.1$ & yes & $23''$\\
3362 & 1.88 & $17.1''\times13.2''$ & $\phantom{0}{-4}^\circ$ & $<5.7$ & -- & & \textit{Upper limit only}\\
3367 & 0.92 & $16.1''\times11.9''$ & $-42^\circ$ & $\phantom{0}6.8\pm0.9$ & no\\
3384 & 0.82 & $15.6''\times12.0''$ & $-38^\circ$ & $\phantom{0}2.1\pm0.8$ & yes & & \textit{Peak intensity}\\
3438 & 0.39 & $15.0''\times12.0''$ & $-26^\circ$ & $10.2\pm0.8$ & yes & $24''$\\
3448 & 0.73 & $14.9''\times10.7''$ & $-77^\circ$ & $12.8\pm1.1$ & yes & $14''$\\
3654 & 0.24 & $13.5''\times10.8''$ & $-43^\circ$ & $64.1\pm0.8$ & yes & $38''$\\
3706 & 1.93 & $14.3''\times11.6''$ & $-63^\circ$ & $10.8\pm3.7$ & yes & $22''$\\
3724 & 1.50 & $18.7''\times14.3''$ & $\phantom{0}{-8}^\circ$ & $<9.6$ & -- & & \textit{Upper limit only}\\
3736 & 0.64 & $20.8''\times14.6''$ & $-23^\circ$ & $<6.1$ & -- & & \textit{Upper limit only}\\
3866 & 3.31 & $25.1''\times14.5''$ & $-22^\circ$ & $\phantom{0}9.7\pm3.5$ & no\\
4409 & 0.46 & $17.3''\times12.0''$ & $\phantom{0}{-3}^\circ$ & $<1.5$ & -- & & \textit{Upper limit only}\\
4436 & 0.33 & $21.0''\times12.2''$ & $-29^\circ$ & $\phantom{0}6.6\pm0.4$ & no\\
4452 & 0.57 & $17.3''\times12.0''$ & $\phantom{0}{-3}^\circ$ & $<6.1$ & -- & & \textit{Upper limit only}\\
4465 & 0.55 & $18.4''\times12.1''$ & $-25^\circ$ & $\phantom{0}1.0\pm0.6$ & no\\
4473 & 3.94 & $16.8''\times12.0''$ & $\phantom{-0}5^\circ$ & $34.7\pm3.9$ & no?\\
4486 & 0.53 & $24.1''\times12.1''$ & $-30^\circ$ & $20.9\pm1.1$ & yes & $31''$\\
4497 & 0.62 & $21.0''\times11.8''$ & $-31^\circ$ & $15.0\pm0.9$ & no\\
4552 & 0.66 & $23.9''\times11.7''$ & $-36^\circ$ & $15.9\pm1.1$ & yes & $22''$ & \textit{Self-calibrated}\\
4580 & 1.00 & $17.5''\times12.2''$ & $-12^\circ$ & $<3.0$ & -- & & \textit{Upper limit only}\\
4584 & 1.21 & $17.4''\times12.1''$ & $-11^\circ$ & $<9.2$ & -- & & \textit{Upper limit only}\\
4589 & 0.55 & $17.8''\times11.7''$ & $-29^\circ$ & $\phantom{0}8.9\pm0.6$ & no\\
4602 & 1.16 & $15.7''\times11.7''$ & $\phantom{-}18^\circ$ & $14.7\pm1.7$ & no\\
4607 & 0.51 & $15.4''\times11.7''$ & $\phantom{-}15^\circ$ & $\phantom{0}5.9\pm0.6$ & no\\
\hline
\end{tabular}
\label{tab:fluxL}
\end{table*}

As mentioned in the previous section, the determination of a spectral index for as many bubbles as possible was critical for this work. Once the flux densities were estimated as described above, the spectral index $\alpha$ is defined as
\begin{equation}
S_\nu\propto\nu^\alpha,
\end{equation}
with an associated error given by
\begin{equation}
\Delta\alpha\simeq\frac{\displaystyle\sqrt{\left(\frac{\Delta S_L}{S_L}\right)^2+\left(\frac{\Delta S_C}{S_C}\right)^2}}{\displaystyle\ln\frac{\nu_C^{\phantom{x}}}{\nu_L}},
\end{equation}
where subscripts $C$ and $L$ refer, respectively, to $6\um{cm}$ and $20\um{cm}$ observations. The error on frequencies was neglected.

\subsection{Results}
\begin{figure}
\includegraphics[width=\columnwidth]{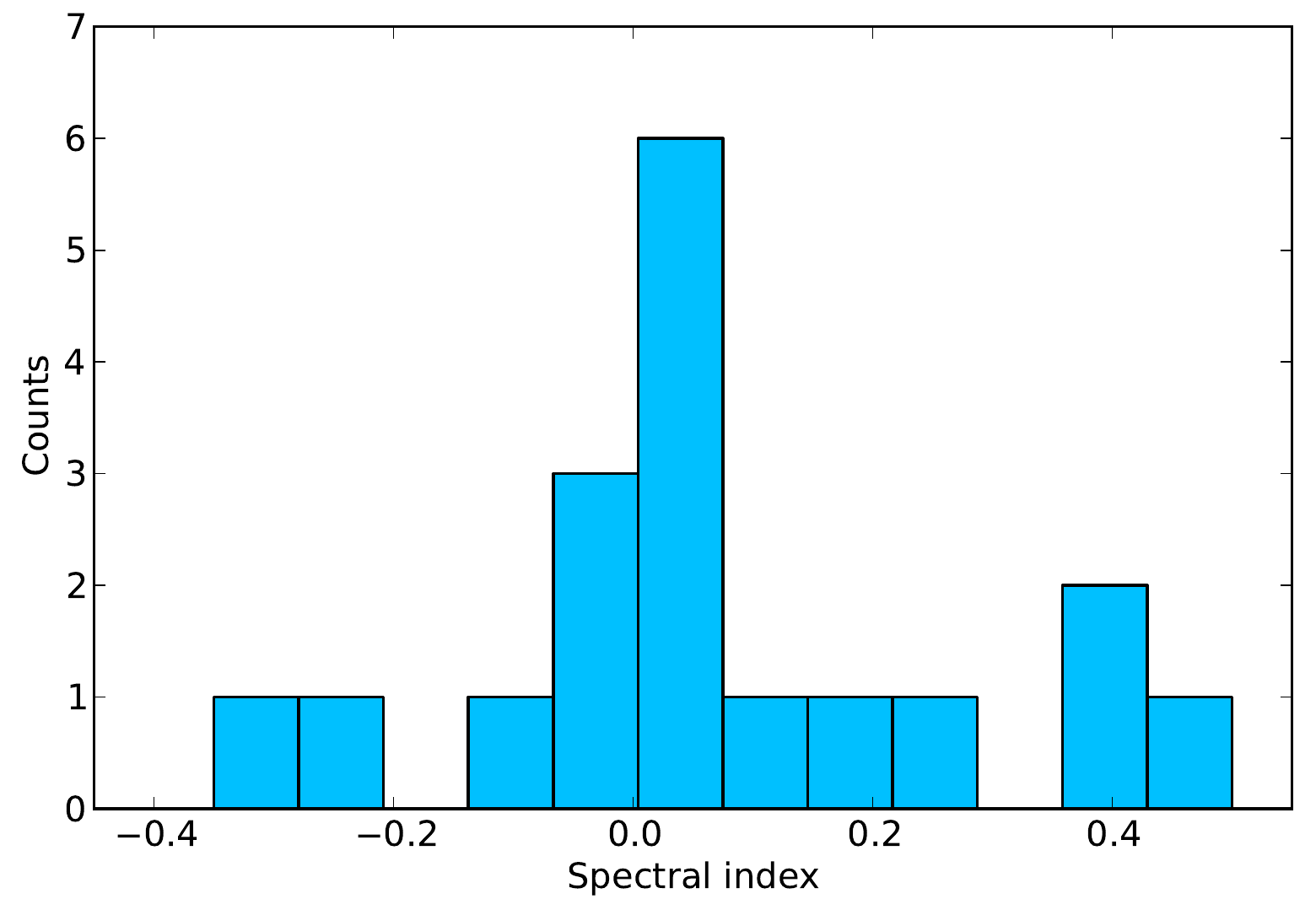}
\caption{Spectral index statistical distribution.}
\label{fig:hist}
\end{figure}

The analysis of the spectral indices, obtained as described above, suggests that many bubbles are free-free emitters, with the majority optically thick at $20\um{cm}$ (see Table \ref{tab:spInPS} and Figure \ref{fig:hist}). Only Bubbles 3367 and 4486 may have spectral index values compatible with non-thermal emission.
\begin{table}
\caption{Spectral index for sources detected in both bands.}
\begin{tabular}{ccccc}\hline
Bubble & Flux density & Flux density & $\alpha$ & Resolved?\\
 & at $20\um{cm}$ (mJy) & at $6\um{cm}$ (mJy)\\\hline
3222 & $21.5\pm3.1$ & $22.9\pm1.5$ & $\phantom{-}0.05\pm0.13$ & no\\
3333 & $\phantom{0}4.0\pm0.2$ & $\phantom{0}6.4\pm0.3$ & $\phantom{-}0.39\pm0.26$ & yes\\
3354 & $12.0\pm1.1$ & $12.3\pm0.1$ & $\phantom{-}0.02\pm0.08$ & yes\\
3367 & $\phantom{0}6.8\pm0.9$ & $\phantom{0}4.7\pm0.9$ & $-0.30\pm0.19$ & no\\
3438 & $10.2\pm0.8$ & $10.5\pm0.1$ & $\phantom{-}0.02\pm0.07$ & yes\\
3448 & $12.8\pm1.1$ & $12.7\pm0.4$ & $-0.01\pm0.08$ & yes\\
3654 & $64.1\pm0.8$ & $59.7\pm0.5$ & $-0.06\pm0.01$ & yes\\
3706 & $10.8\pm3.7$ & $19.6\pm0.8$ & $\phantom{-}0.49\pm0.28$ & yes\\
3866 & $\phantom{0}9.7\pm3.5$ & $10.3\pm0.6$ & $\phantom{-}0.05\pm0.30$ & no\\
4436 & $\phantom{0}6.6\pm0.4$ & $\phantom{0}5.8\pm0.1$ & $-0.11\pm0.05$ & no\\
4465 & $\phantom{0}1.0\pm0.6$ & $\phantom{0}1.6\pm0.1$ & $\phantom{-}0.39\pm0.48$ & no\\
4473 & $34.7\pm3.9$ & $38.1\pm0.8$ & $\phantom{-}0.08\pm0.09$ & no\\
4486 & $20.1\pm1.1$ & $15.5\pm0.9$ & $-0.25\pm0.06$ & yes\\
4497 & $15.0\pm0.9$ & $15.1\pm0.4$ & $\phantom{-}0.01\pm0.06$ & no\\
4552 & $15.9\pm1.1$ & $15.0\pm0.7$ & $-0.04\pm0.07$ & yes\\
4589 & $\phantom{0}8.9\pm0.6$ & $\phantom{0}9.0\pm0.2$ & $\phantom{-}0.01\pm0.06$ & no\\
4602 & $14.7\pm1.7$ & $17.7\pm0.7$ & $\phantom{-}0.15\pm0.10$ & no\\
4607 & $\phantom{0}5.9\pm0.6$ & $\phantom{0}8.2\pm0.4$ & $\phantom{-}0.27\pm0.09$ & no\\
\hline
\end{tabular}
\label{tab:spInPS}
\end{table}

Among all the observed objects, two bubbles, 3654 and 3706, were already classified as PNe \citep{Kerber2003}. These two sources appear resolved in both bands in our images.

As mentioned in Section \ref{sec:det}, not all the bubbles were detected, especially in $L$ band. It is very likely that the bubbles detected in $C$ band but not in $L$ band are characterised by positive spectral indices and also, due to the higher rms in L band, are simply below the detection limit. It is possible to estimate a minimum spectral index for each non-detected bubble assuming an upper limit for their flux density as follows: (1) for point sources in $C$ band the flux density upper limit at $20\um{cm}$ is simply assumed as three times the rms of the respective $L$-band map, (2) for extended sources the size of the source as imaged in $C$ band is reported in number of beams of the $L$-band map and the square-root of this number is multiplied by three times the map rms. Assuming pure black-body emission ($\alpha=2$), a minimum flux density at $20\um{cm}$ was also computed so that, for each bubble, it is possible to define a range of possible $L$-band flux density values. In Table \ref{tab:soloC} we provide the results of this estimate.
\begin{table}
\caption{Bubbles detected only in $C$ band. For flux densities in $L$ band a possible range is provided as described in the text.}
\begin{tabular}{cccccc}\hline
Bubble & \multicolumn{2}{c}{$S(L)$ (mJy)} & $S(C)$ & $\alpha$ & Resolved?\\\cline{2-3}
& min & max & (mJy) & & (mJy)\\\hline
3188 & 0.1 & 4.5 & $\phantom{0}1.0\pm0.2$ & $\gtrsim-1.2$ & no\\
3192 & 0.1 & 2.1 & $\phantom{0}1.2\pm0.6$ & $\gtrsim-0.4$ & no\\
3193 & 0.1 & 2.1 & $\phantom{0}1.4\pm0.6$ & $\gtrsim-0.3$ & no\\
3309 & 0.3 & 5.2 & $\phantom{0}3.4\pm0.4$ & $\gtrsim-0.3$ & yes\\
3313 & 0.4 & 6.9 & $\phantom{0}5.1\pm0.7$ & $\gtrsim-0.3$ & yes\\
3347 & 0.1 & 5.7 & $\phantom{0}1.5\pm0.3$ & $\gtrsim-1.1$ & no\\
3362 & 1.1 & 5.7 & $12.1\pm2.5$ & $\gtrsim+0.6$ & no\\
3724 & 0.3 & 9.6 & $\phantom{0}3.2\pm0.4$ & $\gtrsim-0.9$ & yes\\
3736 & 1.6 & 6.1 & $18.1\pm0.5$ & $\gtrsim+0.9$ & yes\\
4409 & 0.6 & 1.5 & $\phantom{0}7.3\pm0.1$ & $\gtrsim+1.2$ & no\\
4452 & 0.2 & 6.1 & $\phantom{0}2.1\pm0.4$ & $\gtrsim-0.8$ & yes\\
4580 & 0.2 & 3.0 & $\phantom{0}2.0\pm0.4$ & $\gtrsim-0.3$ & no\\
4584 & 0.2 & 9.2 & $\phantom{0}2.1\pm0.5$ & $\gtrsim-1.2$ & yes\\
\hline
\end{tabular}
\label{tab:soloC}
\end{table}

\section{Classification}
The determination of the radio spectral index in the previous section has allowed us to make preliminary hypotheses of the nature of the bubbles. However a multi-wavelength approach is necessary to fully characterise these objects.

In addition to MIPSGAL and GLIMPSE observations, many bubbles were detected in other IR bands, from $1.25\mic{m}$ to $160\mic{m}$.\footnote{Herschel observations detected bubbles also at longer wavelengths, but they will not be discussed in this work.} In particular we took into account data from on-line catalogues of: the 2-Micron All Sky Survey (2MASS) at $1.25\mic{m}$ ($J$ band), $1.65\mic{m}$ ($H$ band) and $2.17\mic{m}$ ($K_s$ band) \citep{Cutri2003}; the Wide-field IR Survey Explorer (WISE) at $3.4\mic{m}$, $4.6\mic{m}$, $12\mic{m}$ and $22\mic{m}$ \citep{Cutri2012}; the Midcourse Space Experiment (MSX) at $8.3\mic{m}$, $12\mic{m}$, $15\mic{m}$ and $21\mic{m}$ \citep{Egan2003}; the IR Astronomical Satellite (IRAS) at $12\mic{m}$, $25\mic{m}$ and $60\mic{m}$; the Japanese satellite AKARI at $9\mic{m}$, $18\mic{m}$, $60\mic{m}$, $90\mic{m}$, $140\mic{m}$ and $160\mic{m}$.

In the Table \ref{tab:synopt} for each bubble listed in the Table \ref{tab:spInPS}, except 3654 and 3706, a brief summary of all the available IR observations will be presented. In the last comment we report a possible classification for each bubble as reported in literature or derived in this work.

Beside the IR archive search, we also looked for possible detections in H$\alpha$ using the SuperCOSMOS H-alpha Survey (SHS; \citealt{Parker2005}). The survey detects all known PNe in Table \ref{tab:obsVLA} (except 3558 and 3654, not covered by the survey), but also bubbles 3193, 4436, 4602 and 4607. Our radio spectral index analysis has shown that these four bubbles are thermal emitters (see Tables \ref{tab:spInPS}, \ref{tab:soloC} and \ref{tab:synopt}). If we assume that the H$\alpha$ emission is a good tracer of the radio free-free continuum, the detection of these four bubbles in SHS corroborates our classification. However only the Bubble 4602 is clearly detected in H$\alpha$, while the other three nebulae appear very faint and barely visible (we cannot even exclude a fake detection). We therefore cautiously avoid a quantitative analysis in this moment.

In the following subsections, we will make use of this information to attempt a classification of the bubbles whose nature is still uncertain.

\begin{landscape}
\begin{table}
\caption{Synoptic table of IR observations. Legend: `C' only central source, `N' only diffuse emission, `B' both central source and diffuse emission, `P' point source due to low resolution, `--' no source detected. In the last column the `?' indicates a candidate while `\textit{RadTh}' that we can only state that we are observing a radio thermal emitter.}
\begin{tabular}{cccccccl}\hline
Bubble & 2MASS & WISE & IRAC\footnote{From GLIMPSE.} & MSX & IRAS & AKARI & Comments\\
& $J$/$H$/$K_s$ & [3.4]/[4.6]/[12]/[22] & [3.6]/[4.5]/[5.8]/[8] & [8.3]/[12]/[15]/[21] & [12]/[25]/[60] & [9]/[18]/[65]/[90]/[140]/[160]\\\hline
3222 & --/C/C & C/C/N/N & C/C/B/B & P/P/P/P & --/P/-- & --/--/--/--/-- & PN? \citep{Urquhart2009}\\
3333 & --/--/-- & --/--/--/N & --/--/--/-- & --/--/--/-- & --/--/-- & --/--/--/--/-- & \textit{RadTh} \textbf{(This work)}\\
3354 & --/--/-- & --/--/N/N & --/--/N/N & --/--/--/-- & N/--/P & --/--/P/--/P/P & H \textsc{ii} region? \citep{Anderson2011}\\ 
3367 & --/C/C & C/C/N/N & C/C/C/N & --/--/--/-- & --/--/-- & --/P/--/--/--/-- & PN? \textbf{(This work)}\\
3438 & C/C/C & C/C/C/N & C/C/C/C & P/P/P/P & P/P/-- & P/--/--/--/--/-- & \textit{RadTh} \textbf{(This work)}\\

3448 & C/C/C & C/C/N/N & C/C/C/N & --/--/--/-- & --/P/P & --/P/--/--/--/-- & PN? \citep{Gvaramadze2010}\\
3866 & --/--/-- & --/--/--/-- & --/--/--/-- & --/--/--/-- & --/--/-- & --/--/--/--/--/-- & PN? \citep{Anderson2011}\\
4436 & --/--/-- & --/--/N/N & --/--/--/-- & --/--/--/-- & --/P/P & --/P/--/--/--/-- & PN? \textbf{(This work)}\\
4465 & --/--/-- & --/--/N/N & --/--/--/-- & --/--/--/-- & --/--/-- & --/--/--/--/--/-- & \textit{RadTh} \textbf{(This work)}\\
4473 & --/--/-- & --/N/N/N & --/N/N/N & --/--/--/--  & --/P/-- & --/P/--/--/--/-- & PN? \textbf{(This work)}\\
4486 & --/--/-- & --/--/N/N & --/--/--/-- & --/--/--/-- & --/--/-- & --/--/--/--/--/--\\
4497 & --/--/-- & --/--/N/N & --/--/--/-- & --/--/--/-- & --/--/-- & --/--/--/--/--/-- & \textit{RadTh} \textbf{(This work)}\\
4552 & --/--/-- & --/--/N/N & --/--/--/-- & --/--/--/-- & --/--/-- & --/--/--/--/--/-- & \textit{RadTh} \textbf{(This work)}\\
4589 & --/--/-- & --/--/N/N & --/--/--/-- & --/--/--/-- & --/--/-- & --/P/--/--/--/-- & \textit{RadTh} \textbf{(This work)}\\
4602 & --/--/-- & N/N/N/N & N/N/N/N & P/--/P/P & --/P/P & --/P/--/P/P/-- & PN? \citep{Kohoutek2001}\\
4607 & --/--/-- & --/--/N/N & --/--/--/N & --/--/--/-- & --/--/-- & --/--/--/--/--/-- & \textit{RadTh} \textbf{(This work)}\\\hline
\end{tabular}
\label{tab:synopt}
\end{table}

\end{landscape}

\subsection{Radio emission characterization}
\label{sec:radiochar}
In Section \ref{sec:detsrc} we discussed the derivation of the radio spectral index between $20\um{cm}$ and $6\um{cm}$ for all those bubbles whose flux density is well determined. We found that most of the bubbles have a positive or slightly negative spectral index, indicating that we are very likely observing thermal free-free emission typically in optically thick regime, with a large amount of sources presenting a spectral index of 0. This behaviour was somehow expected, since the majority of the already classified bubbles are PNe (see Table \ref{tab:obsVLA}). Furthermore also other kinds of evolved stars (such as LBV or WR) are characterized by a radio free-free emission, with only SNR showing clear non-thermal features.

\begin{figure*}
\begin{center}
\includegraphics[width=10cm]{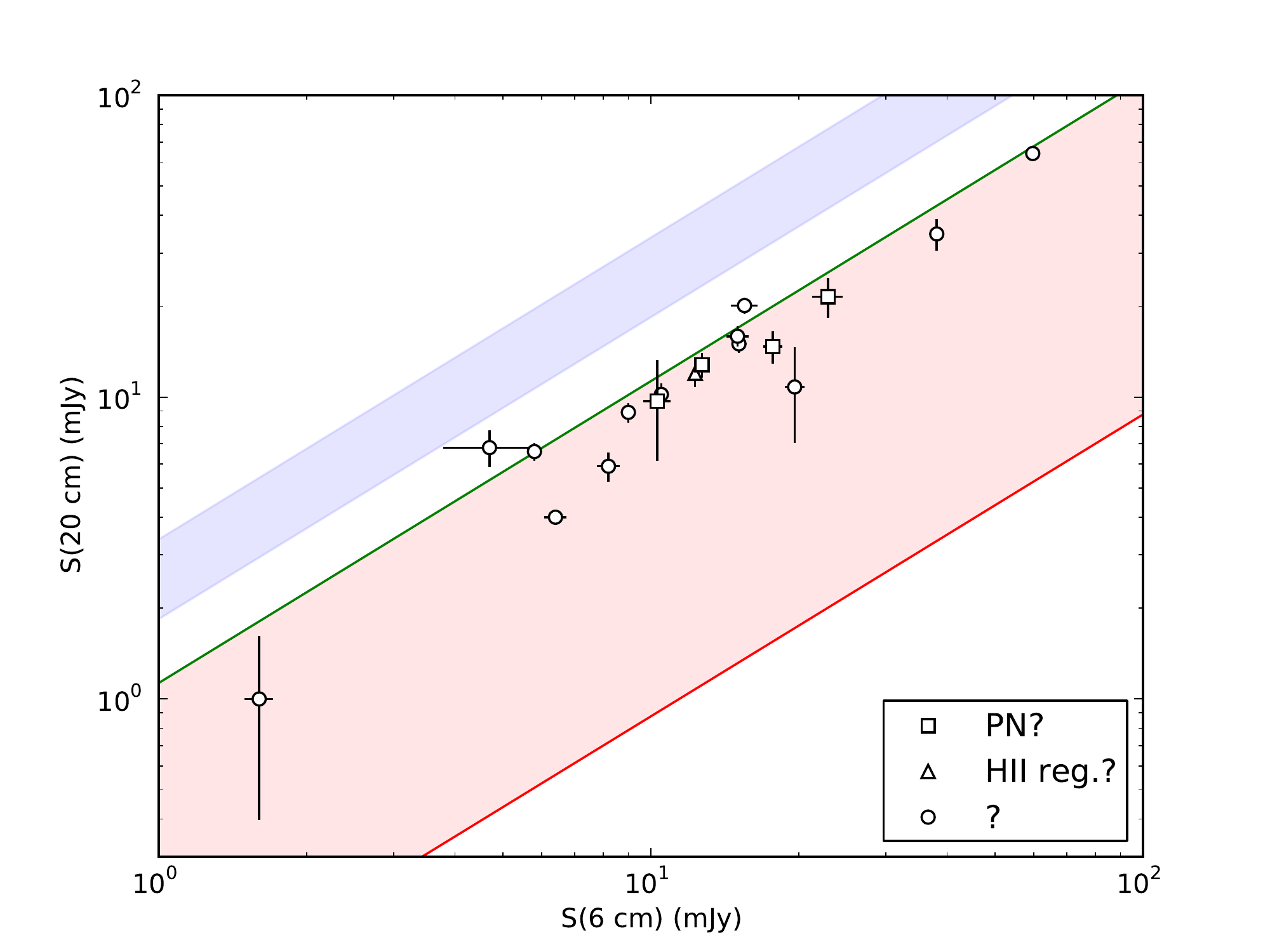}
\caption{Flux densities comparison at $20\um{cm}$ and at $6\um{cm}$. The red (lower) area delimits the range of expected values for free-free emission, with the red line (bottom) representing a pure black-body emission ($\alpha=2$) and the green (top) a pure optically thin free-free emission ($\alpha=-0.1$). The blue (upper) area delimited spectral indices between $-0.5$ and $-1$, typical of an optically thin synchrotron emission. Noticeably, the majority of the points lie close to the green line.}
\label{fig:radio_color}
\end{center}
\end{figure*}

For 5 bubbles, a potential classification is available from the literature, according to which 4 are PNe candidate (denoted as squares in Figure \ref{fig:radio_color}) and 1 is a H \textsc{ii} region candidate (denoted as triangles in Figure \ref{fig:radio_color}). For these sources, the spectral index derived from our analysis is consistent with the existing classification.

Two sources, i.e. Bubbles 3367 and 4486, are characterized by rather negative spectral index values. Their spectral indices were estimated as $-0.30$ and $-0.25$ respectively, values too low to be ascribed to pure free-free emission. However, the errors associated with these measurements are significant, so the thermal emission hypothesis cannot be entirely ruled out.

\subsection{Relation between radio and MIPS 24 micron emission}
\label{sec:radioMIPSGAL}
The emission at $24\mic{m}$ and $6\um{cm}$ have a different origins. In fact, the emission at $24\mic{m}$ originates both from warm thermal dust emission, and from gas forbidden lines, such as [O \textsc{iv}] at $25.89\mic{m}$ \citep{Flagey2011}. The radio emission at $6\um{cm}$, instead, originates from either thermal free-free emission or synchrotron emission. However it was shown by several authors that a strong correlation between mid-/far-IR and radio emission exists (\citealt{deJong1985}; \citealt{Helou1985}; \citealt{Pinheiro2011}).

\begin{figure*}
\begin{center}
\includegraphics[width=10cm]{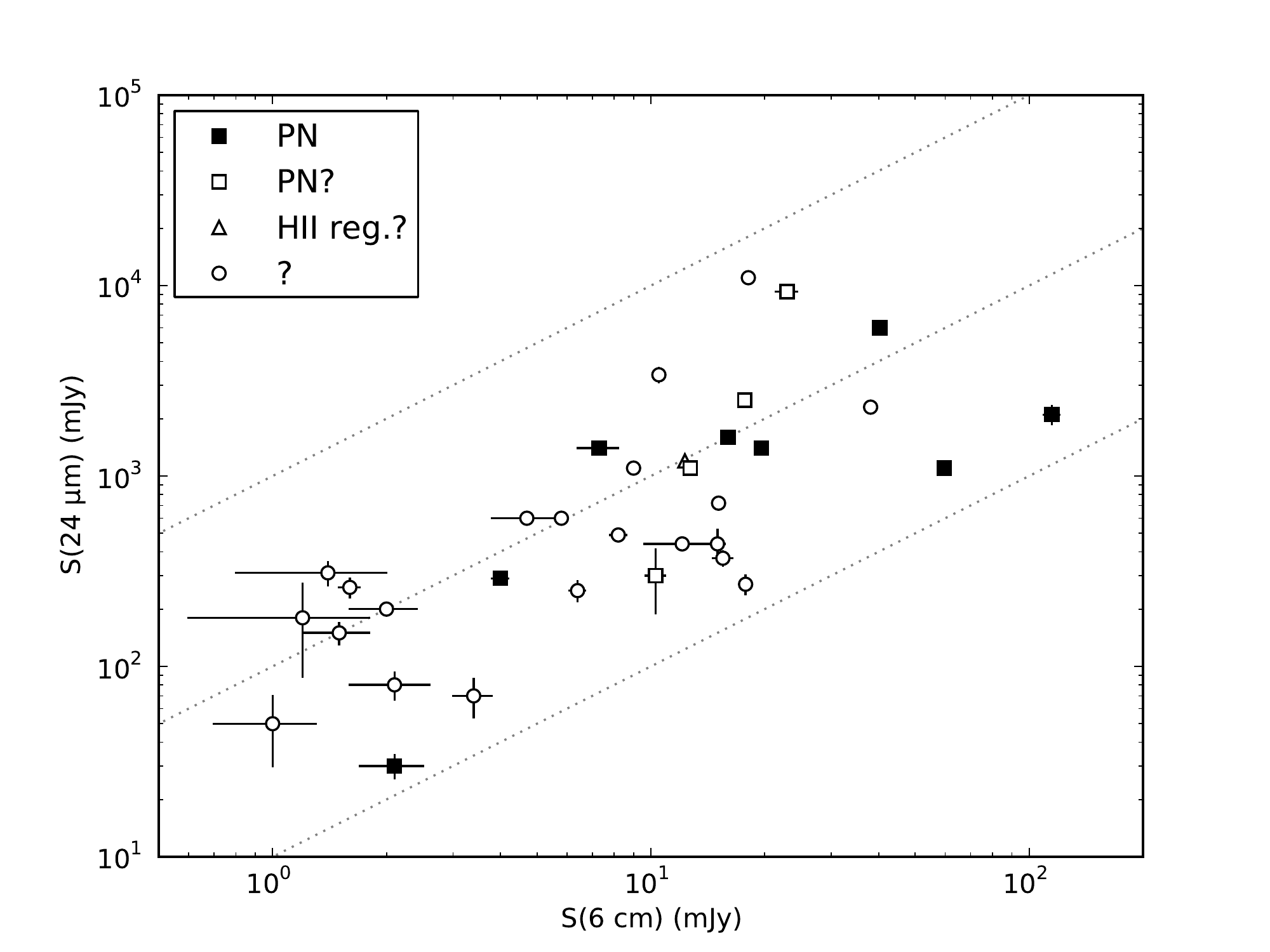}
\caption{Correlation between MIPSGAL flux densities at $24\mic{m}$ and radio data at $6\um{cm}$ from our EVLA observations. The grey dotted lines represent flux density ratios of 10, 100 and 1000.}
\label{fig:MIra}
\end{center}
\end{figure*}

In Figure \ref{fig:MIra} the flux density at $24\mic{m}$ from MIPSGAL plotted against the flux density at $6\um{cm}$ from our observations (Table \ref{tab:fluxC}), for all the bubbles with measured 6-cm flux density with the only exception of the Bubble 3313 (see below). The figure evidences a clear correlation between the emission in the two bands. If we defined for each bubble the quantity
\begin{equation}
q=\log\frac{S_{\mathrm{IR}}}{S_{\mathrm{ra}}}
\end{equation}
we find that $\overline{q}=1.9\pm0.4$, where the error is computed as the standard deviation of the distribution. A linear fit to the ensemble of the $\log S_\mathrm{IR}$ vs. $\log S_\mathrm{ra}$ values retrieves
\begin{equation}
\log S_{\mathrm{IR}}=0.9\log S_{\mathrm{ra}}+2.0
\end{equation}
from which, despite the small size of the sample, it is clear that the relation is almost perfectly linear (0.9 instead of 1), therefore the mean value $\overline{q}$ is a good representation of $\log(S_\mathrm{IR}/S_\mathrm{ra})$.

Bubble 3313 has a much higher $S_\mathrm{IR}/S_\mathrm{ra}$ value ($\sim\!4000$) with respect to the rest of the sample. At $24\mic{m}$ this source appears very extended (about $80''$) and might be interacting with Bubble 3312 (\citealt{Gvaramadze2010}; \citealt{Wachter2010}). Spectroscopic near-IR studies of the central sources of these two bubbles reveal that both can be classified as WR stars of the same spectral type WN9h \citep{Burgemeister2013}. Our radio observations at $6\um{cm}$ show a very faint irregular nebula around the central star of Bubble 3313, less extended than the 24-$\umu$m nebula, with no emission around the other bubble or in any other region where the 24-$\umu$m emission is present (see Figure A17 in appendix A). Despite the fact that this bubble is detected in the MAGPIS 20-cm tile, no emission is visible from our maps at $20\um{cm}$. Indeed it is possible that the extended emission  is below our detection limit (especially at $20\um{cm}$) and/or that it was resolved out (especially at $6\um{cm}$). For these reasons the flux density computation is not considered reliable enough and the bubble was not included in this part of the analysis.

Although the emission at $24\mic{m}$ is well correlated with the emission at $6\um{cm}$, we cannot use this effect to classify our sources. For example, if we compute $\overline{q}$ for 8 known PNe, we find a value of $1.7\pm0.4$ which 
is consistent with the value for the whole sample.

\subsection{Relation between radio and IRAS 25 micron emission}
\label{sec:radioIRAS}
Combining our radio observations with IRAS archive data it is possible to discriminate whether a source is a PN candidate or not. Although IRAS poor resolution did not allow us to resolve individual PNe, its sensitivity was enough to detect these objects at least at the distance of the galactic center \citep{Pottasch1988}.

Unfortunately, only few bubbles studied here have archival IRAS fluxes, and none has a flux density determination in more than two bands. Using the IRAS Point Source Catalogue and archival VLA 6-cm data \citep{Becker1994} for a sample 
of known PNe and H \textsc{ii} regions, we were able to generate color plots useful for our classification purposes.

\begin{figure*}
\begin{center}
\includegraphics[width=10cm]{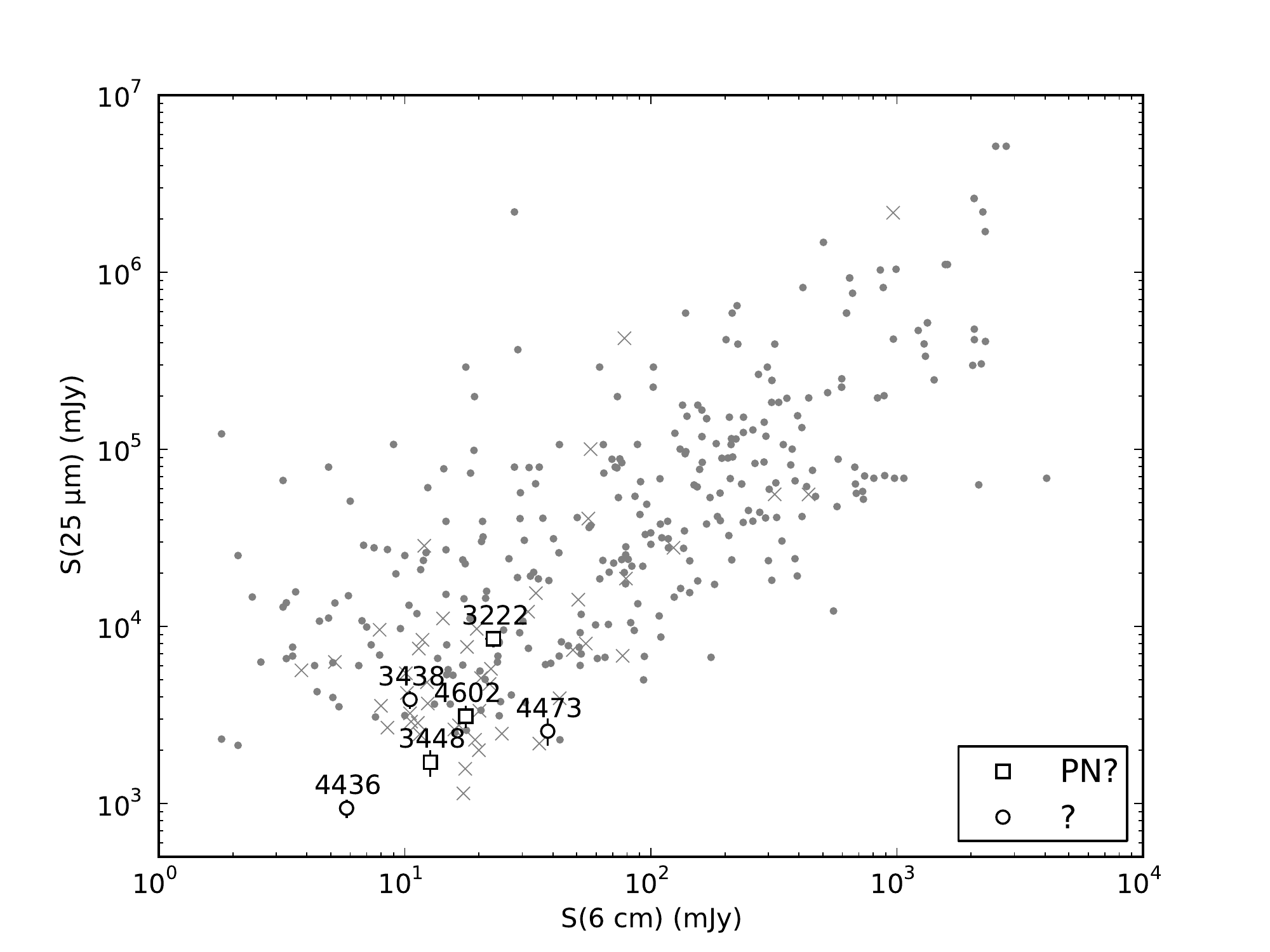}
\caption{Correlation between IRAS flux densities at $25\mic{m}$ and radio data at $6\um{cm}$. Small crosses are archive PNe and small points archive H \textsc{ii} regions; larger markers represent our bubbles with IRAS archive values and our radio data. It is possible to notice that the two flux densities are well-correlated and that the PNe are usually characterized by lower flux density values.}
\label{fig:I25ra}
\end{center}
\end{figure*}

As a first step, it is important to notice that, following the discussion in Section \ref{sec:radioMIPSGAL}, the IRAS flux densities at $25\mic{m}$ are well-correlated with the radio flux densities at $6\um{cm}$ (Figure \ref{fig:I25ra}).
This plot is quite similar to Figure \ref{fig:MIra}. It is however interesting to notice how the two plots span a different range of values in flux parameter space, with the MIPSGAL and EVLA observations extending the coverage towards lower flux densities. We also notice that, though PNe and H \textsc{ii} regions partly overlap in Figure \ref{fig:I25ra}, H \textsc{ii} regions become dominant at very high flux densities. All our 6 bubbles, for which both flux density values are available, are located in the lower-left region of the plot, so they are all compatible both with PNe and H \textsc{ii} regions.

\begin{figure*}
\begin{center}
\includegraphics[width=10cm]{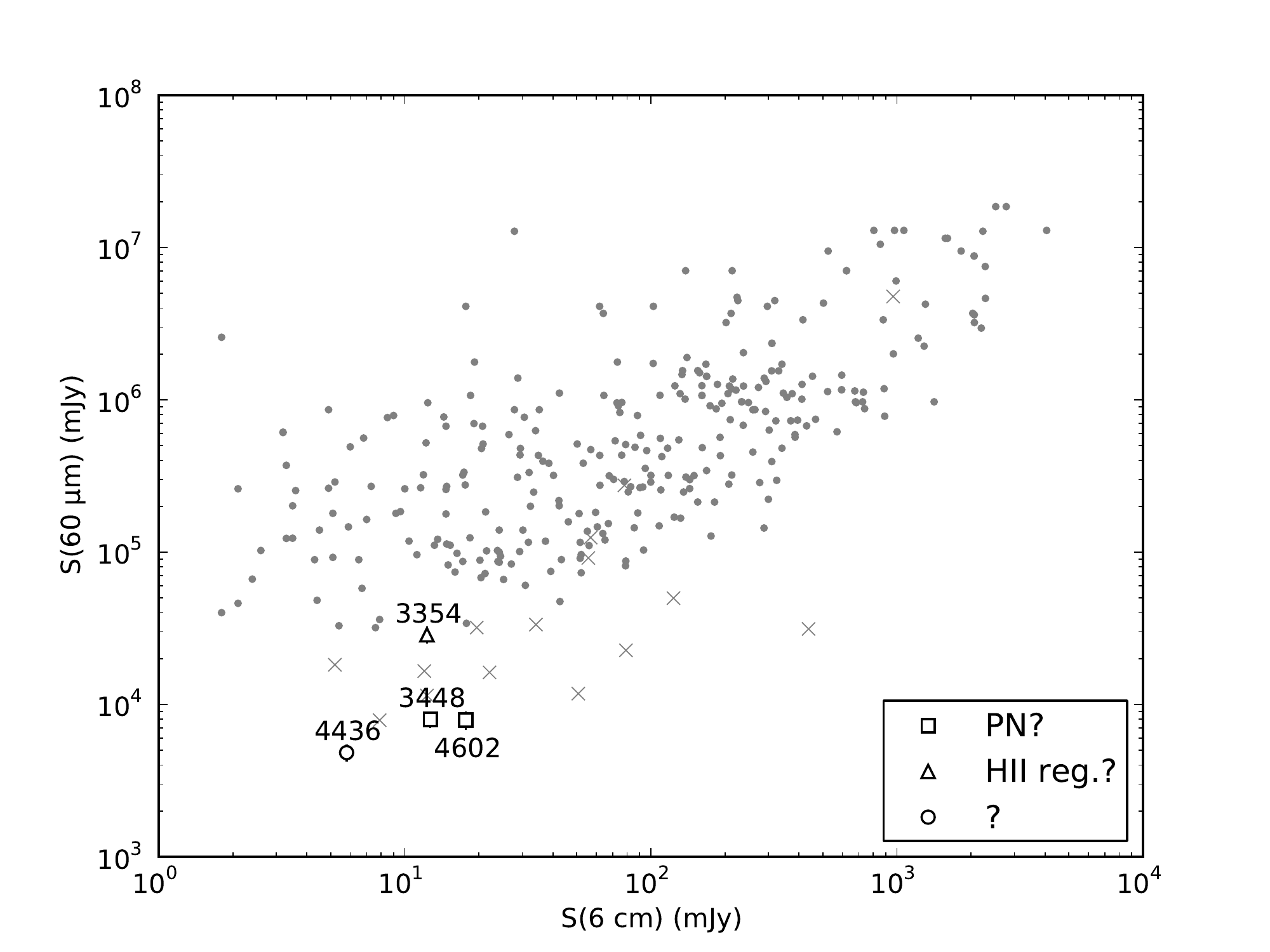}
\caption{Correlation between IRAS flux densities at $60\mic{m}$ and radio data at $6\um{cm}$. Small crosses are archive PNe and small points archive H \textsc{ii} regions; larger markers represent our bubbles with IRAS archive values and our radio data. It is possible to notice that the PN population is characterized by a lower value of the two flux density ratio and is well-separated from the H \textsc{ii} regions.}
\label{fig:I60ra}
\end{center}
\end{figure*}

\begin{figure*}
\begin{center}
\includegraphics[width=10cm]{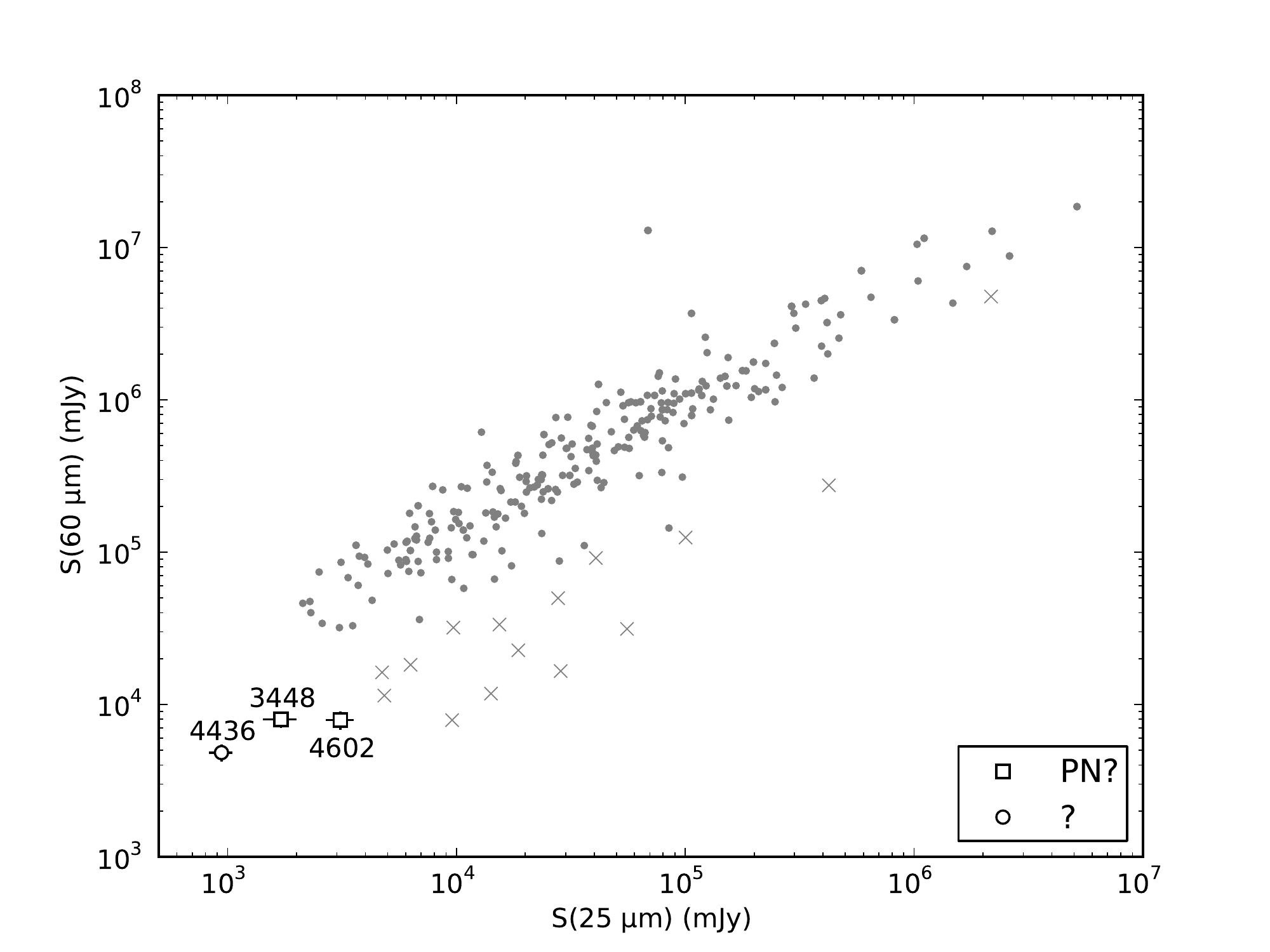}
\caption{Correlation between IRAS flux densities at $60\mic{m}$ and at $25\mic{m}$. Small crosses are archive PNe and small points archive H \textsc{ii} regions; larger markers represent our bubbles with IRAS archive values. Also in this plot it is possible to notice that the PN population is characterized by a lower value of the two flux density ratio and is well-separated from the H \textsc{ii} regions.}
\label{fig:I60I25}
\end{center}
\end{figure*}

A more interesting result can be obtained by plotting the IRAS flux density values at $60\mic{m}$ against the radio flux densities at $6\um{cm}$ (Figure \ref{fig:I60ra}).
In this plot it is still evident how IR and radio flux density values correlate but it is also possible to notice how PNe represent a population clearly separated from other H \textsc{ii} regions (despite some exceptions). From this plot, we might be tempted to classify Bubble 4436 as a PN candidate. However, this hypothesis is not supported by the distribution 
of IRAS 60 micron vs. 25 micron fluxes (Figure \ref{fig:I60I25}).
In this case, PNe still occupy a well-defined and separate region of space with respect to H \textsc{ii} regions, but bubbles and PNe do not share the same region in the plot, with bubbles having a much lower flux density than both PNe and H \textsc{ii} regions. Indeed, their low surface brightness is likely the reason these sources were not detected by the IRAS survey. Therefore, it is difficult to say which classification is more appropriate for Bubble 4436, given its outlier behaviour when compared to already classified objects.

Using all IRAS bands combined with 6-cm data, we also generated color-color diagrams. However, none of them was useful for our classification attempt, since no particular trend was observed.

\subsection{The importance of GLIMPSE data}
\label{sec:claGLIMPSE}
As we discussed in the introduction, one of the characteristic of bubbles is that they are mostly detected only at $24\mic{m}$. The GLIMPSE survey, in fact, failed to detect extended emission for the majority of the bubbles, despite the great sensitivity of IRAC. However, in seven cases, a faint nebular emission appears in the GLIMPSE data and for five of these we performed aperture photometry using the Aperture Photometry Tool\footnote{http://www.aperturephotometry.org}. For Bubbles 3222 and 4607 it was impossible to derive a reliable flux density: in fact the first nebula is very small and dominated by its central source while the second is faint and immersed in a confused fore- and background. To this end, we subtracted foreground point-sources,  performed an interpolation of the empty pixels using the information from the surrounding background, and then estimated the sky background as the median value of a sufficiently large region in proximity of the source. In addition to aperture photometry, when the central source is visible within the bubble, we extracted  point-source photometry from the online GLIMPSE catalogue.

Information on the nature of a source detected in all IRAC bands come directly from the 3-color image obtained by superposition  of the monochromatic maps at $8\mic{m}$, $5.8\mic{m}$ and one among the other two bands. As discussed in \citet{Murphy2010}, PNe usually appear red, while H \textsc{ii} regions appear either yellow or white. This is due, for H \textsc{ii} regions, to PAH emission (yellow) or broad-band thermal emission by dust (white) \citep{Cohen2011}. An inspection of the GLIMPSE 3-color images for Bubbles 3367, 3448, 4473 and 4602, reveals a red color for all of them. Of these, two, namely 3448 and 4602, are classified in the literature as PN candidates (\citealt{Kohoutek2001}; \citealt{Gvaramadze2010}), while nothing is found about the nature of the other two. From what emerges from this discussion and follows in the next section, it can be concluded that Bubbles 3367 and 4473 could also be considered PN candidates. It is remarkable, in particular, how Bubble 4473 morphologically resembles Bubble 4602 in the GLIMPSE images. On the other hand, Bubble 3354, classified as H \textsc{ii} by \citet{Anderson2011}, shows the expected yellow appearance.
\begin{figure*}
\begin{center}
\includegraphics[width=5cm]{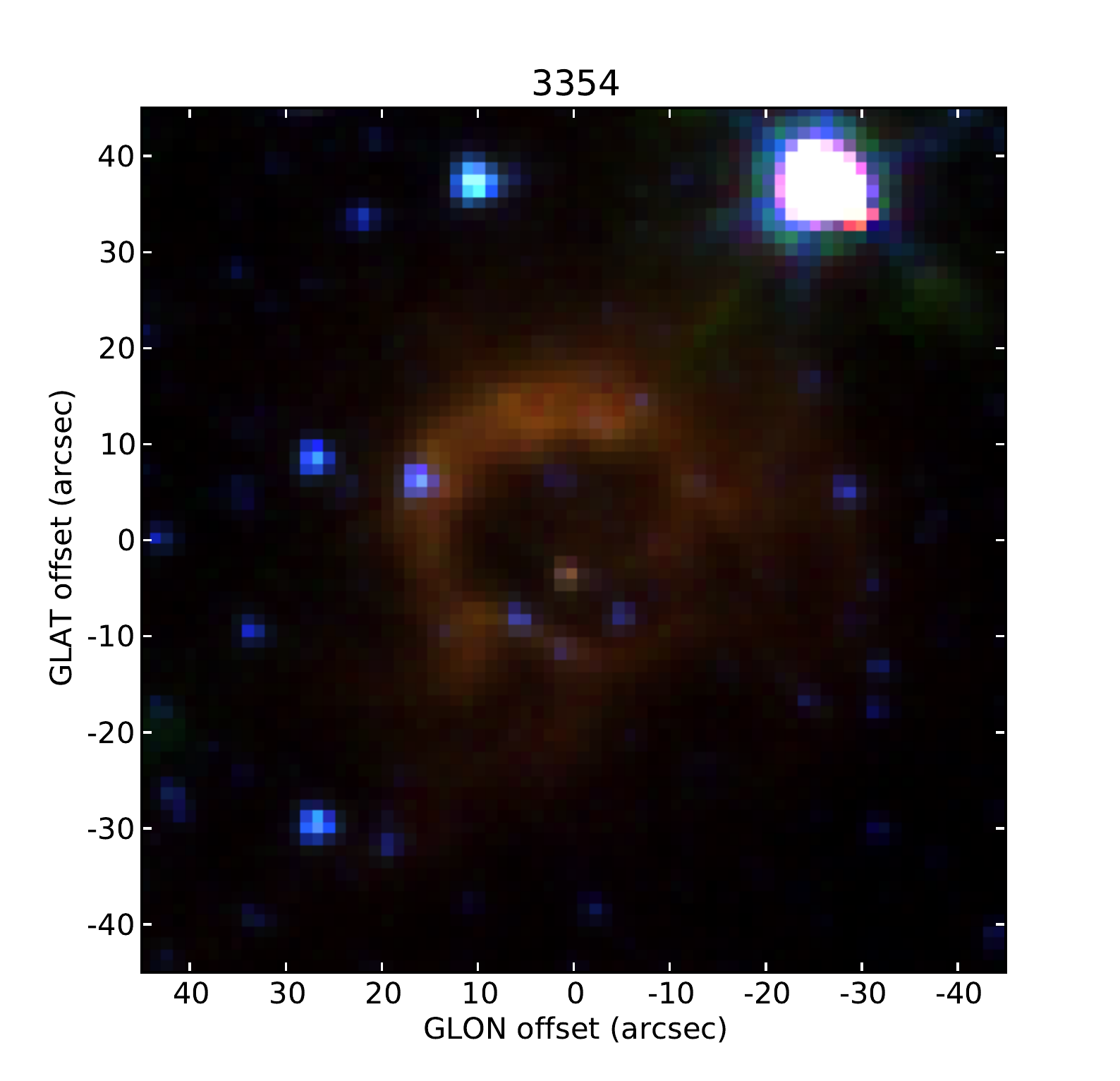}
\includegraphics[width=5cm]{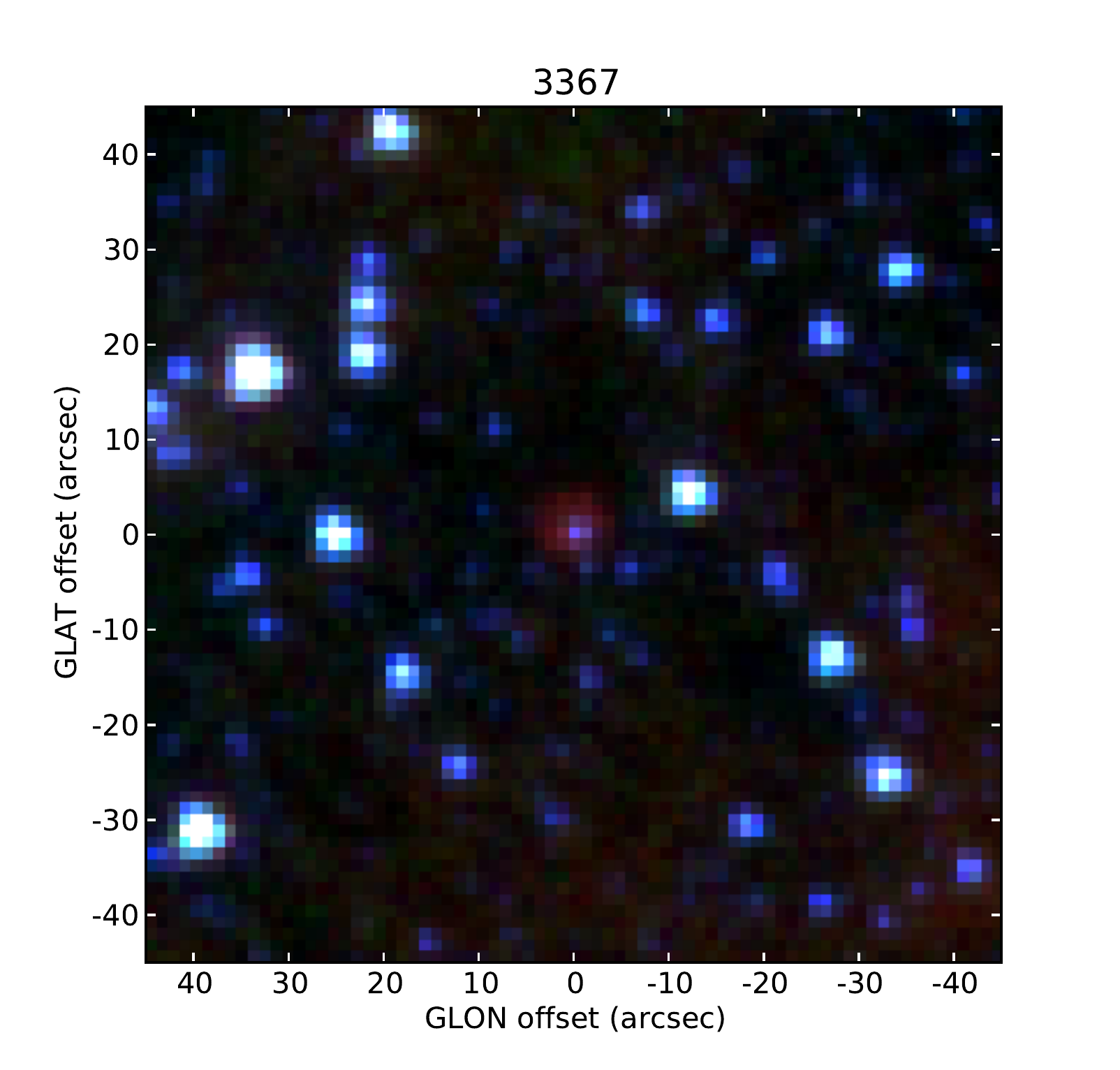}
\includegraphics[width=5cm]{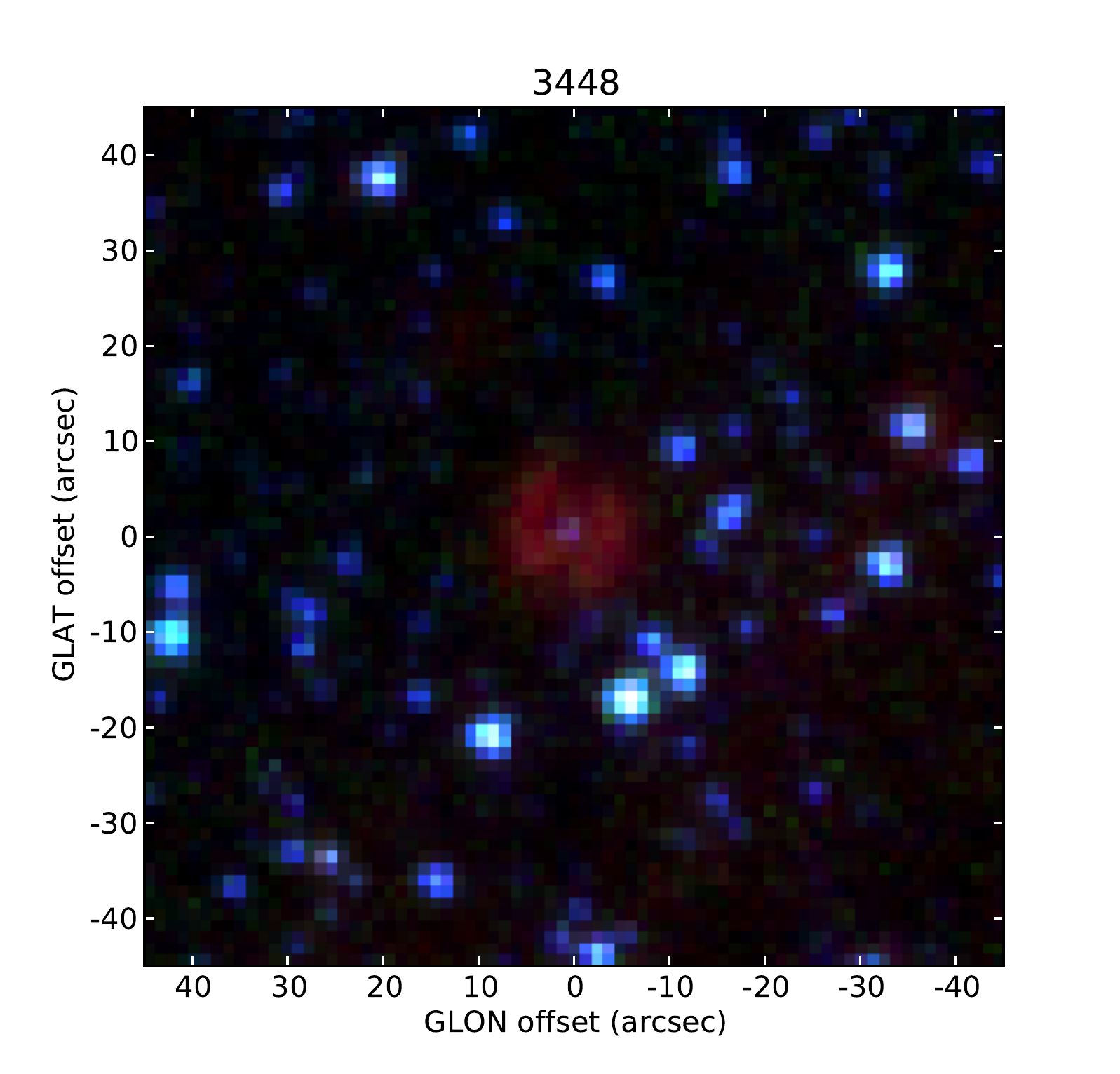}\\
\includegraphics[width=5cm]{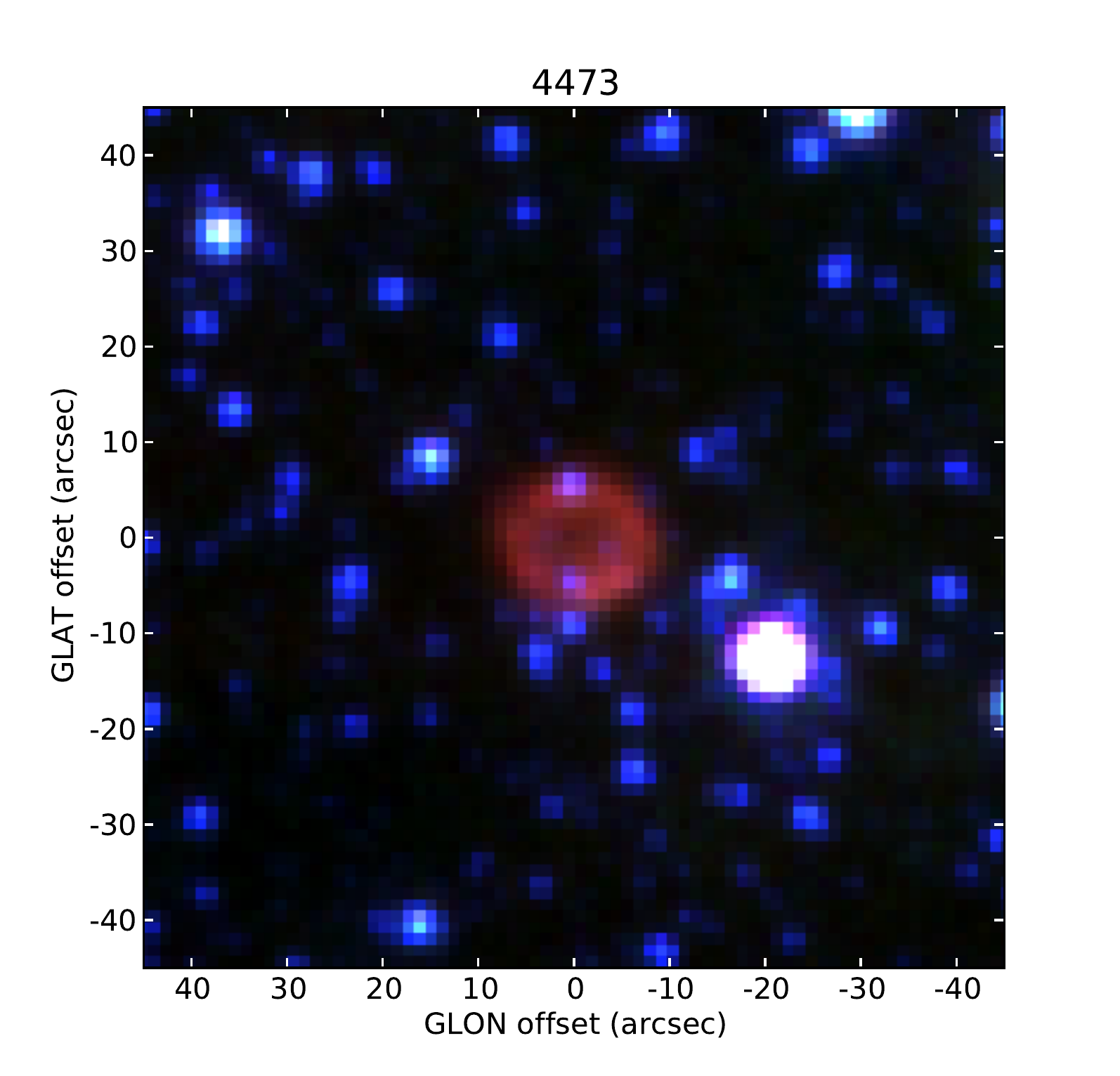}
\includegraphics[width=5cm]{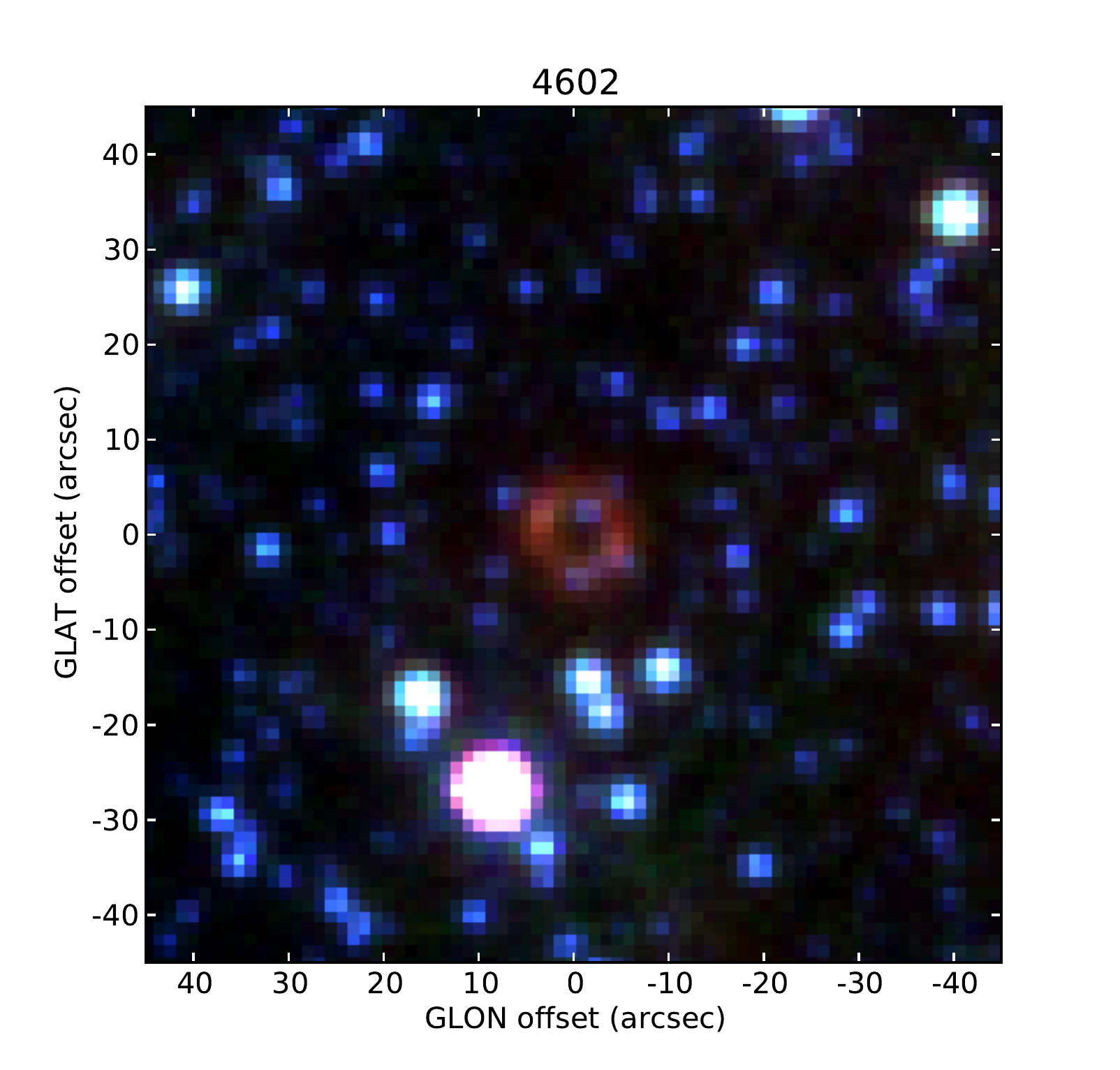}
\caption{Three-color superposition of GLIMPSE tile cut-outs at $3.6\mic{m}$ (blue), $5.8\mic{m}$ (green) and $8\mic{m}$ (red) for Bubbles 3354, 3367, 3448, 4473 and 4602. It is remarkable how Bubble 3354 appears different in shape and color with respect to the others and how 4473 and 4602 are morphologically and chromatically similar.}
\label{fig:GLIMPSEimg}
\end{center}
\end{figure*}

\begin{figure*}
\begin{center}
\includegraphics[width=10cm]{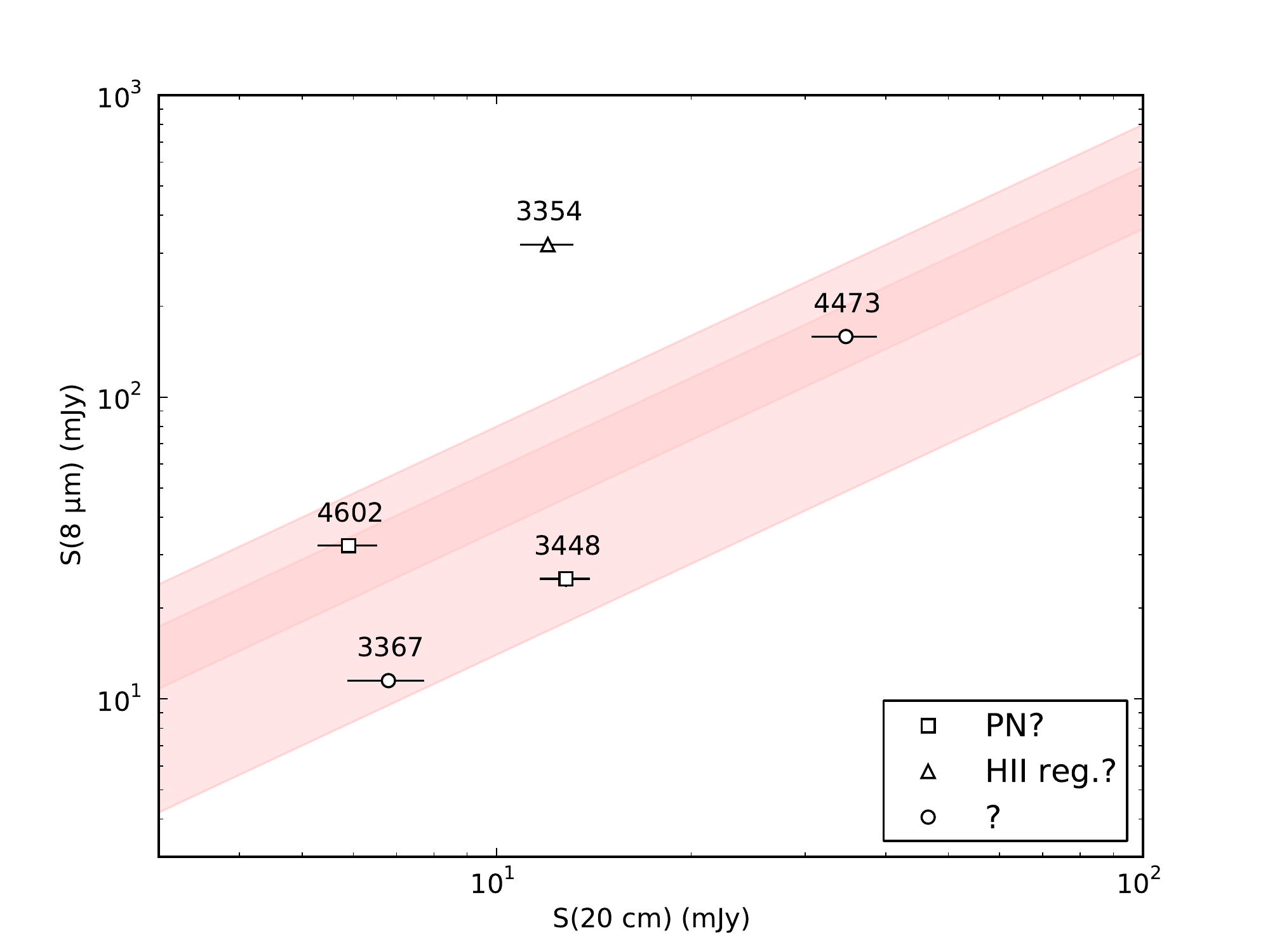}
\caption{GLIMPSE flux densities at $8\mic{m}$ against radio flux densities at $20\um{cm}$ derived from our observations. The coloured area represents the ratio interval where PNe are usually located according to \citet{Cohen2011}, with a confidence level of $1\sigma$ (darker area) and $3\sigma$ (lighter area).}
\label{fig:I8R20}
\end{center}
\end{figure*}

All the 5 bubbles considered show a nebular emission at $8\mic{m}$, while for only one (Bubble 4602) this nebular emission is detected in all four bands. It was shown that the ratio between the flux density at $8\mic{m}$ and at $20\um{cm}$ ranges in a well-determined interval and that different kinds of PNe are characterized by different values of this ratio \citep{Cohen2011}. In Figure \ref{fig:I8R20} we plot the GLIMPSE flux densities against the radio values from our data.
It is possible to notice how Bubble 3354 clearly does not satisfy the selection criterion, in agreement with a classification as an H \textsc{ii} region and not a PN. The other 4 bubbles all lie inside the area where PNe should be found. In particular the unclassified Bubble 4473 is very close to the median ratio value of 4.7, with a calculated ratio of 4.5. These 4 bubbles can be divided in two groups, according to their ratio value: the first group comprises Bubbles 4473 and 4602 and the second group Bubbles 3367 and 3448. We have already talked about the morphological similarities of Bubbles 4473 and 4602: the result found here may suggest that these two objects could share many of their physical characteristics. The other two bubbles appear different from the first two. Indeed, for Bubble 3367 no morphological consideration can be done, while Bubble 3448 seems to have bipolar structure. If all these bubbles will be confirmed to be PNe their morphological and physical differences may be due to intrinsic properties or their evolutionary stage.

\section{Summary and conclusions}
The classification of bubbles is very complicated and a definitive answer on this topic is far from being given here. However, from this analysis it has clearly emerged that the multi-wavelength approach that we presented is a powerful tool for achieving a sensible classification.

For at least 21 bubbles, previously unclassified, the spectral index analysis suggests that they are thermal free-free emitters.

Important results have been obtained when our radio data have been combined with archival data from IR observation with Spitzer and IRAS. We have shown that correlation and color-color plots can help to discriminate among different 
types of objects.

A word of caution is necessary concerning the IR-radio correlation. Although we have demonstrated that such a correlation 
-- which is known to characterize various classes of astronomical objects -- holds true also for Galactic bubbles, yet it cannot be used alone for classification purposes.

We have discussed the morphology of the bubbles at different wavelengths, considering a peculiar shape as indicative of some kind of circumstellar envelope. These considerations are applicable only to few sources. Indeed, many bubbles are barely resolved and their lack of significant feature may be both an intrinsic property or an instrumental limit.

\section*{Acknowledgments}
This work is based on observations made with the Very Large Array of the National Radio Astronomy Observatory, a facility of the National Science Foundation operated under cooperative agreement by Associated Universities Inc., and on data products from the \textit{Spitzer Space Telescope}, which is operated by the Jet Propulsion Laboratory, California Institute of Technology under a contract with NASA. Archive search made use of the SIMBAD database and the VizieR catalogue access tool, operated by the Centre de Donn\'ees astronomique de Strasbourg.

\appendix
\section{Images}
In this appendix we show radio contour plots from our data at $6\um{cm}$ and $20\um{cm}$ for all the bubbles listed in Table \ref{tab:synopt}. For eight of them we also present a superposition of MIPSGAL $24\mic{m}$ and radio contour at $6\um{cm}$. The poor resolution of radio observations, along with a very elongated beam in some cases, only allows us to state that the radio emission is usually co-spatial with the IR, with the important exception of Bubble 3354.

\newpage
\begin{figure*}
\begin{center}
\includegraphics[width=7cm]{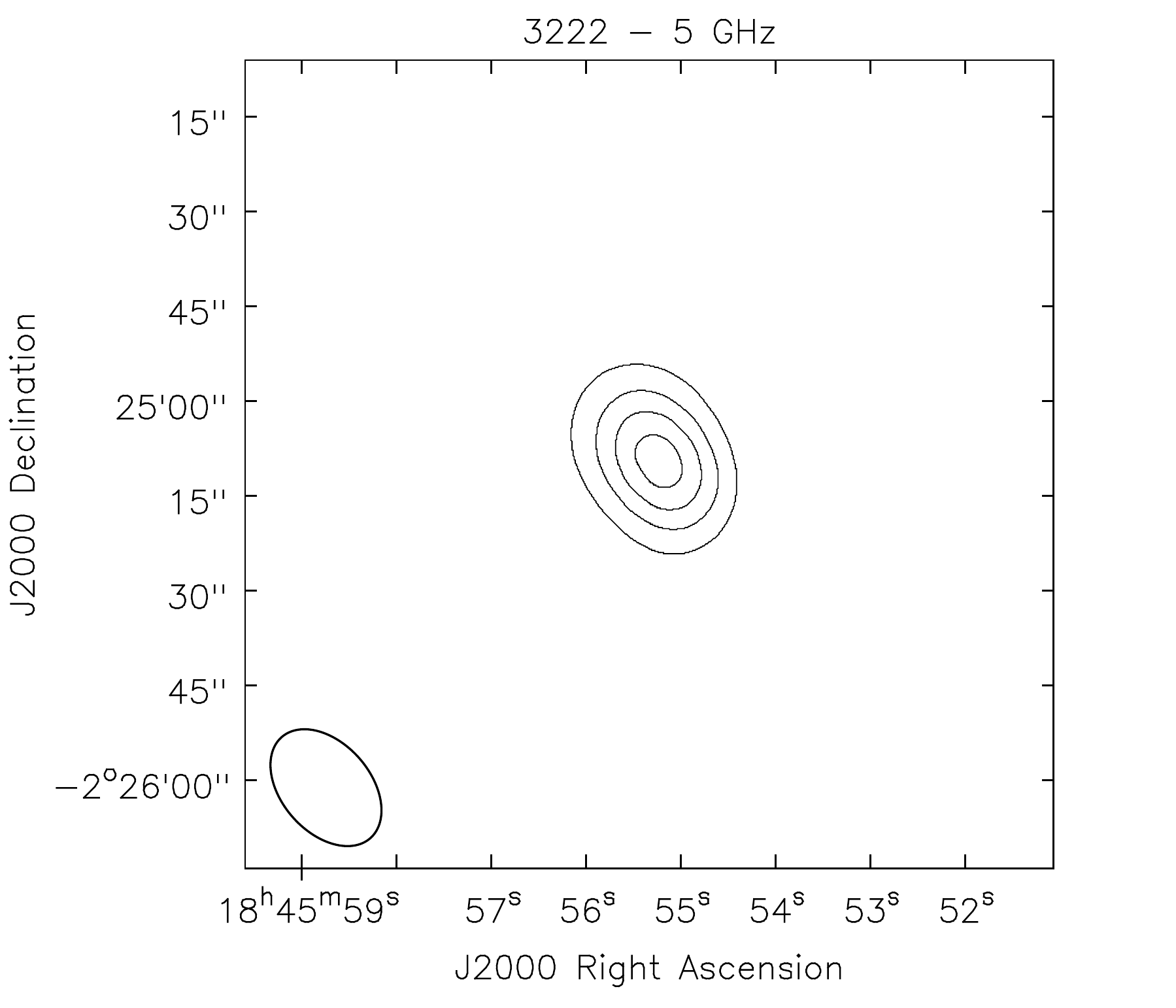}
\includegraphics[width=7cm]{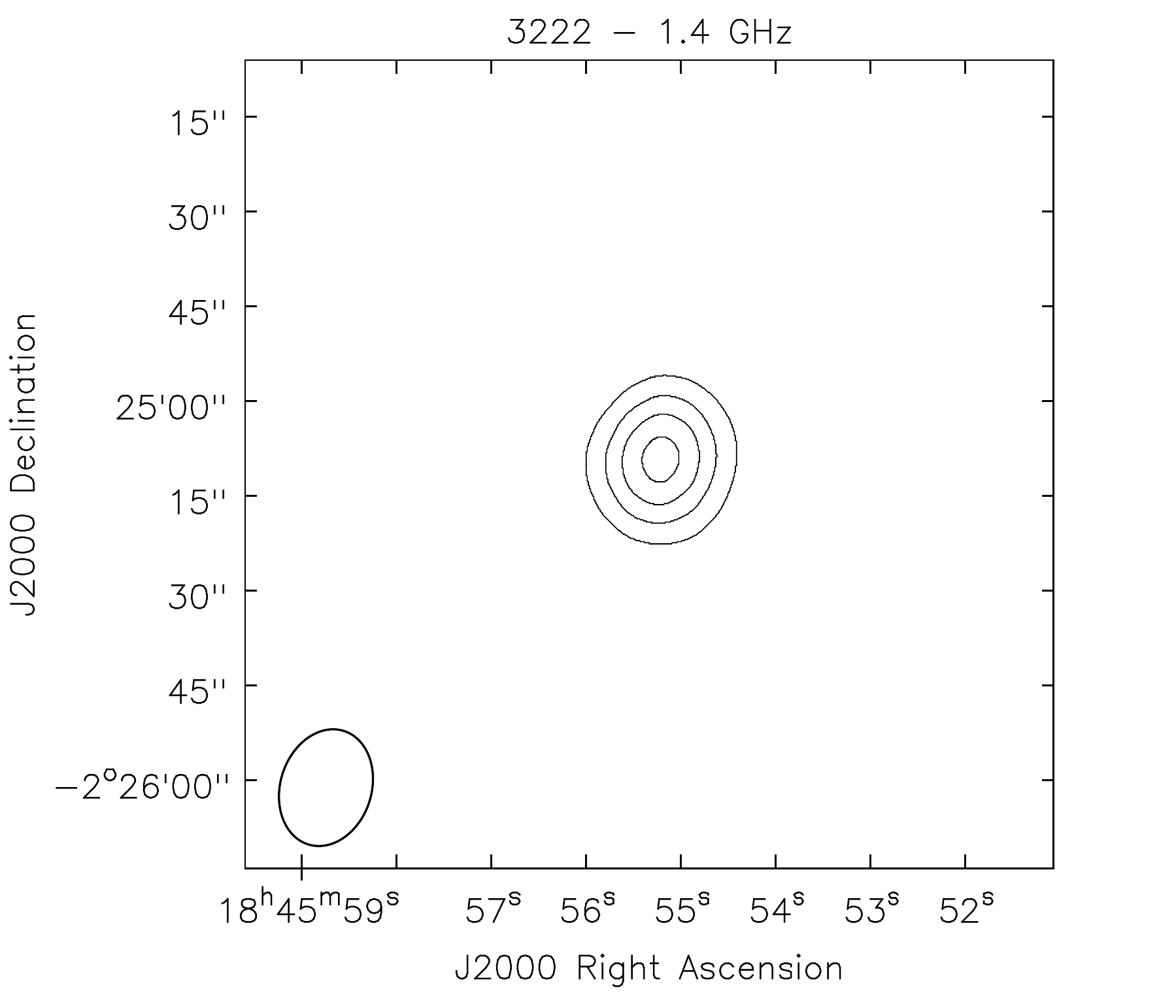}
\caption{Radio contours are 10, 15, 20 and $25\um{mJy/beam}$ (left) and 5, 10, 15, and $20\um{mJy/beam}$ (right).}
\label{fig:3222}
\end{center}
\end{figure*}

\begin{figure*}
\begin{center}
\includegraphics[width=7cm]{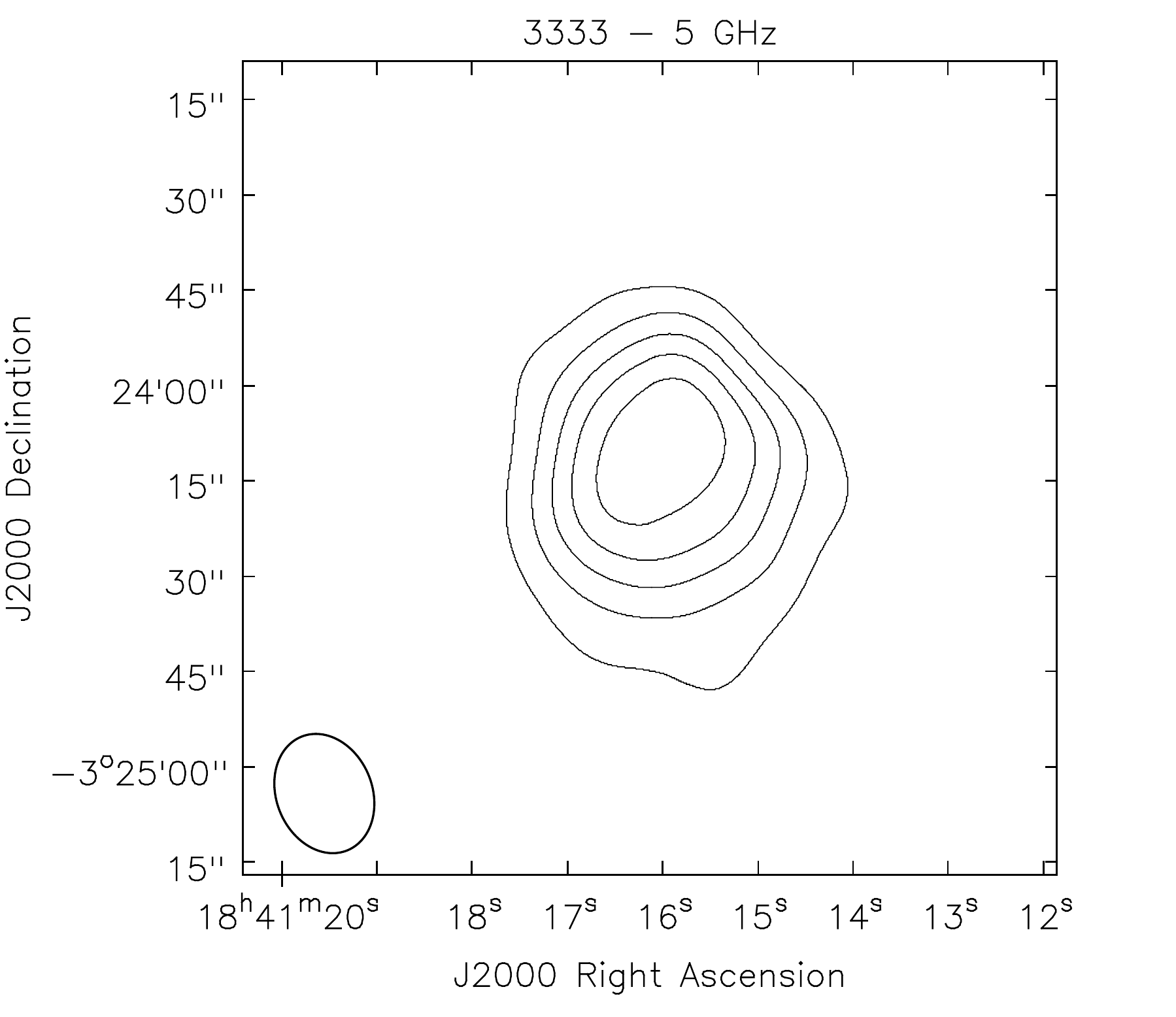}
\includegraphics[width=7cm]{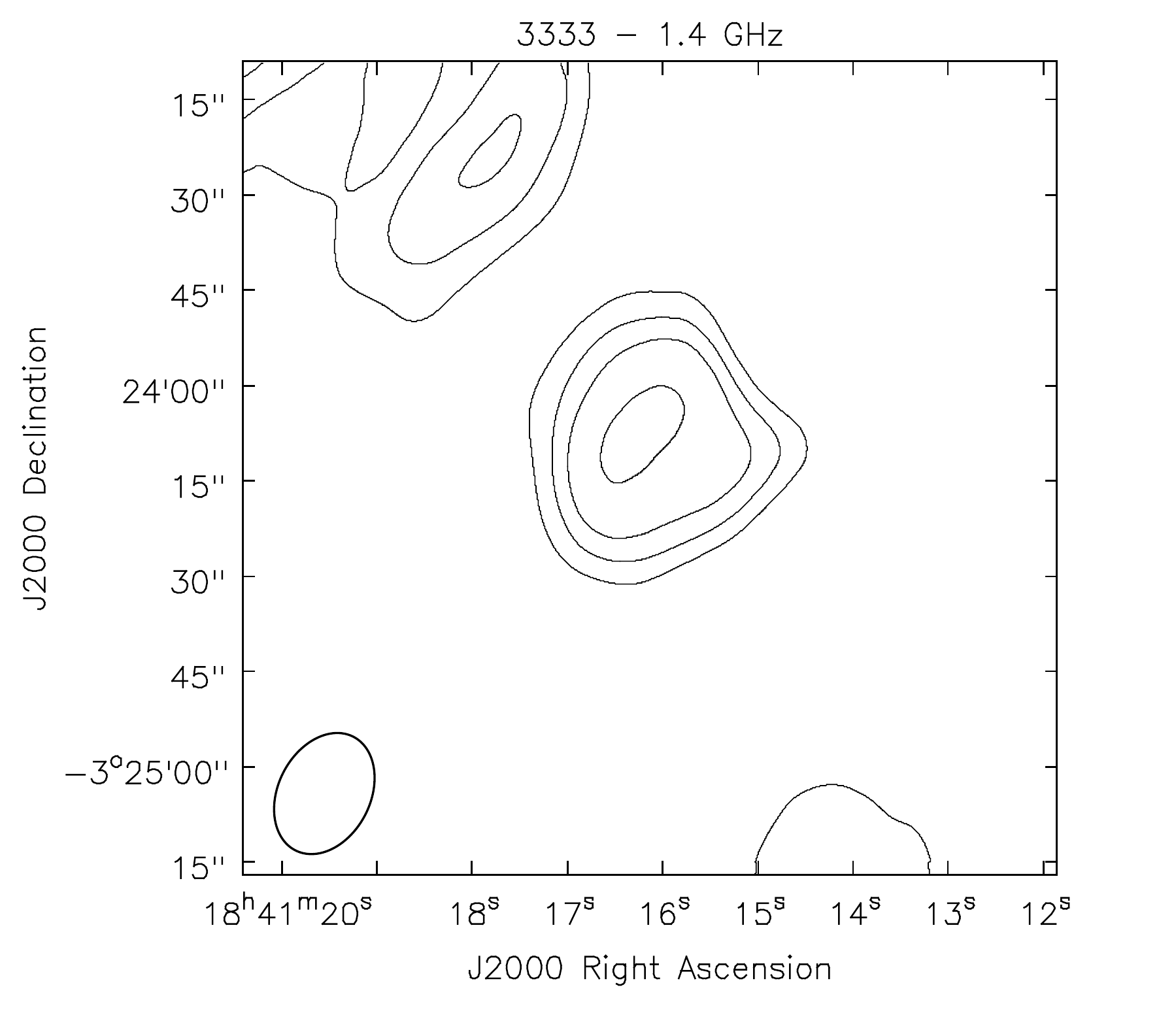}
\caption{Radio contours are 0.35, 0.70, 1.05 and $1.75\um{mJy/beam}$ (left) and 0.5, 1.0, 1.5, 2.0 and $2.5\um{mJy/beam}$ (right).}
\label{fig:3333}
\end{center}
\end{figure*}

\begin{figure*}
\begin{center}
\includegraphics[width=7cm]{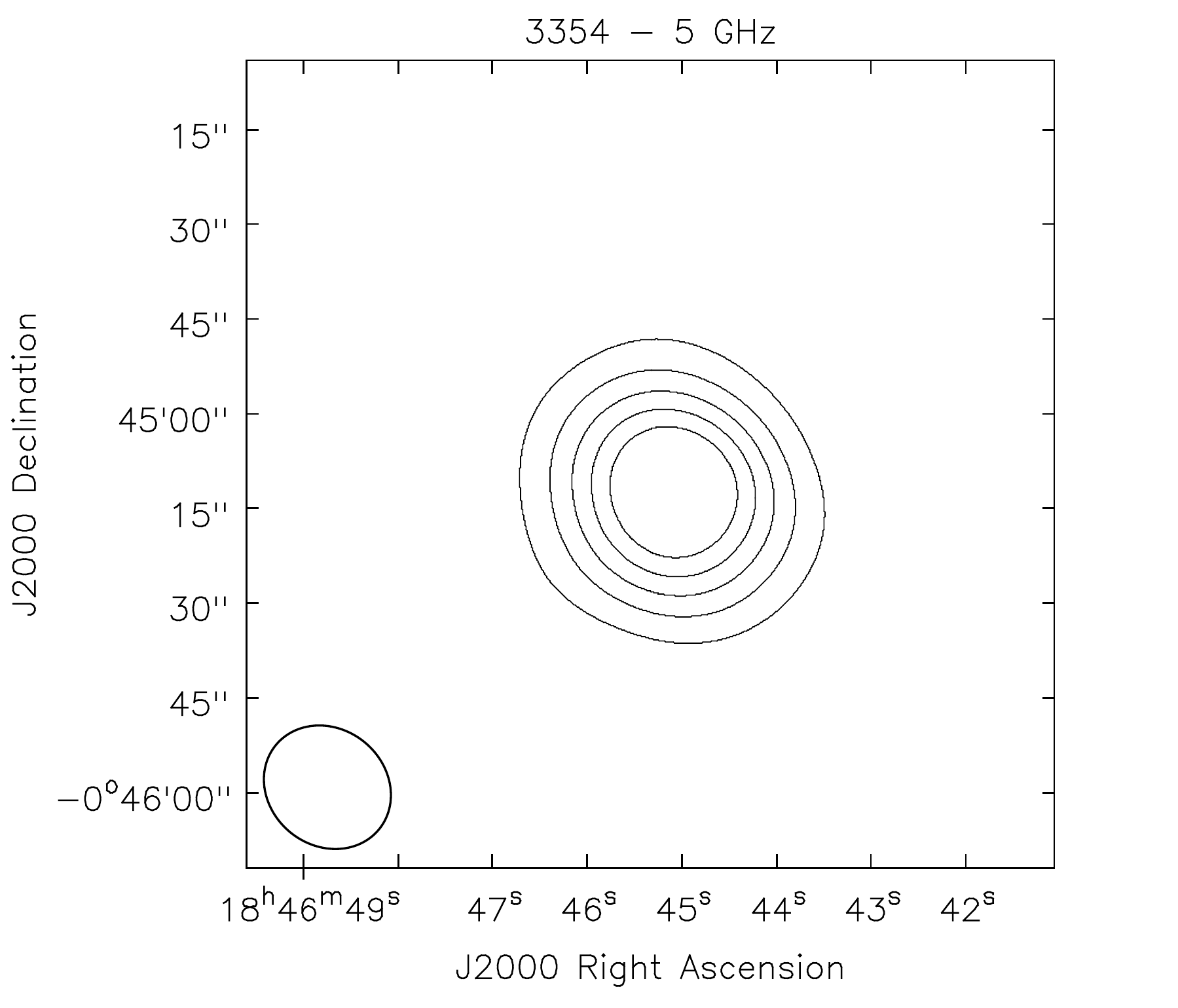}
\includegraphics[width=7cm]{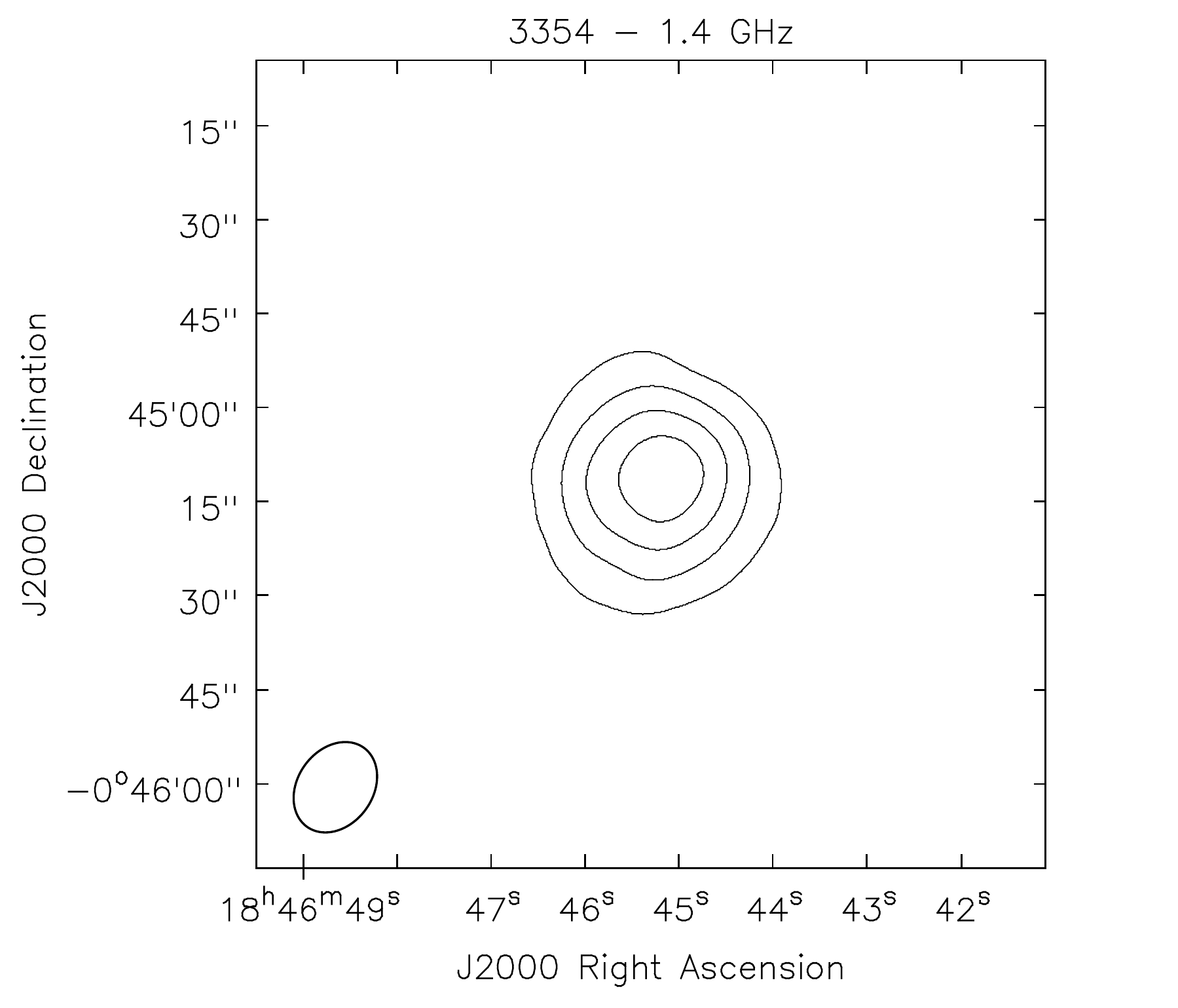}
\caption{Radio contours are 1, 2, 3, 4 and $5\um{mJy/beam}$ (left) and 1, 2, 3, and $4\um{mJy/beam}$ (right).}
\label{fig:3354}
\end{center}
\end{figure*}

\begin{figure*}
\begin{center}
\includegraphics[width=7cm]{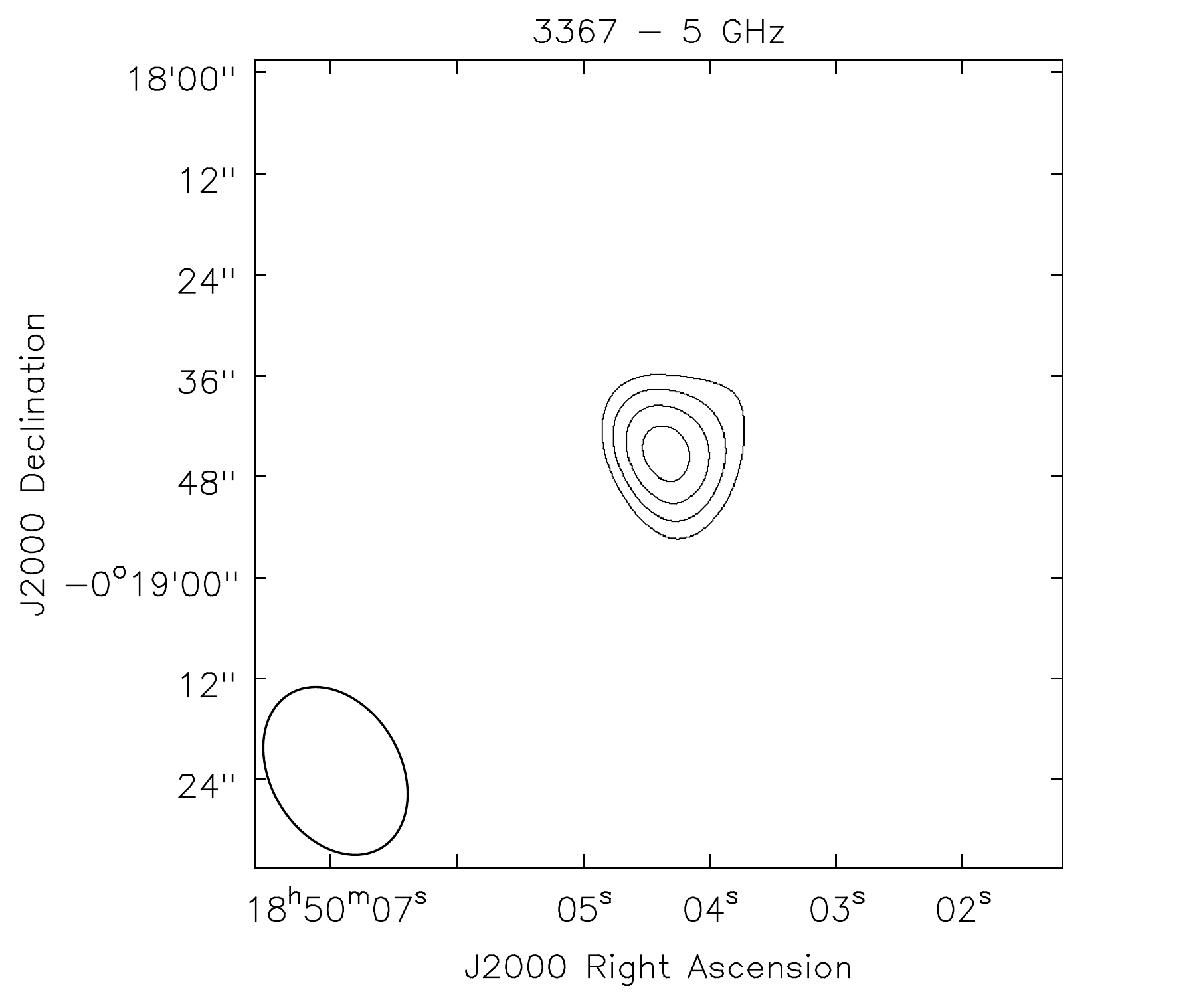}
\includegraphics[width=7cm]{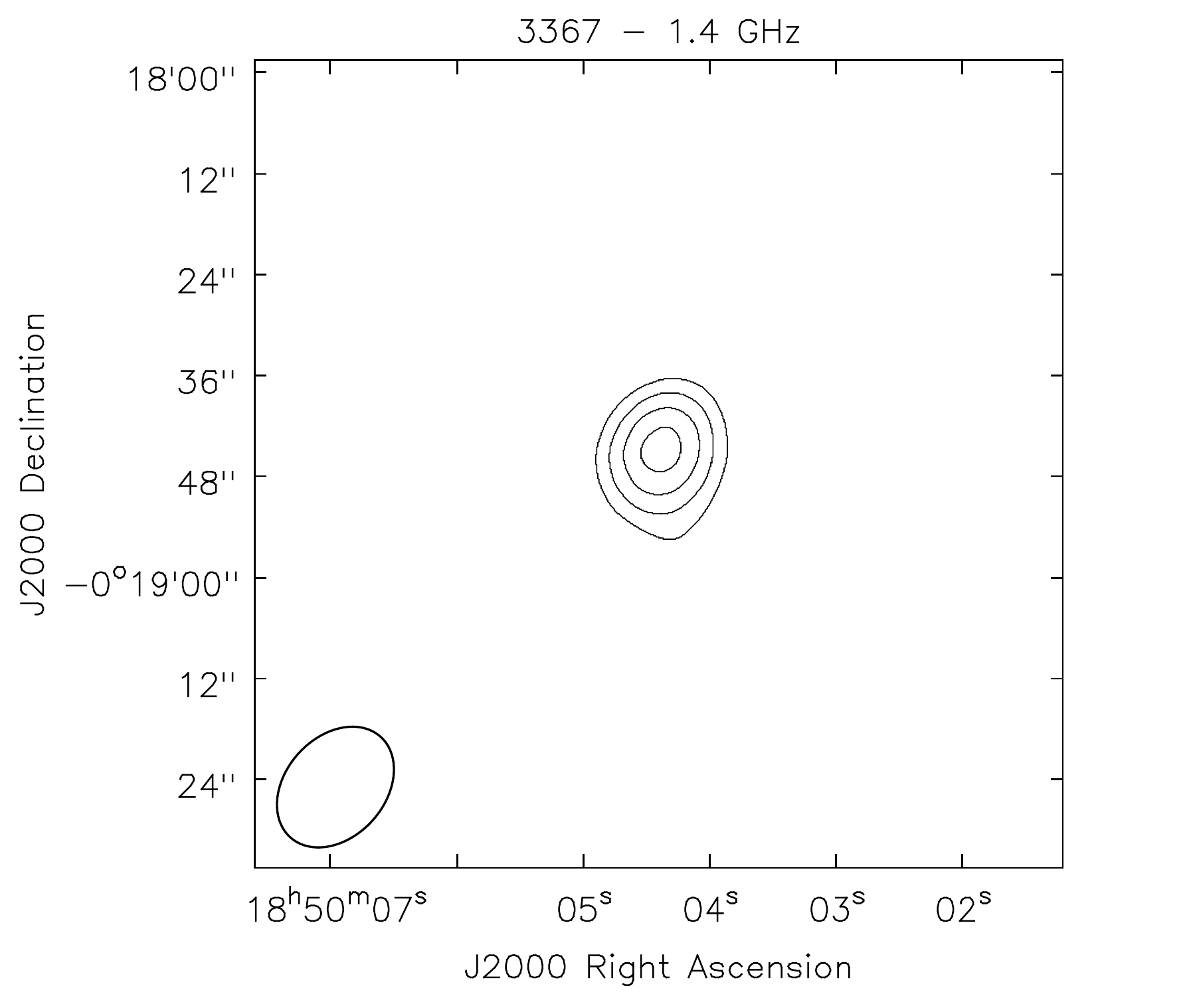}
\caption{Radio contours are 2.5, 3.5, 4.5 and $5.5\um{mJy/beam}$ (left) and 3, 5, 7 and $9\um{mJy/beam}$ (right).}
\label{fig:3367}
\end{center}
\end{figure*}

\begin{figure*}
\begin{center}
\includegraphics[width=7cm]{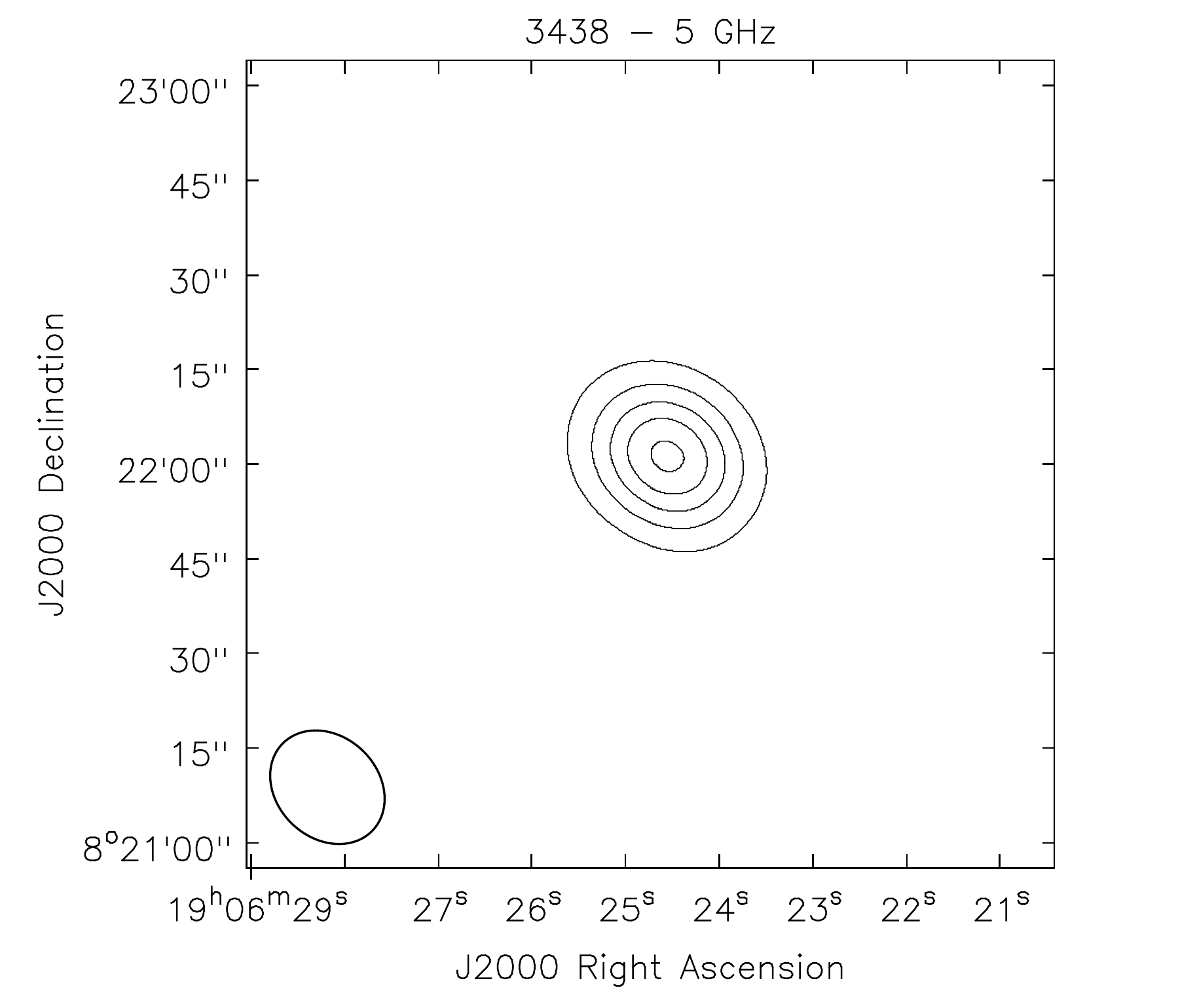}
\includegraphics[width=7cm]{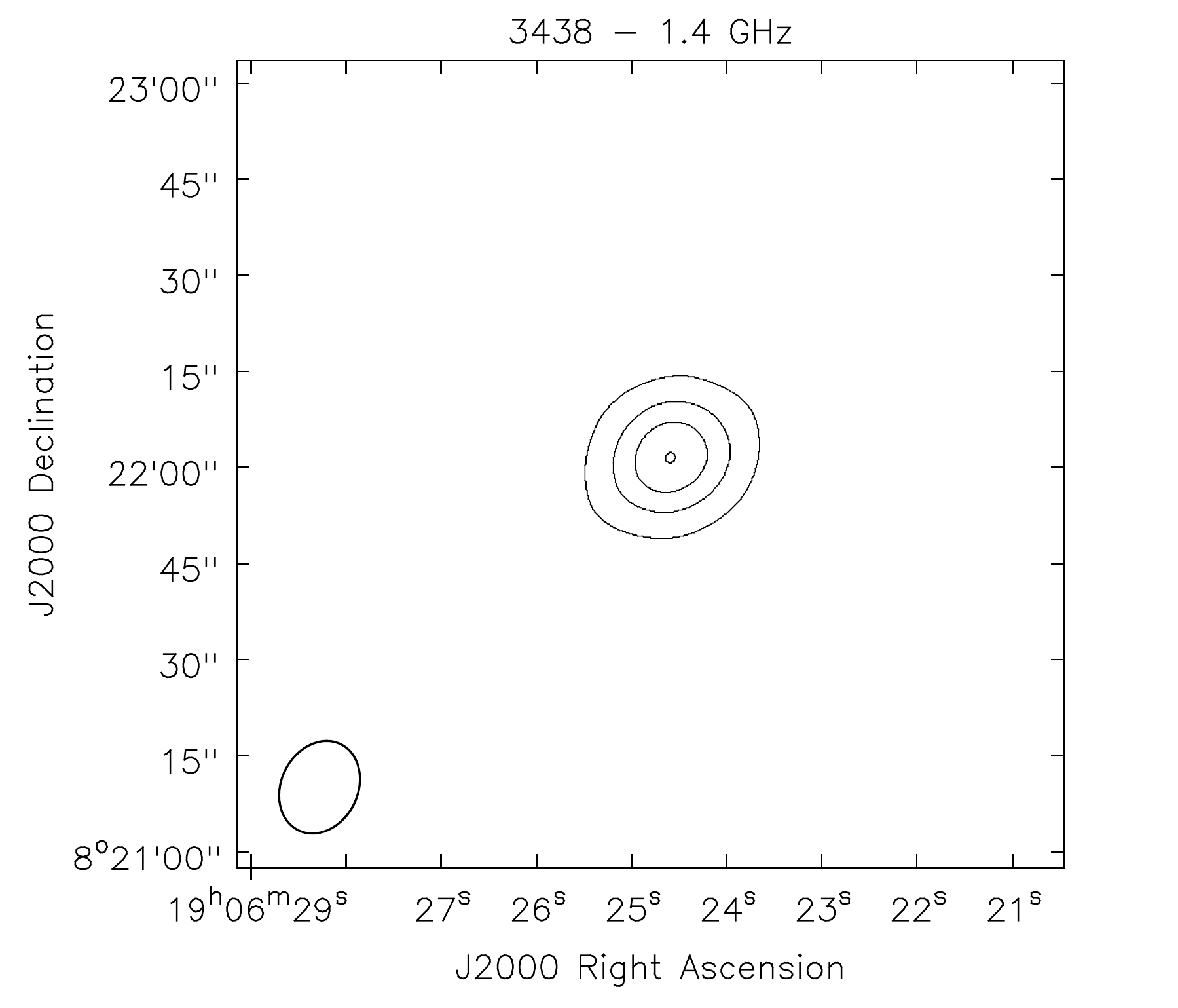}
\caption{Radio contours are 2, 4, 6, 8 and $10\um{mJy/beam}$ (left) and 1.5, 3, 4.5, and $6\um{mJy/beam}$ (right).}
\label{fig:3438}
\end{center}
\end{figure*}

\begin{figure*}
\begin{center}
\includegraphics[width=7cm]{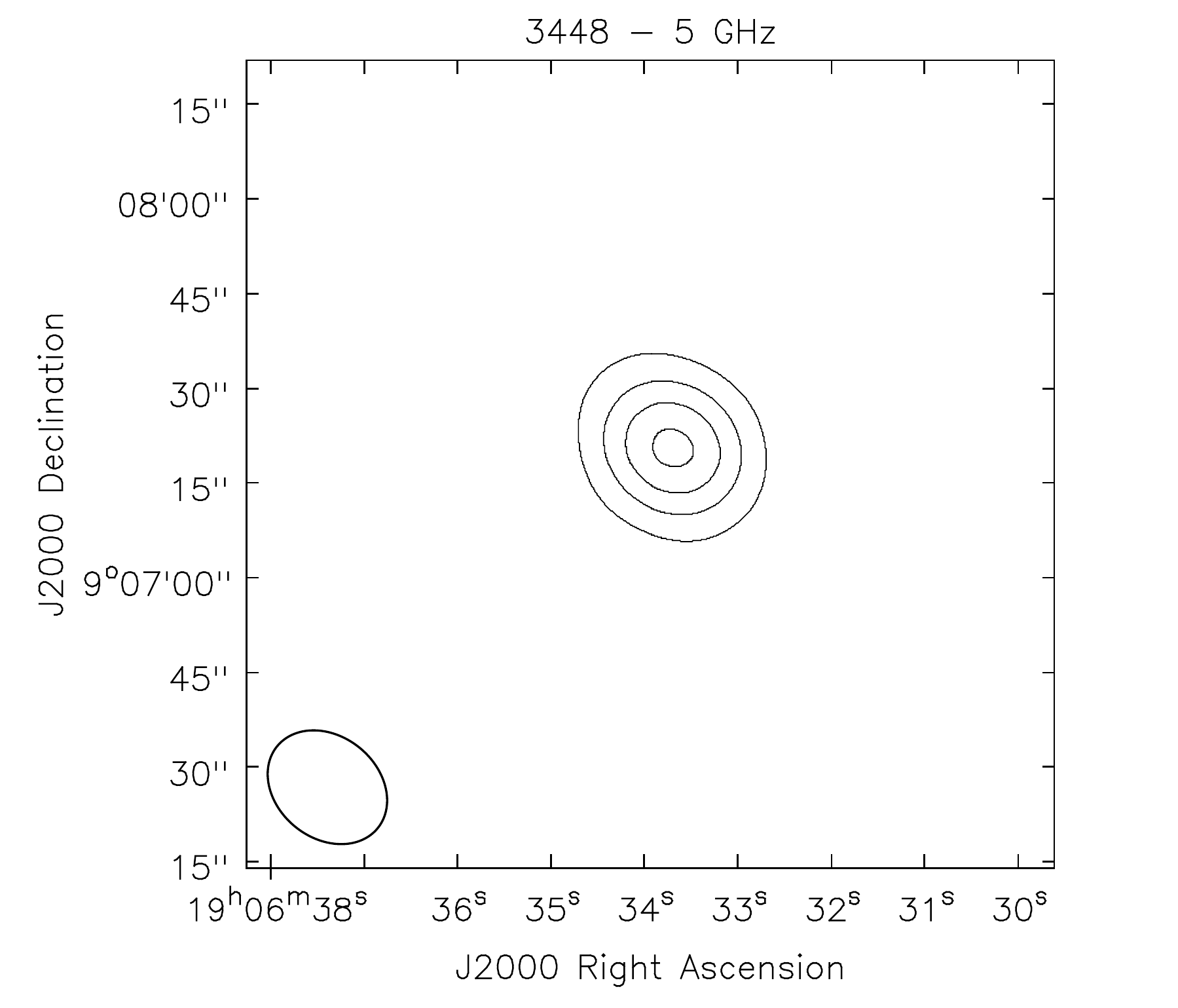}
\includegraphics[width=7cm]{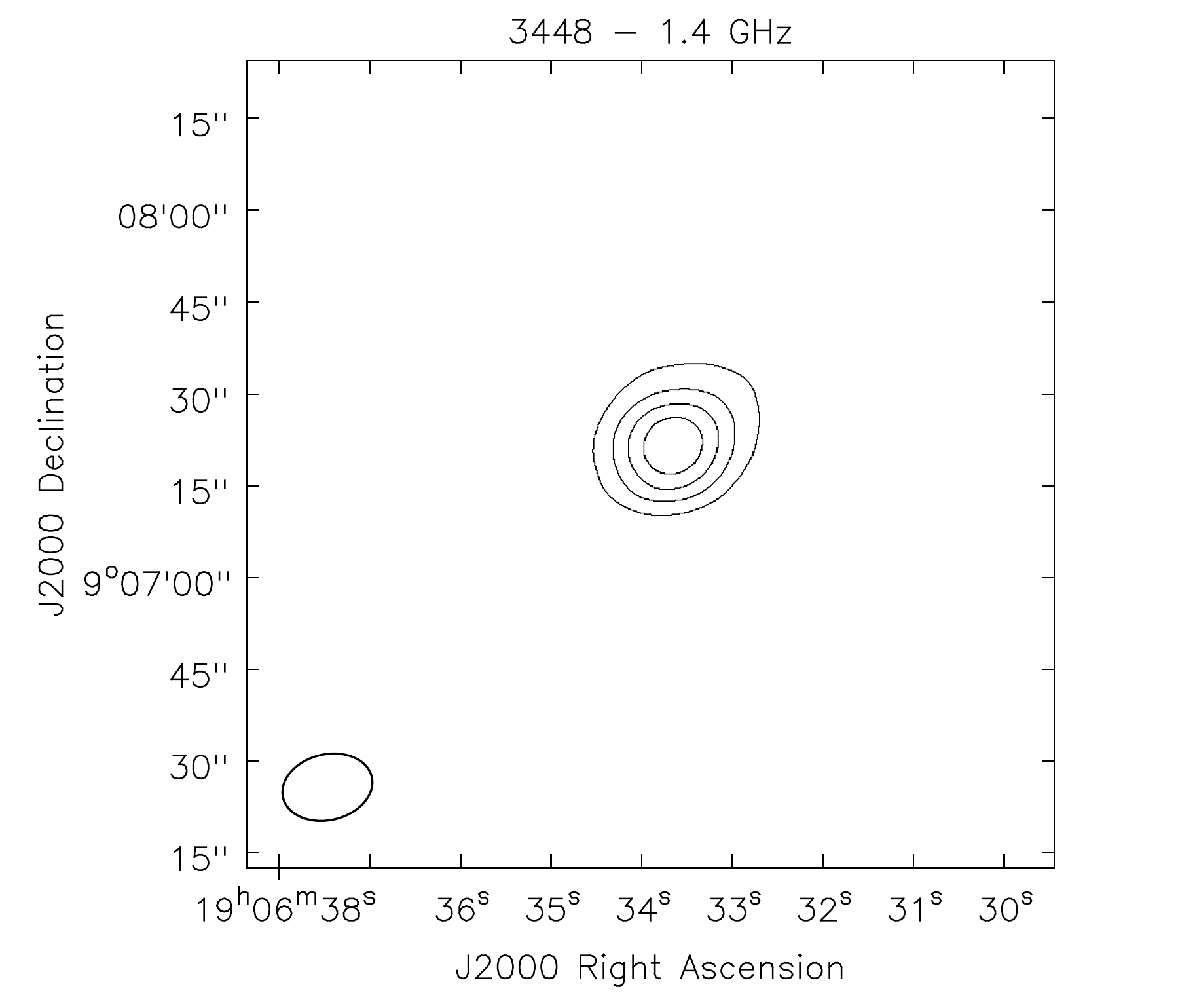}
\caption{Radio contours are 3, 6, 9 and $12\um{mJy/beam}$ (left) and 2, 4, 6 and $8\um{mJy/beam}$ (right).}
\label{fig:3448}
\end{center}
\end{figure*}

\begin{figure*}
\begin{center}
\includegraphics[width=7cm]{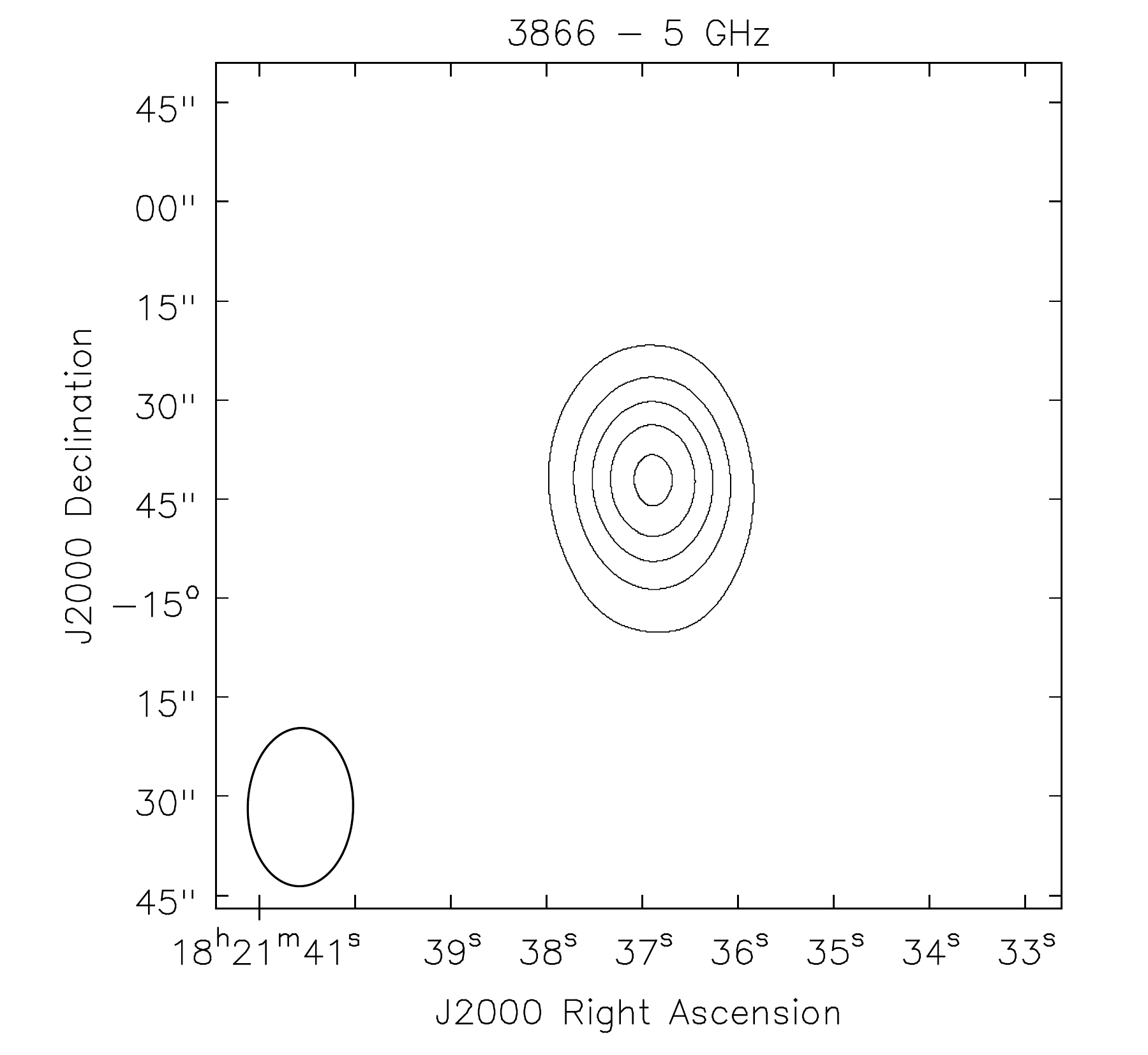}
\includegraphics[width=7cm]{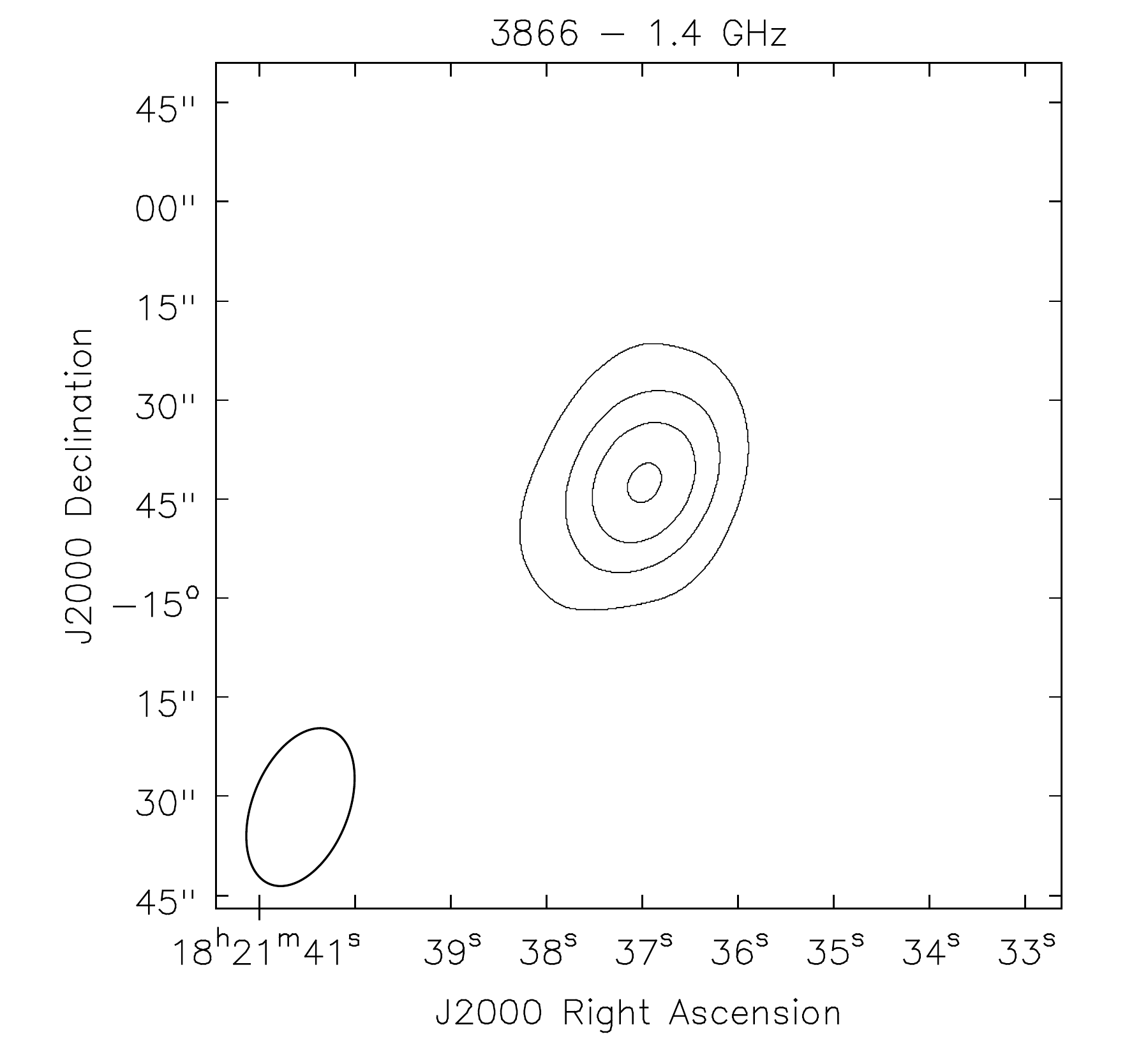}
\caption{Radio contours are 2, 4, 6, 8 and $10\um{mJy/beam}$ (left) and 3, 6, 9, and $12\um{mJy/beam}$ (right).}
\label{fig:3866}
\end{center}
\end{figure*}

\begin{figure*}
\begin{center}
\includegraphics[width=7cm]{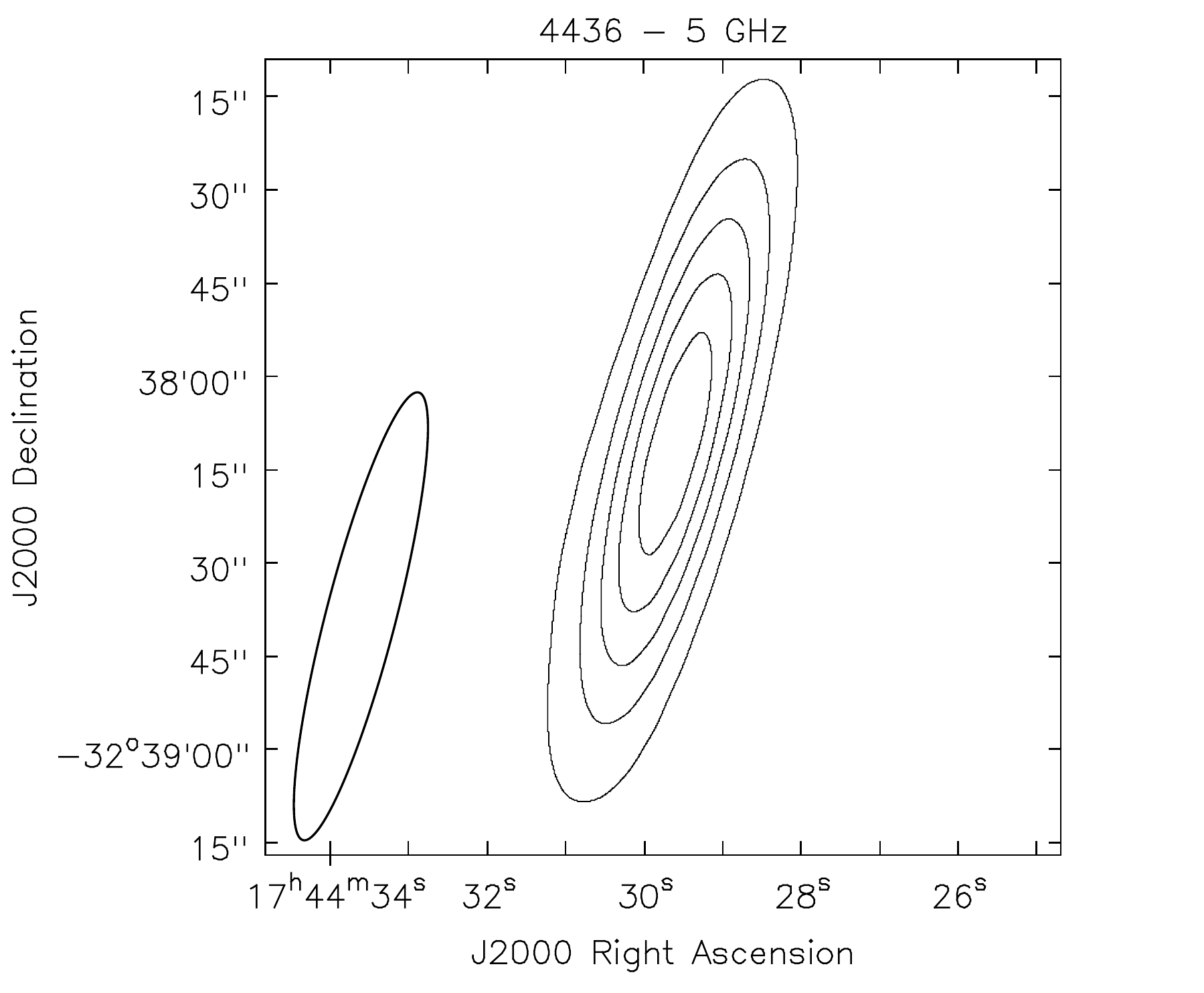}
\includegraphics[width=7cm]{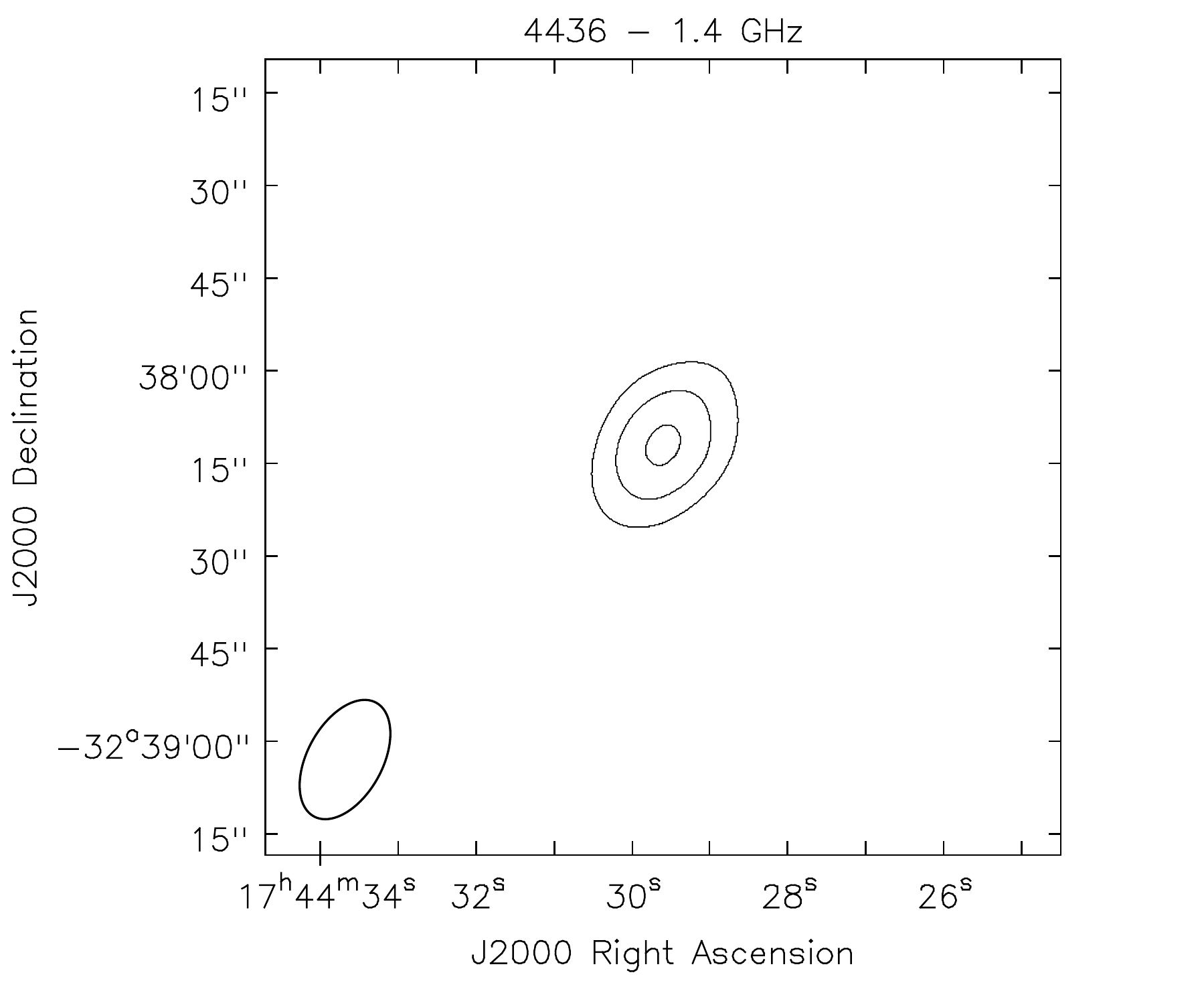}
\caption{Radio contours are 1, 2, 3, 4 and $5\um{mJy/beam}$ (left) and 2, 4 and $6\um{mJy/beam}$ (right).}
\label{fig:4436}
\end{center}
\end{figure*}

\begin{figure*}
\begin{center}
\includegraphics[width=7cm]{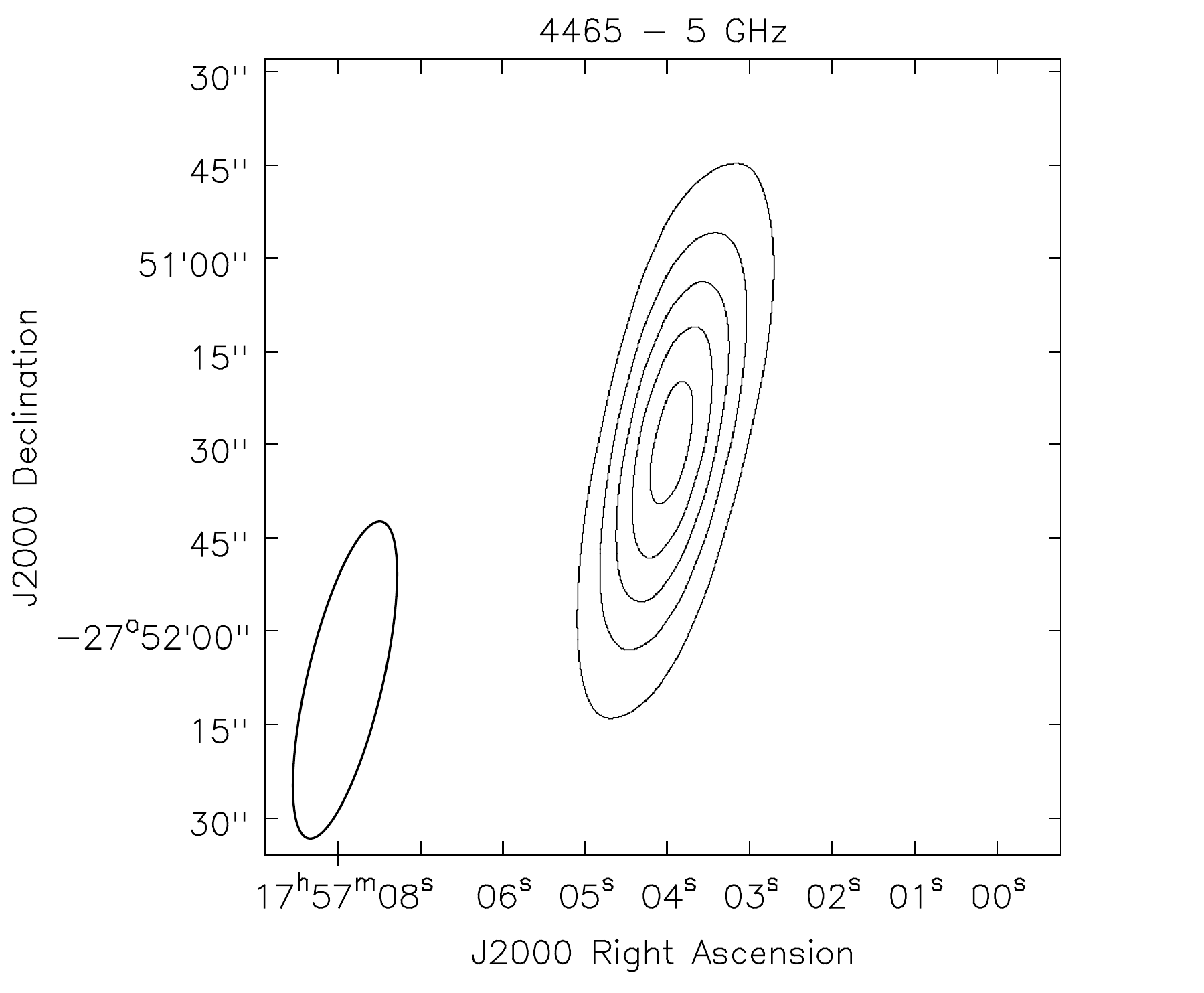}
\includegraphics[width=7cm]{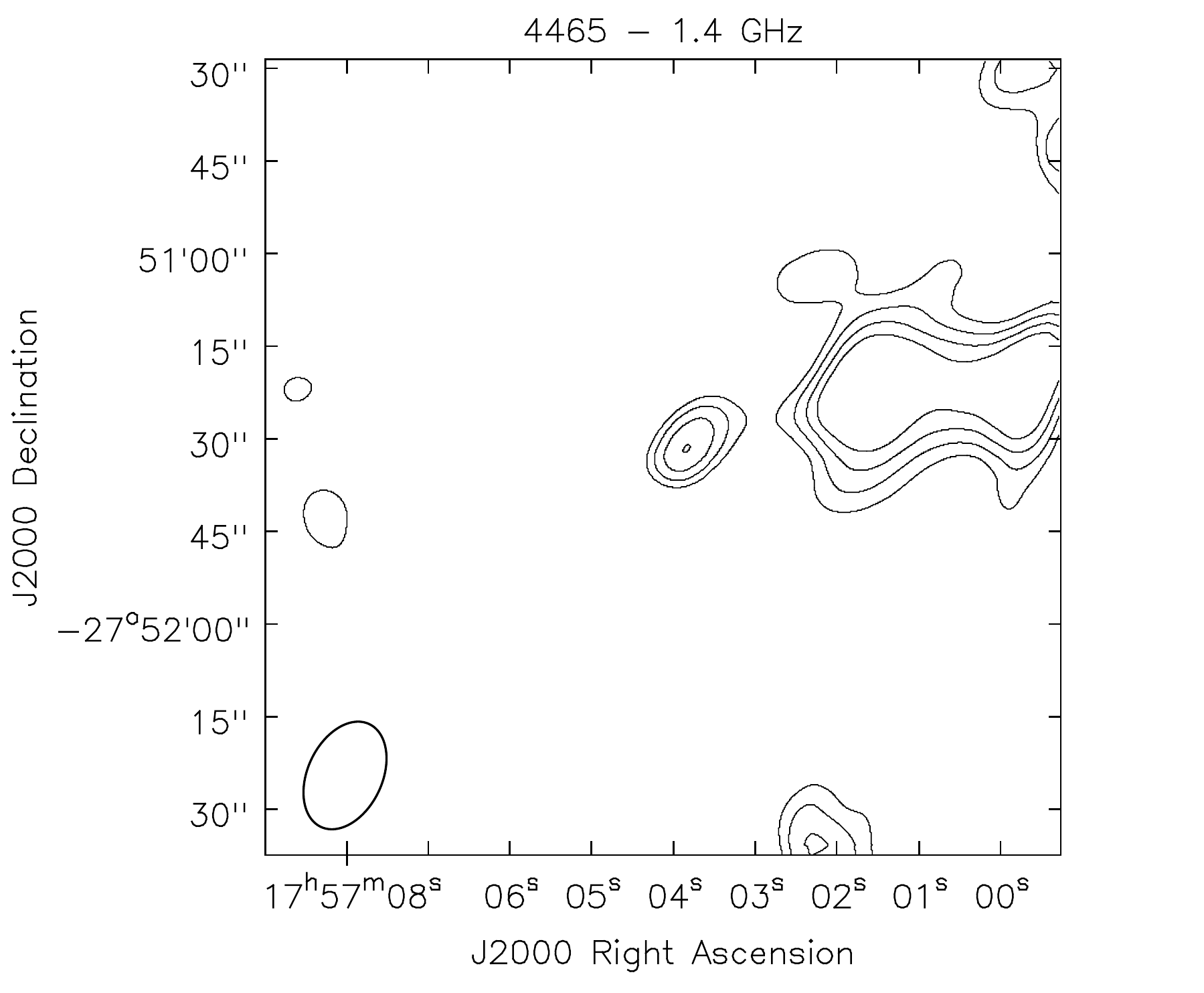}
\caption{Radio contours are 0.3, 0.6, 0.9, 1.2 and $1.5\um{mJy/beam}$ (left) and 0.7, 0.8, 0.9, and $1.0\um{mJy/beam}$ (right).}
\label{fig:4465}
\end{center}
\end{figure*}

\begin{figure*}
\begin{center}
\includegraphics[width=7cm]{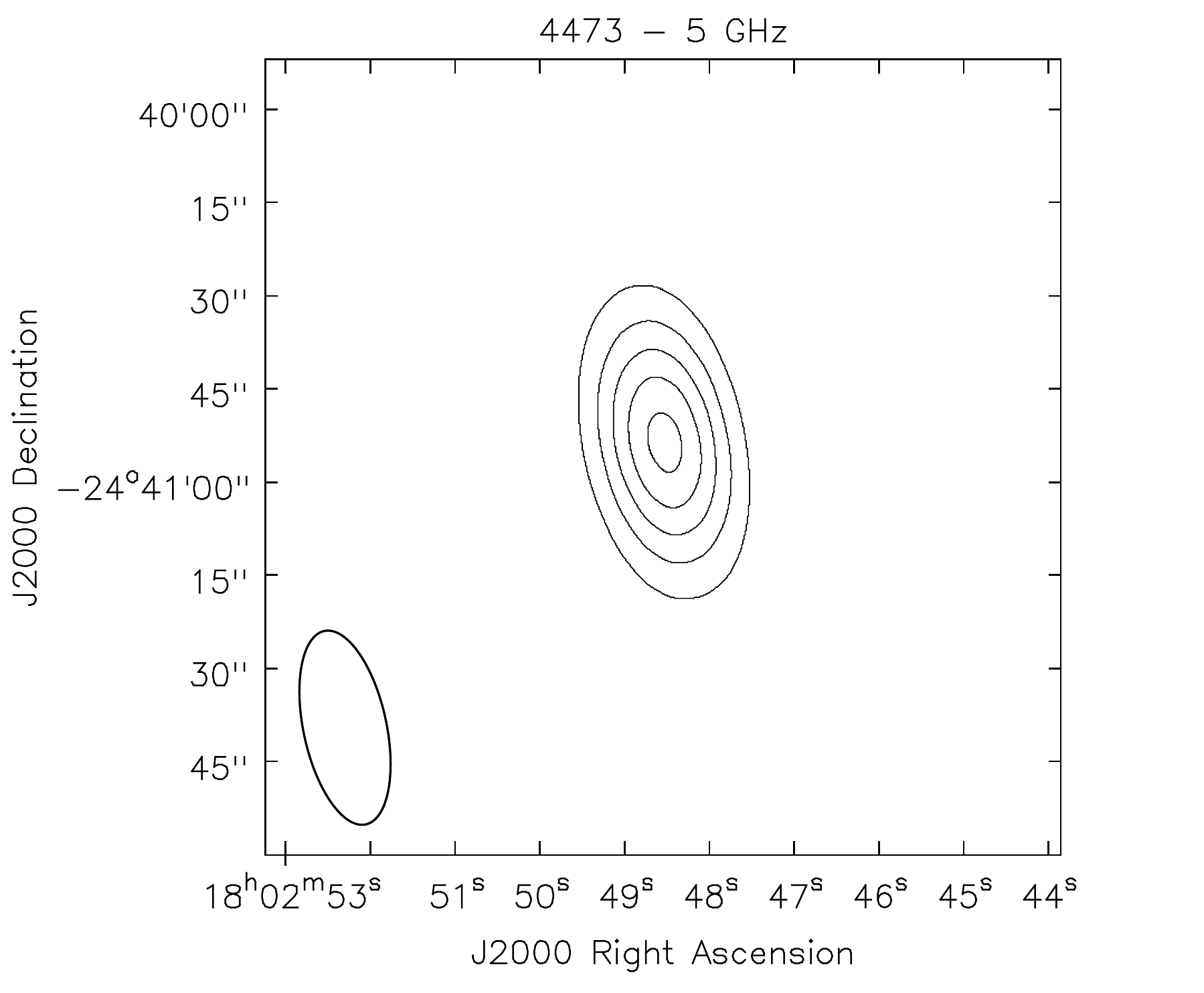}
\includegraphics[width=7cm]{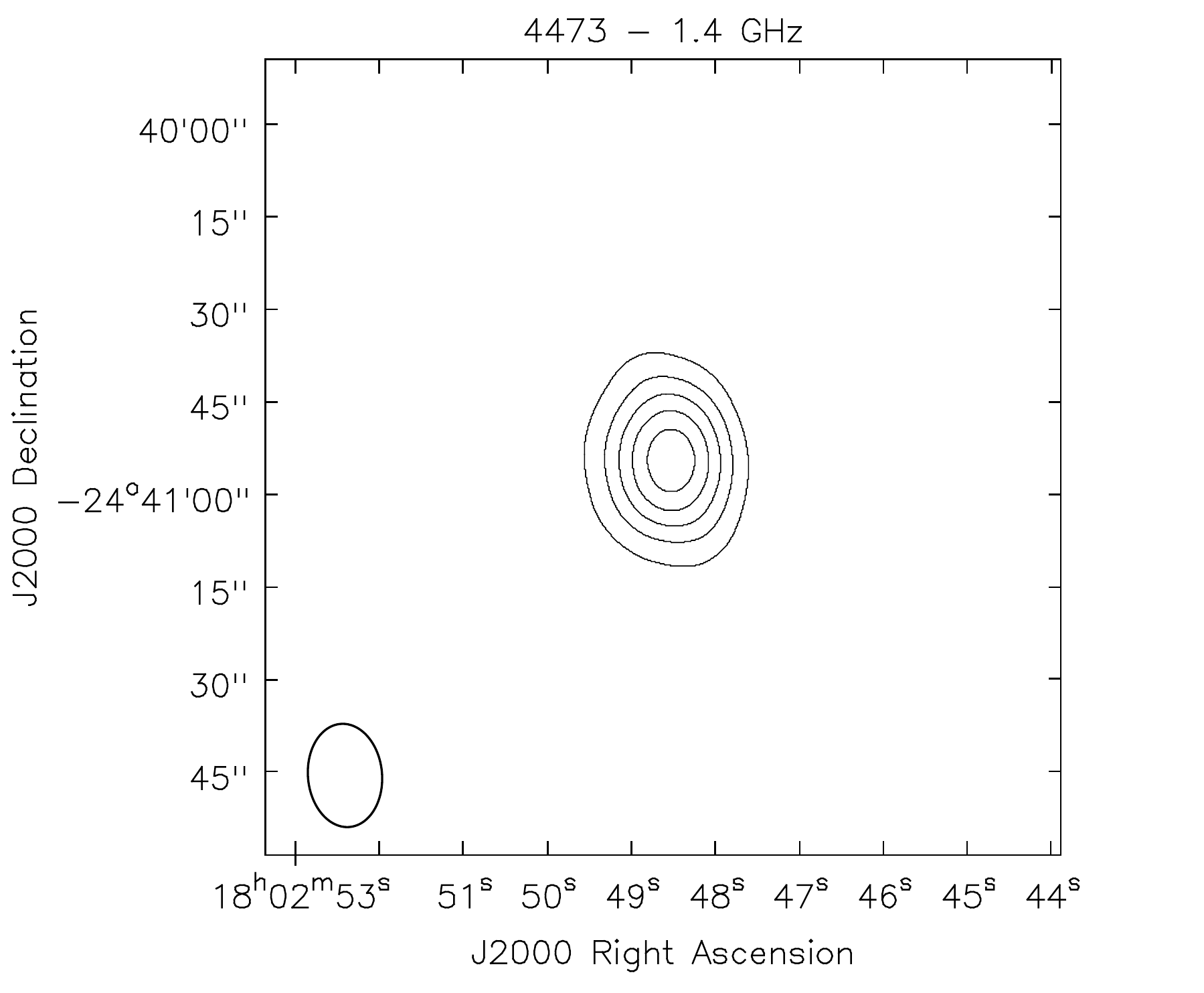}
\caption{Radio contours are 7, 14, 21, 28 and $35\um{mJy/beam}$ (left) and 6, 12, 18, 24 and $30\um{mJy/beam}$ (right).}
\label{fig:4473}
\end{center}
\end{figure*}

\begin{figure*}
\begin{center}
\includegraphics[width=7cm]{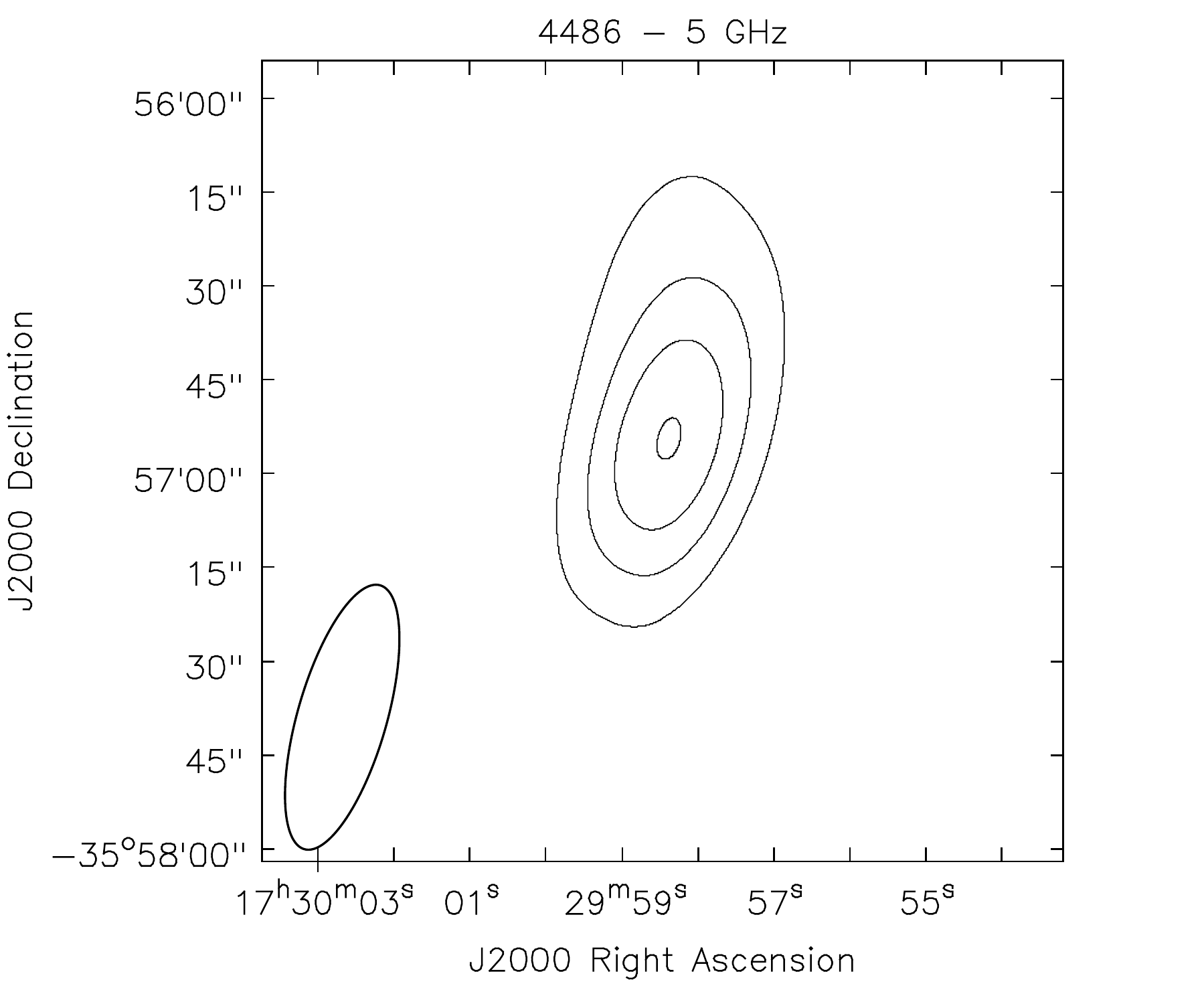}
\includegraphics[width=7cm]{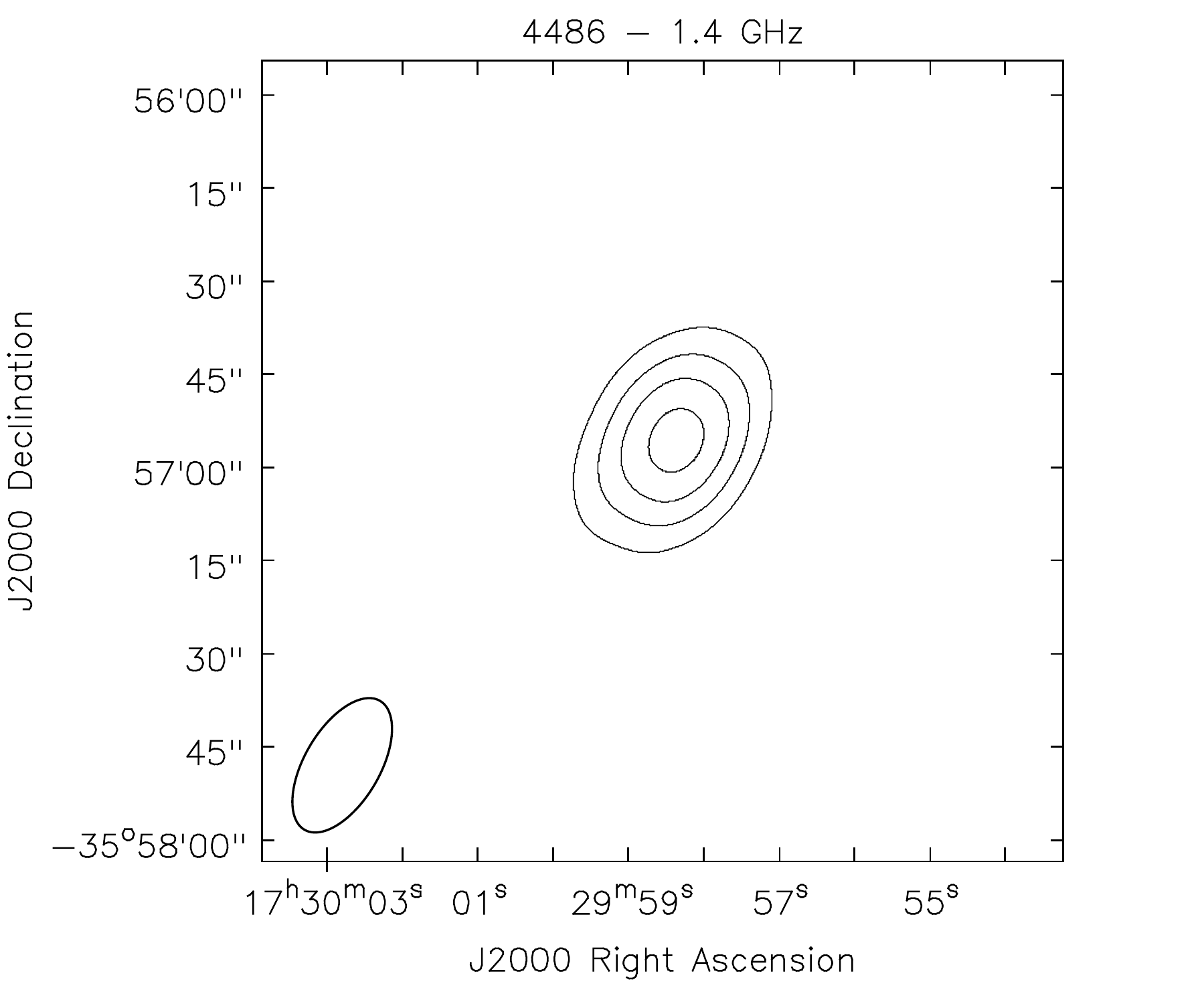}
\caption{Radio contours are 4, 8, 12 and $16\um{mJy/beam}$ (left) and 3, 6, 9, and $12\um{mJy/beam}$ (right).}
\label{fig:4486}
\end{center}
\end{figure*}

\begin{figure*}
\begin{center}
\includegraphics[width=7cm]{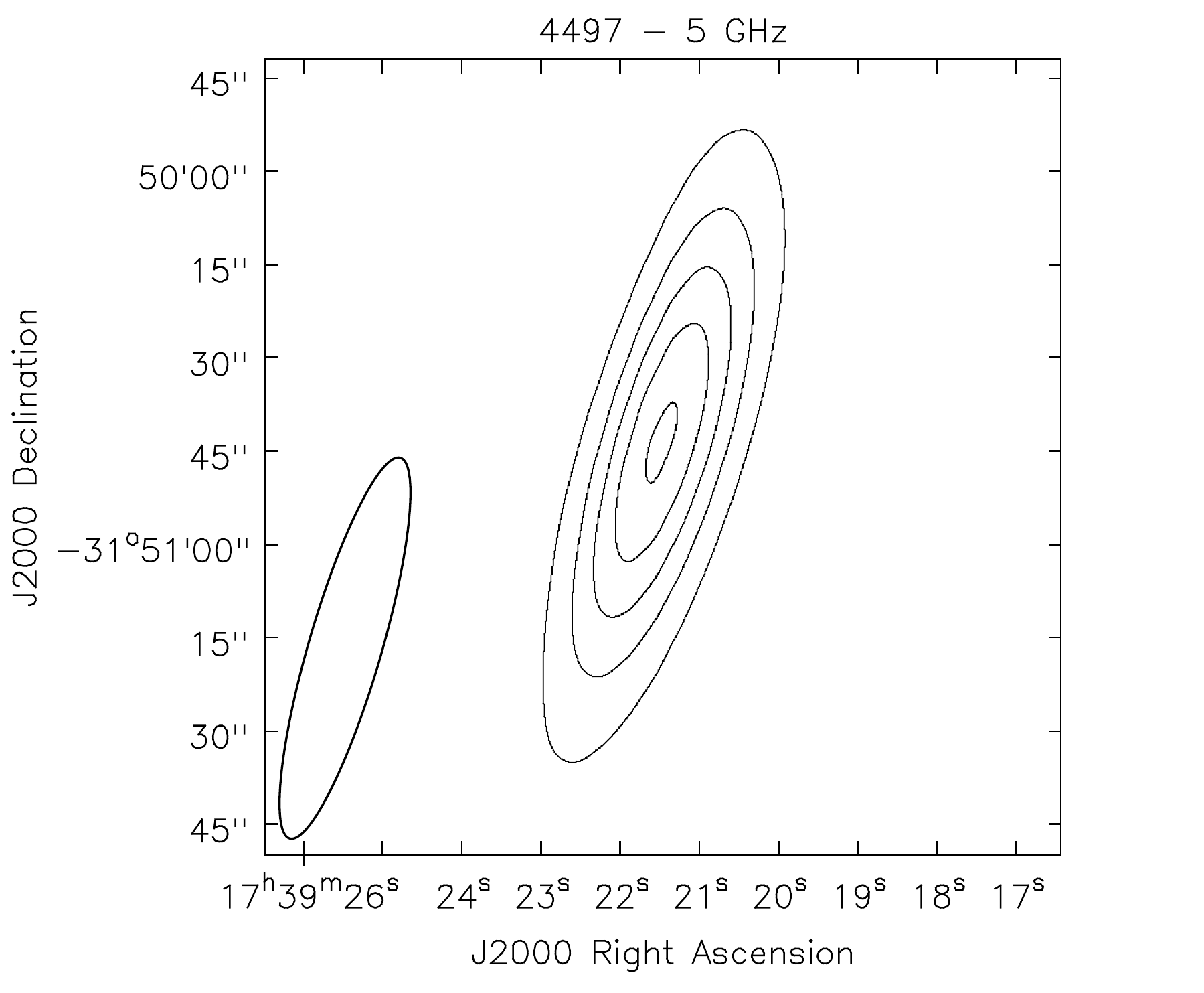}
\includegraphics[width=7cm]{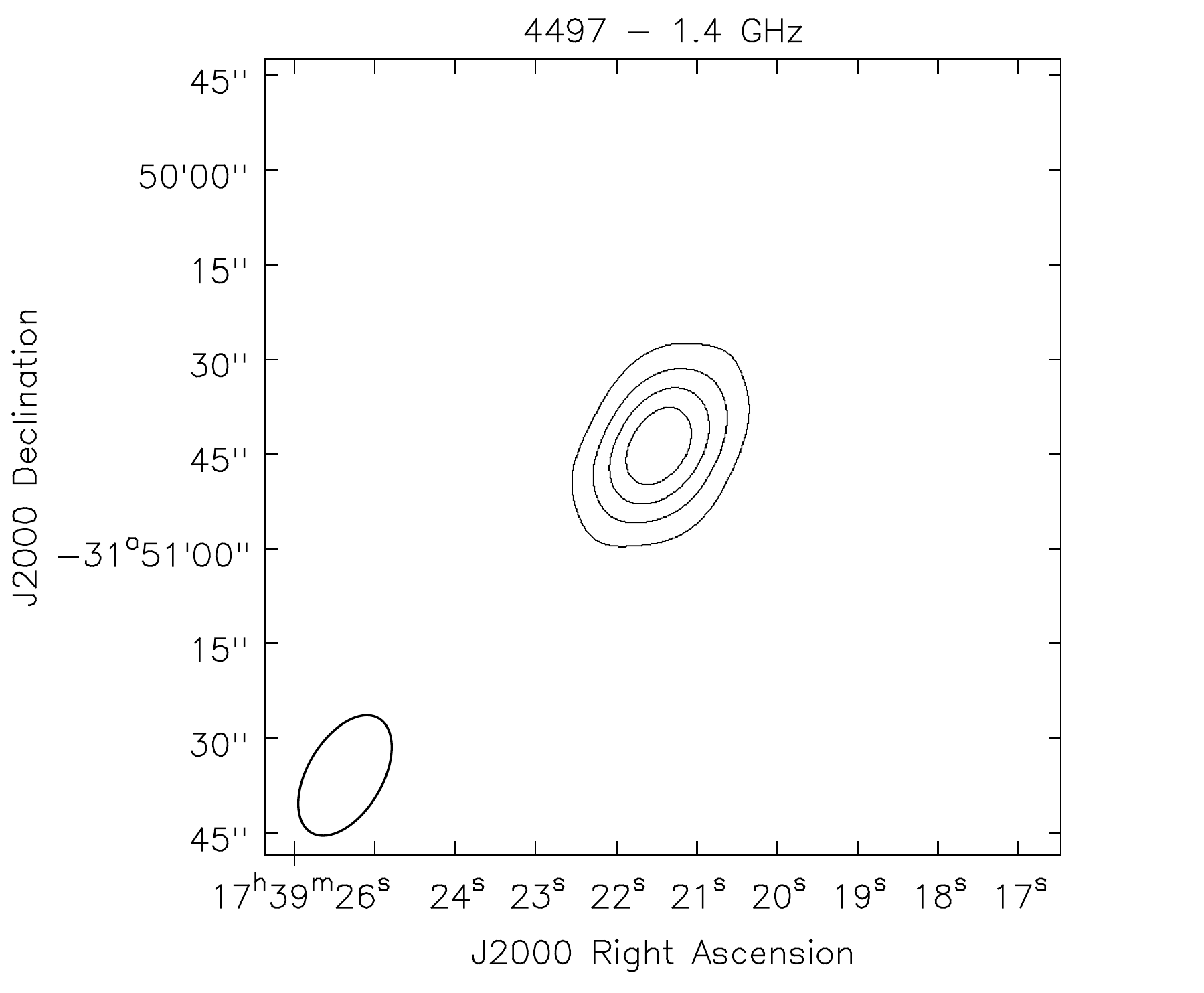}
\caption{Radio contours are 3, 6, 9, 12 and $15\um{mJy/beam}$ (left) and 3, 6, 9 and $12\um{mJy/beam}$ (right).}
\label{fig:4497}
\end{center}
\end{figure*}

\begin{figure*}
\begin{center}
\includegraphics[width=7cm]{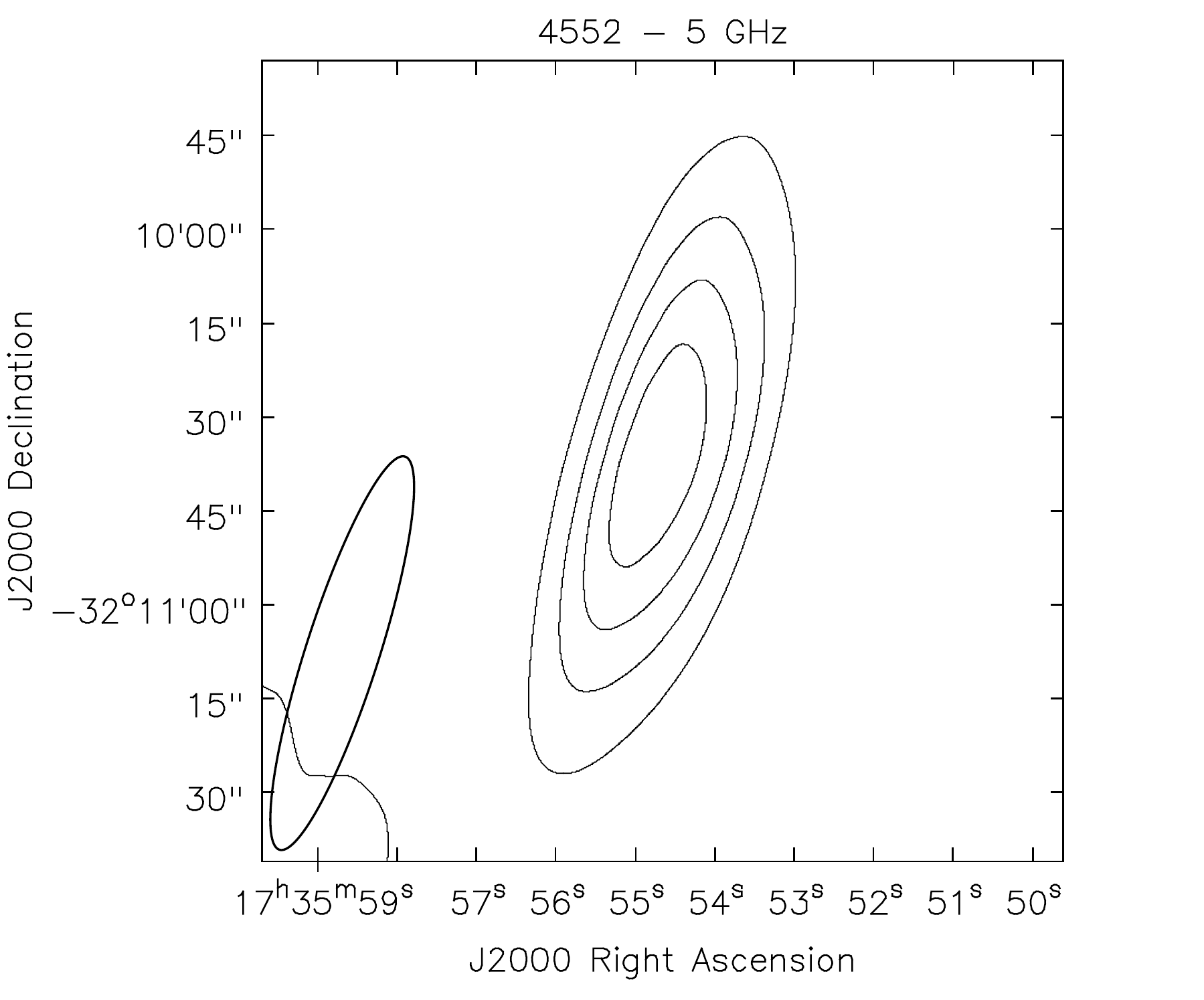}
\includegraphics[width=7cm]{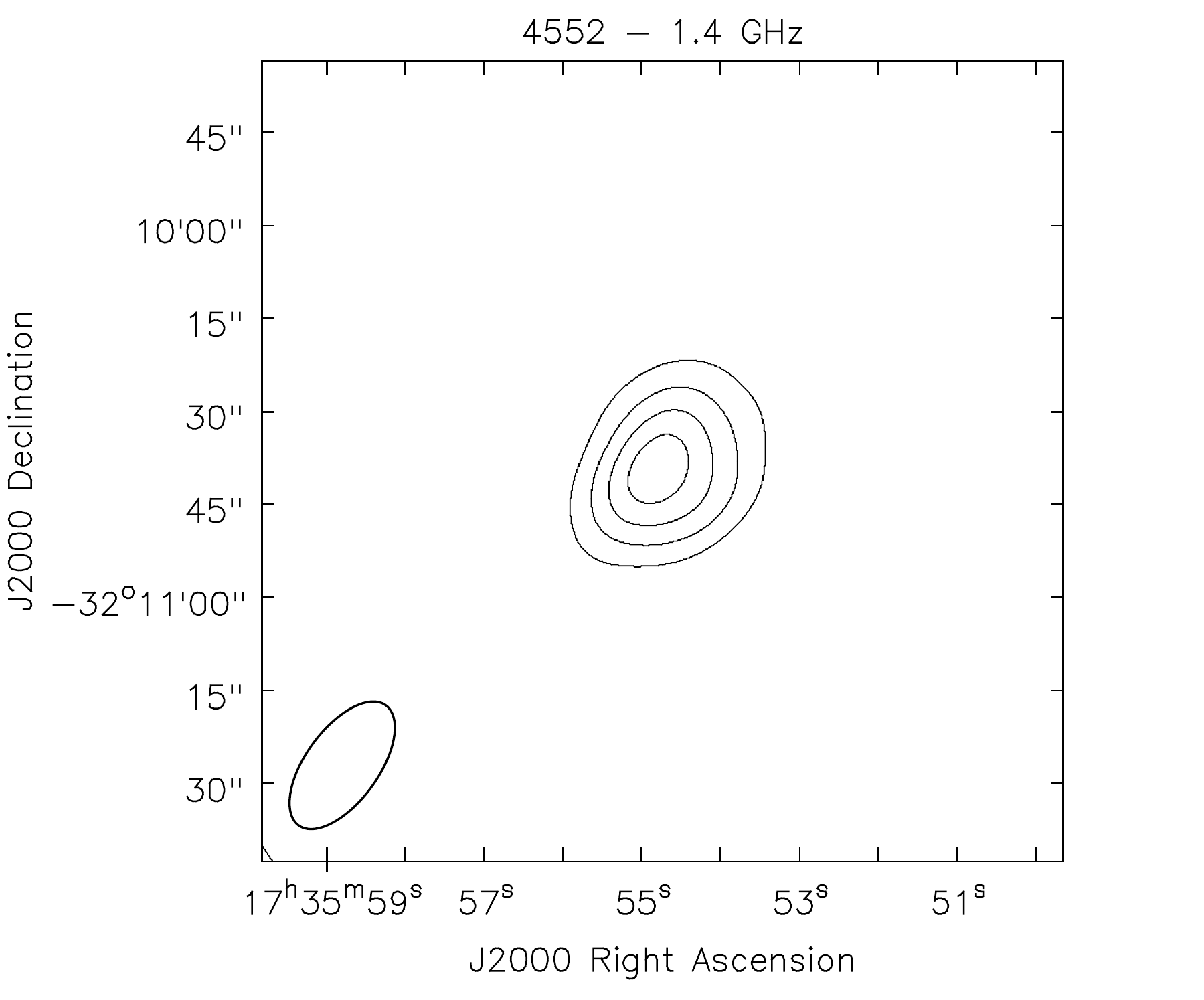}
\caption{Radio contours are 3, 6, 9 and $12\um{mJy/beam}$ (left) and 2.5, 5, 7.5, and $10\um{mJy/beam}$ (right).}
\label{fig:4552}
\end{center}
\end{figure*}

\begin{figure*}
\begin{center}
\includegraphics[width=7cm]{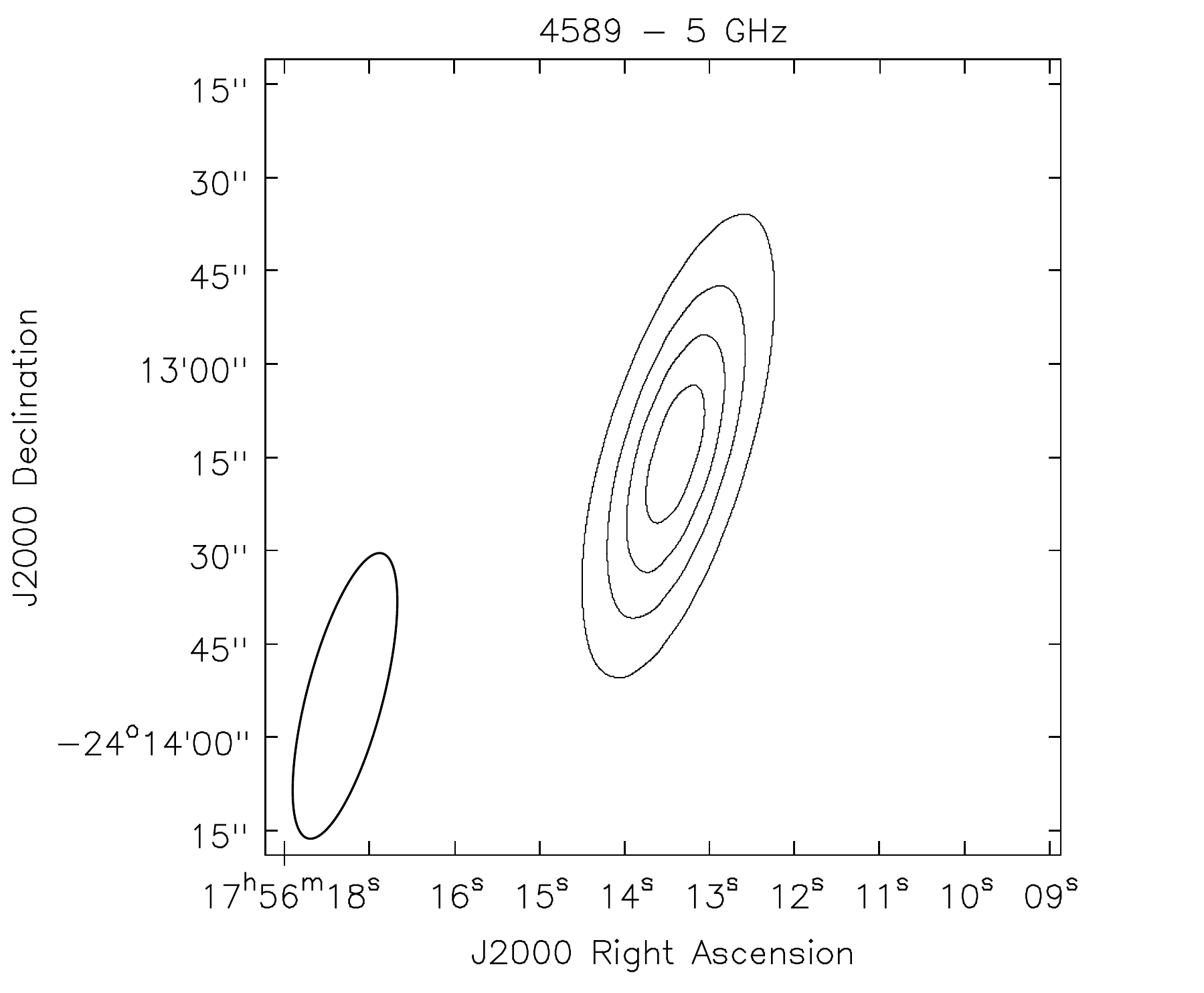}
\includegraphics[width=7cm]{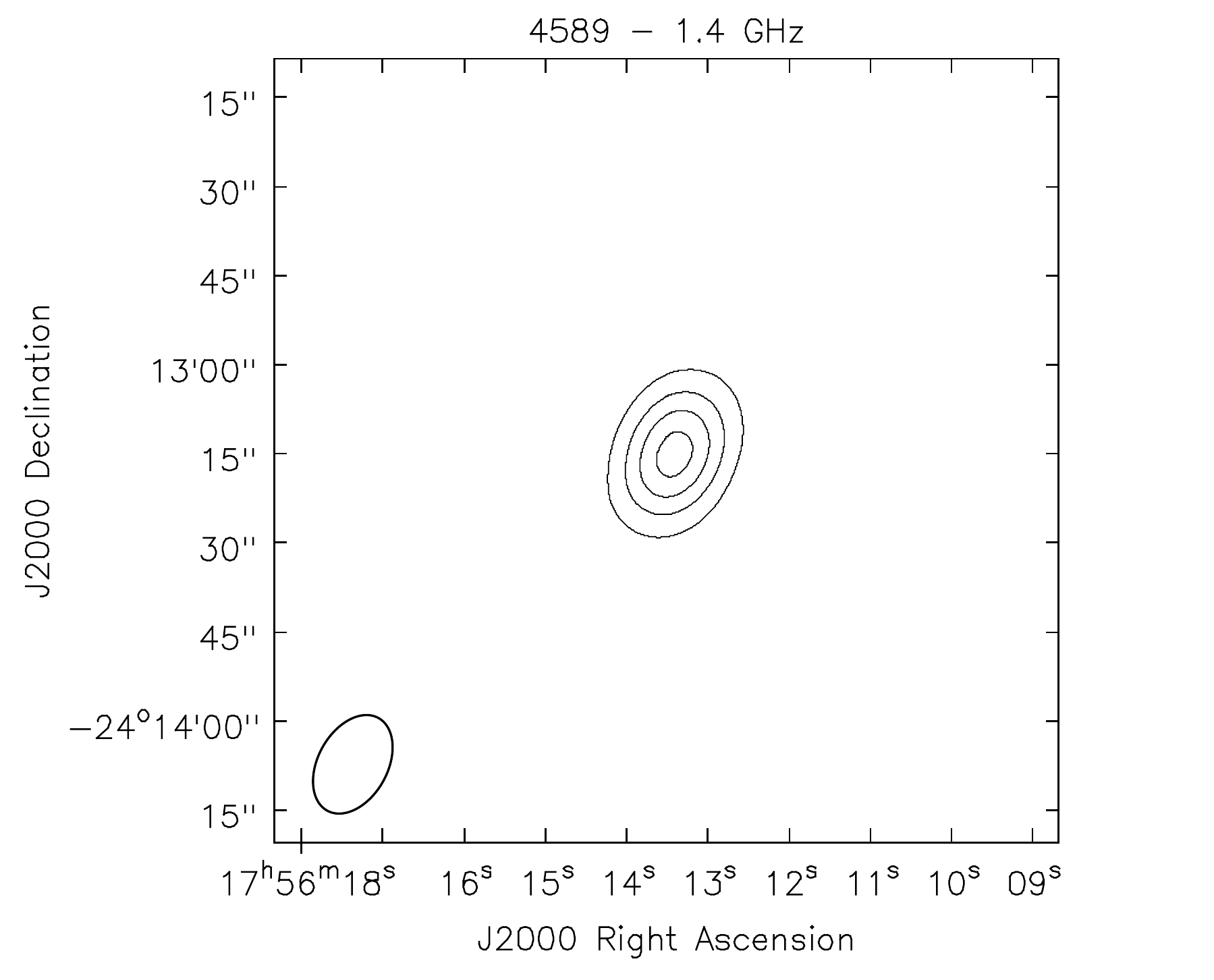}
\caption{Radio contours are 2, 4, 6 and $8\um{mJy/beam}$ (left) and 2, 4, 6 and $8\um{mJy/beam}$ (right).}
\label{fig:4589}
\end{center}
\end{figure*}

\begin{figure*}
\begin{center}
\includegraphics[width=7cm]{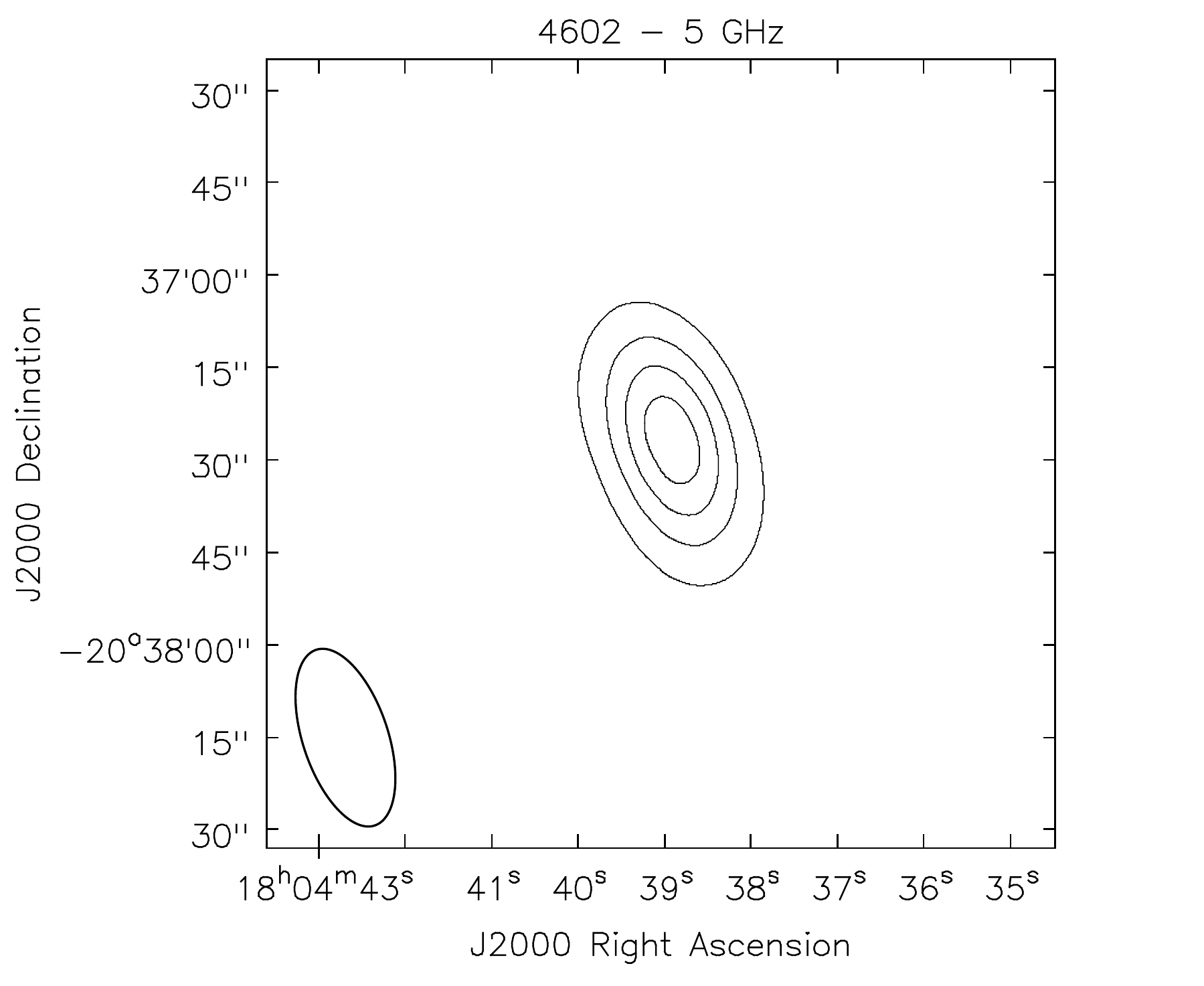}
\includegraphics[width=7cm]{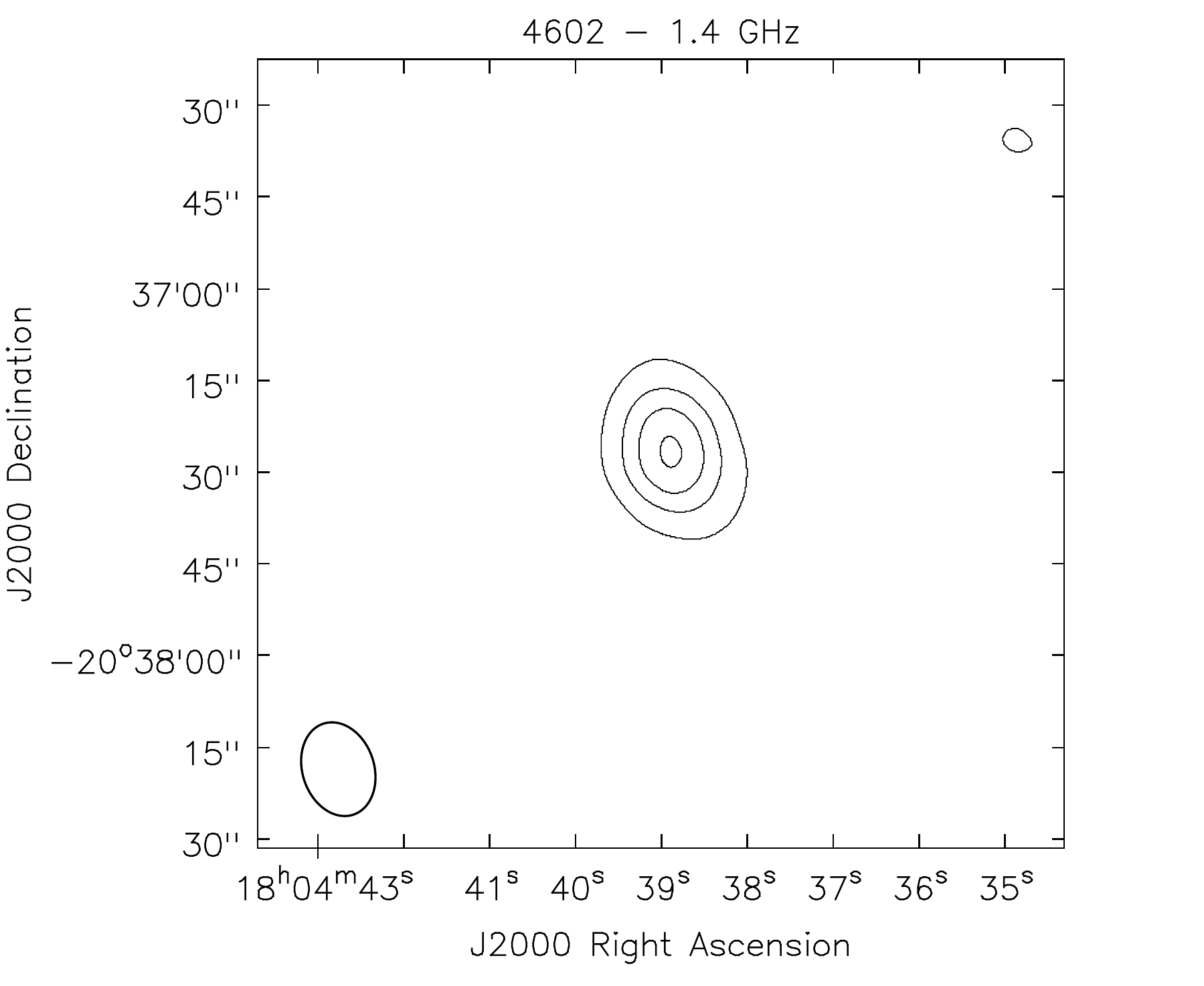}
\caption{Radio contours are 4, 8, 12 and $16\um{mJy/beam}$ (left) and 4, 8, 12, and $16\um{mJy/beam}$ (right).}
\label{fig:4602}
\end{center}
\end{figure*}

\begin{figure*}
\begin{center}
\includegraphics[width=7cm]{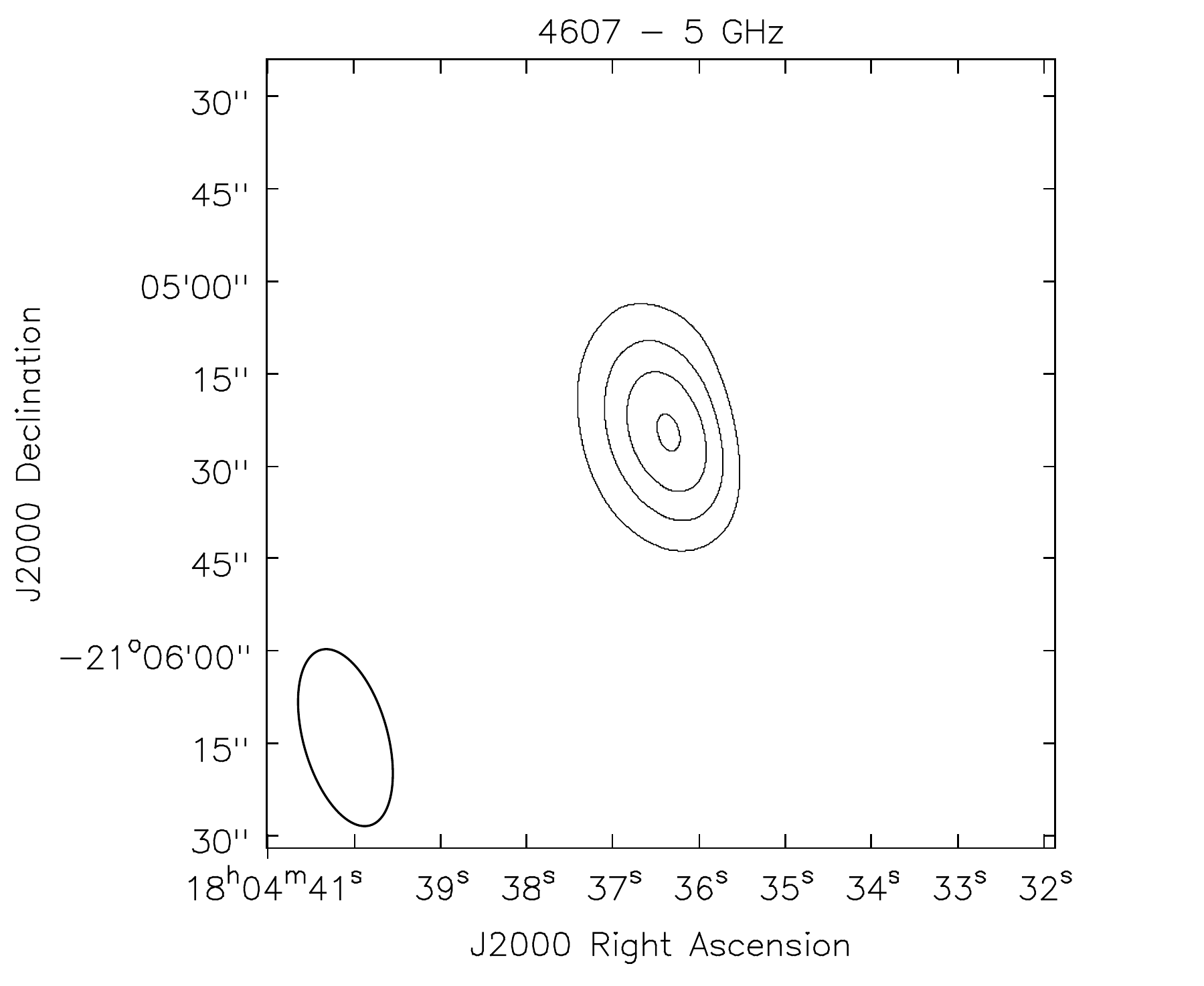}
\includegraphics[width=7cm]{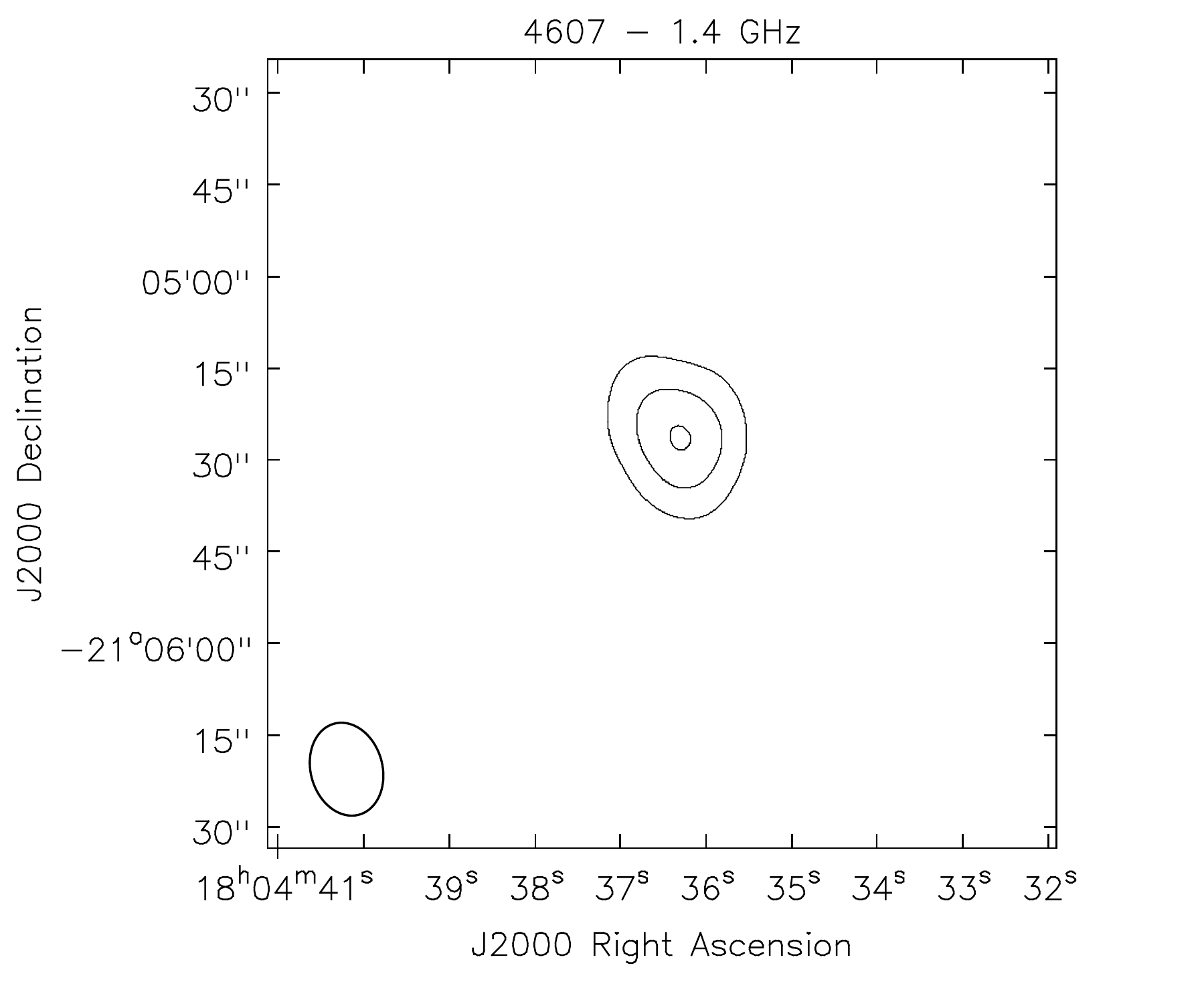}
\caption{Radio contours are 4, 6, 8 and $10\um{mJy/beam}$ (left) and 2, 4 and $6\um{mJy/beam}$ (right).}
\label{fig:4607}
\end{center}
\end{figure*}

\begin{figure*}
\begin{center}
\includegraphics[width=11cm]{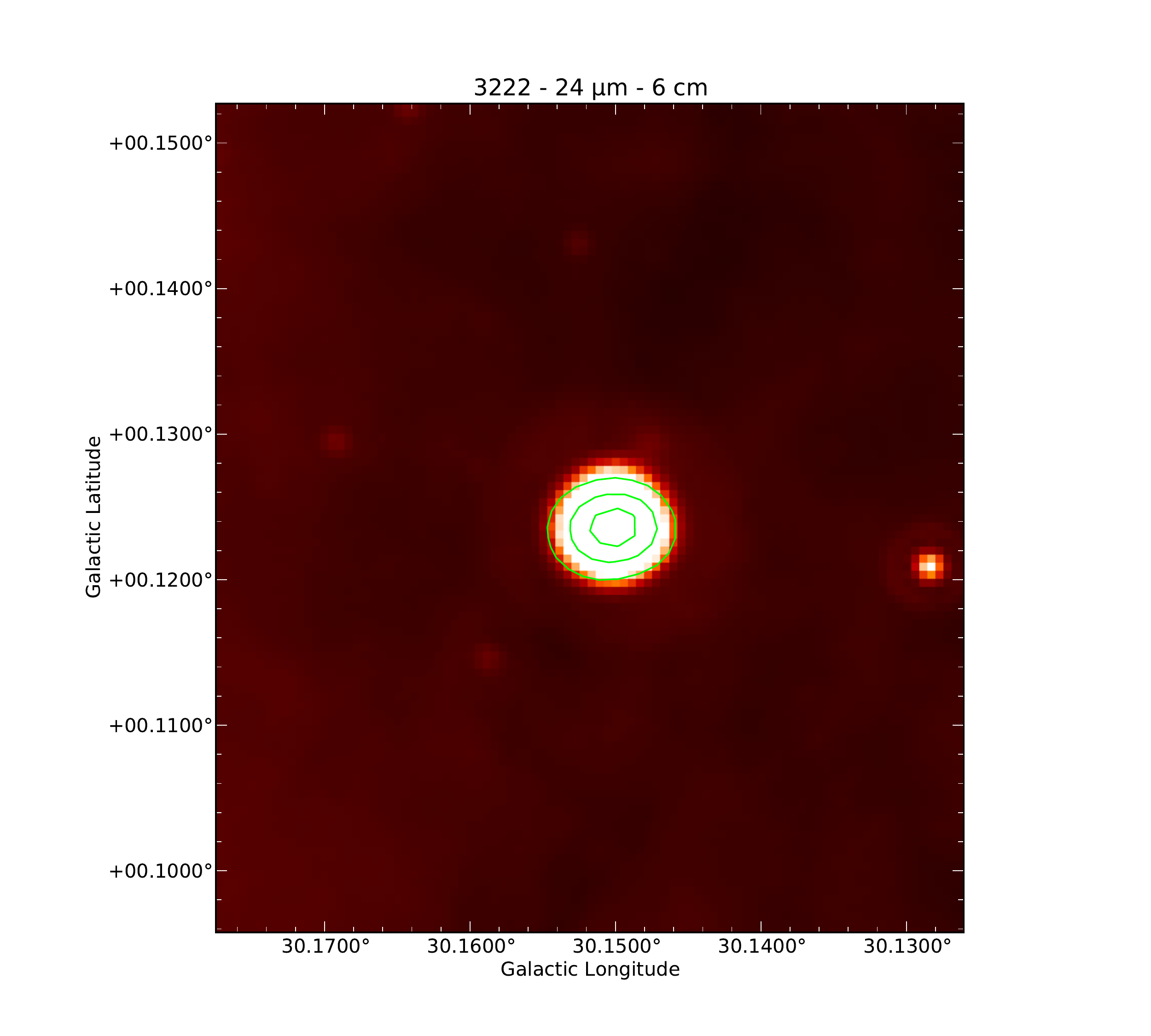}
\includegraphics[width=11cm]{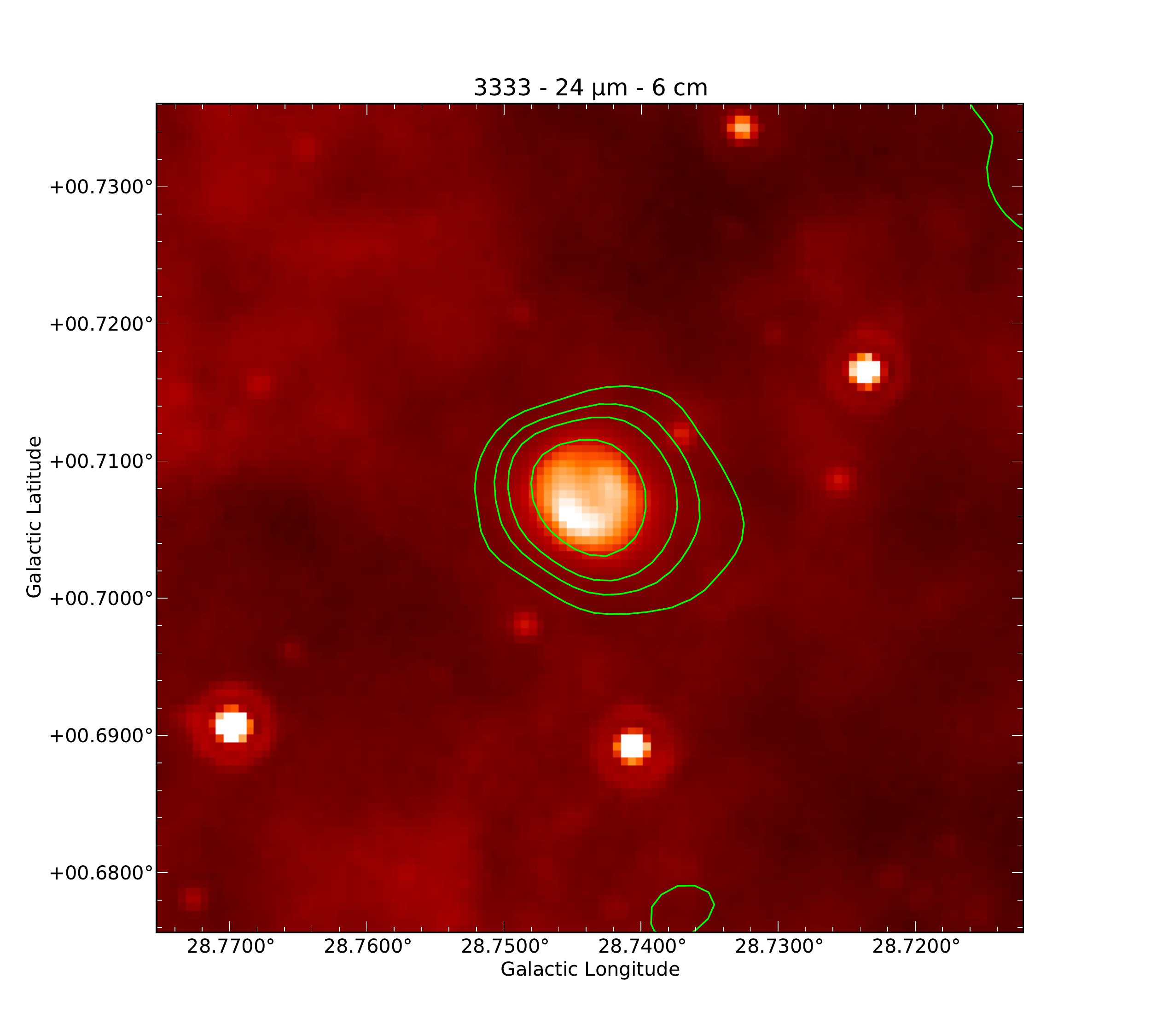}
\end{center}
\end{figure*}
\newpage
\begin{figure*}
\begin{center}
\includegraphics[width=11cm]{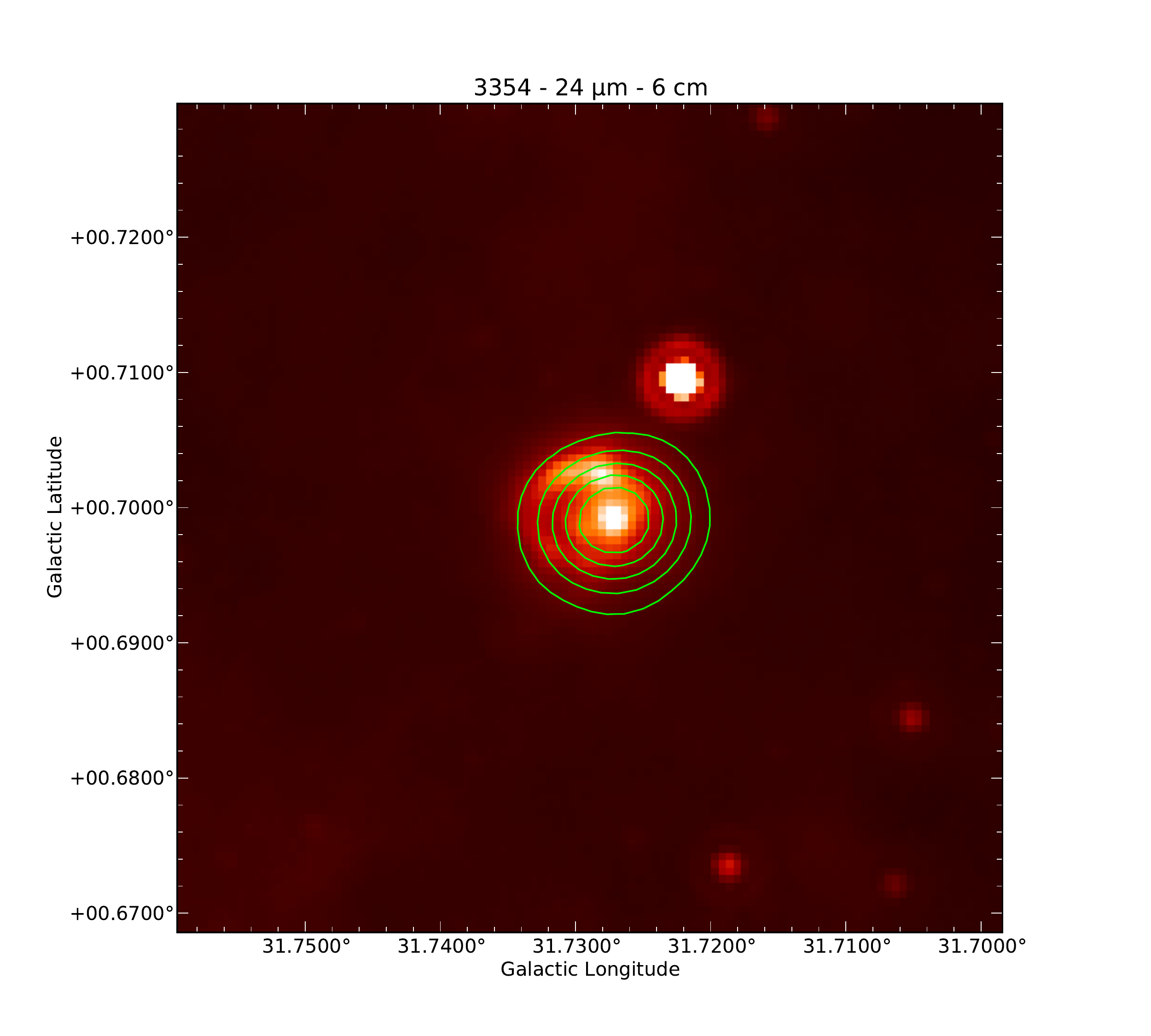}
\includegraphics[width=11cm]{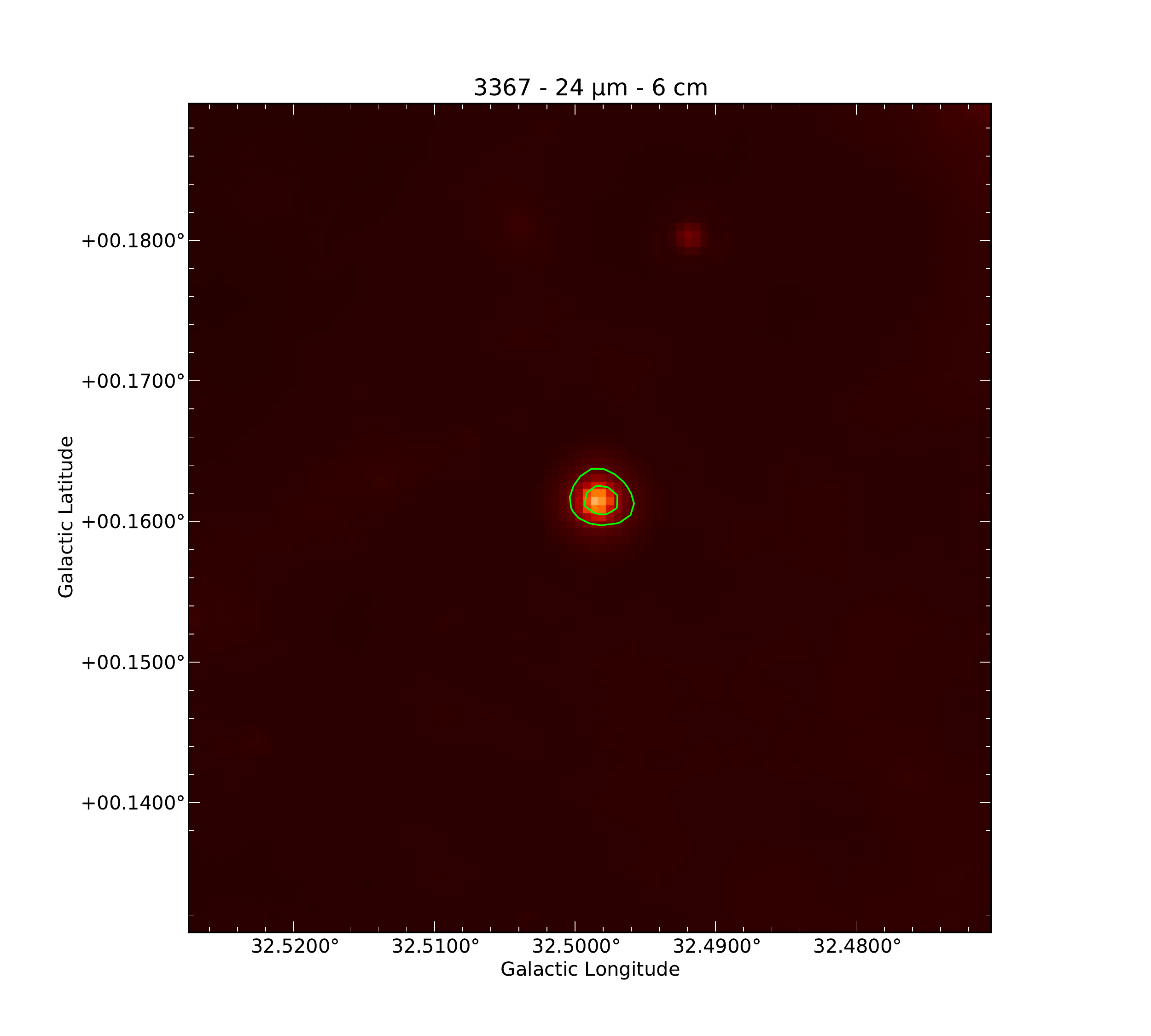}
\end{center}
\end{figure*}

\begin{figure*}
\begin{center}
\includegraphics[width=11cm]{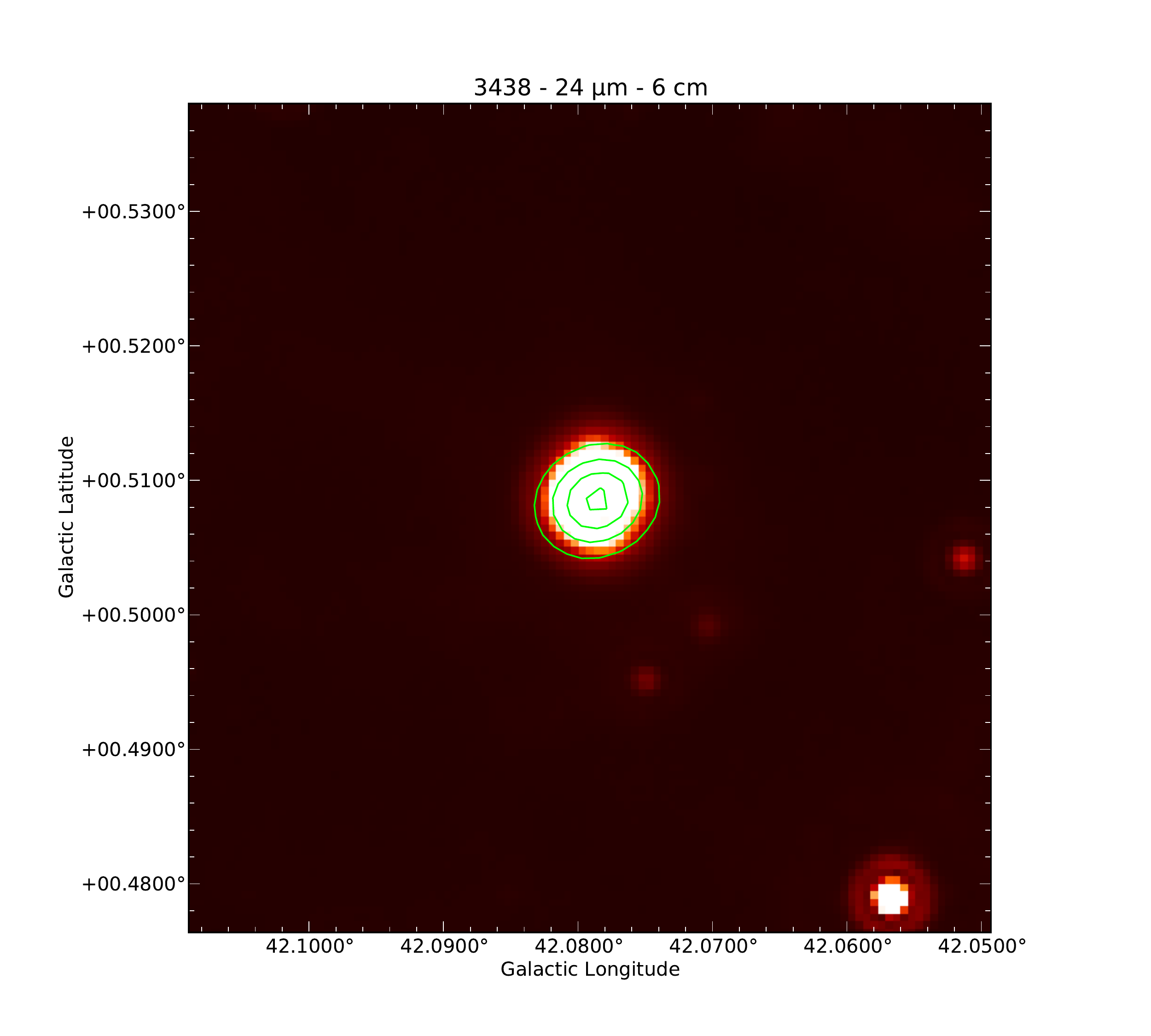}
\includegraphics[width=11cm]{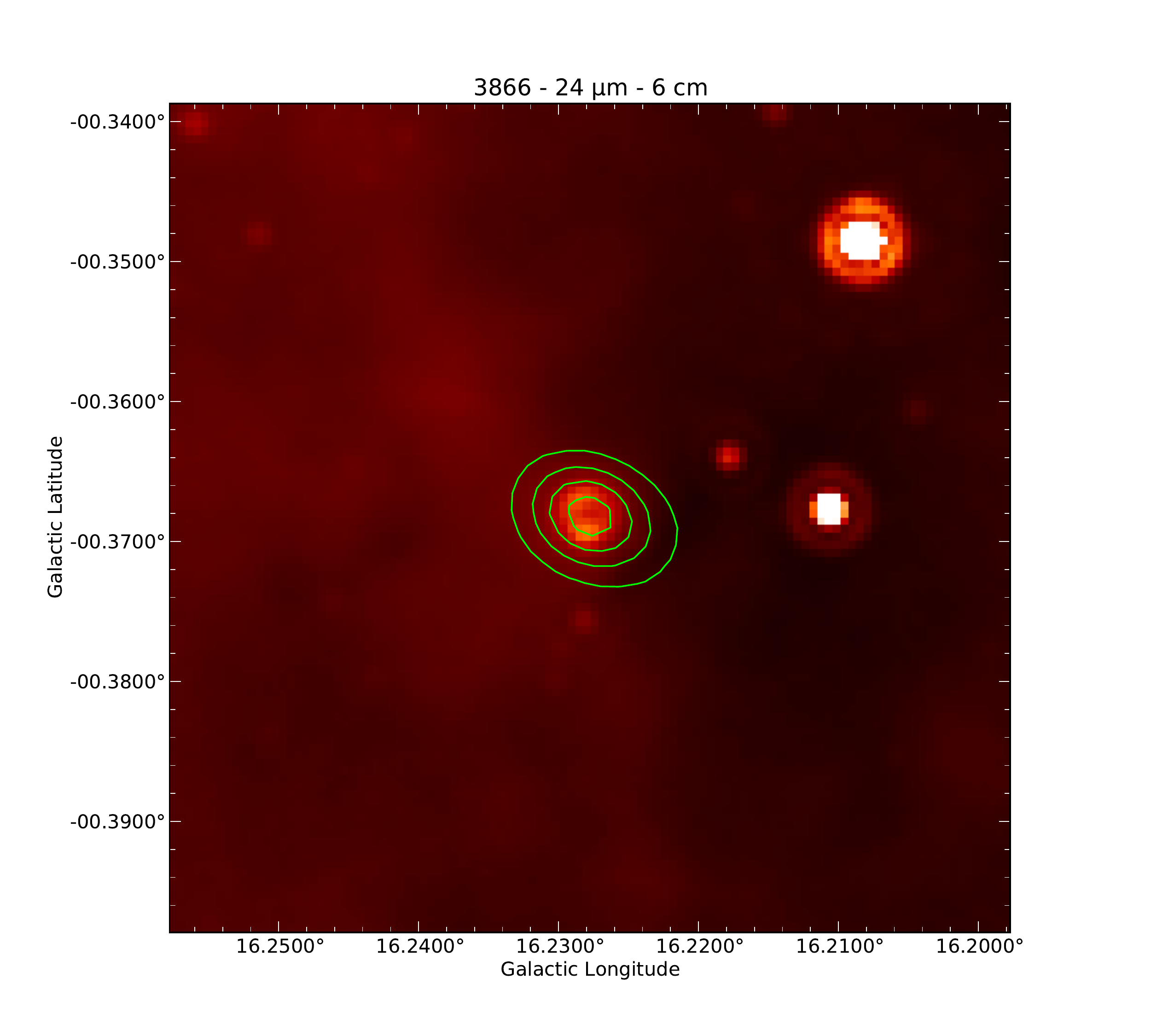}
\end{center}
\end{figure*}

\begin{figure*}
\begin{center}
\includegraphics[width=11cm]{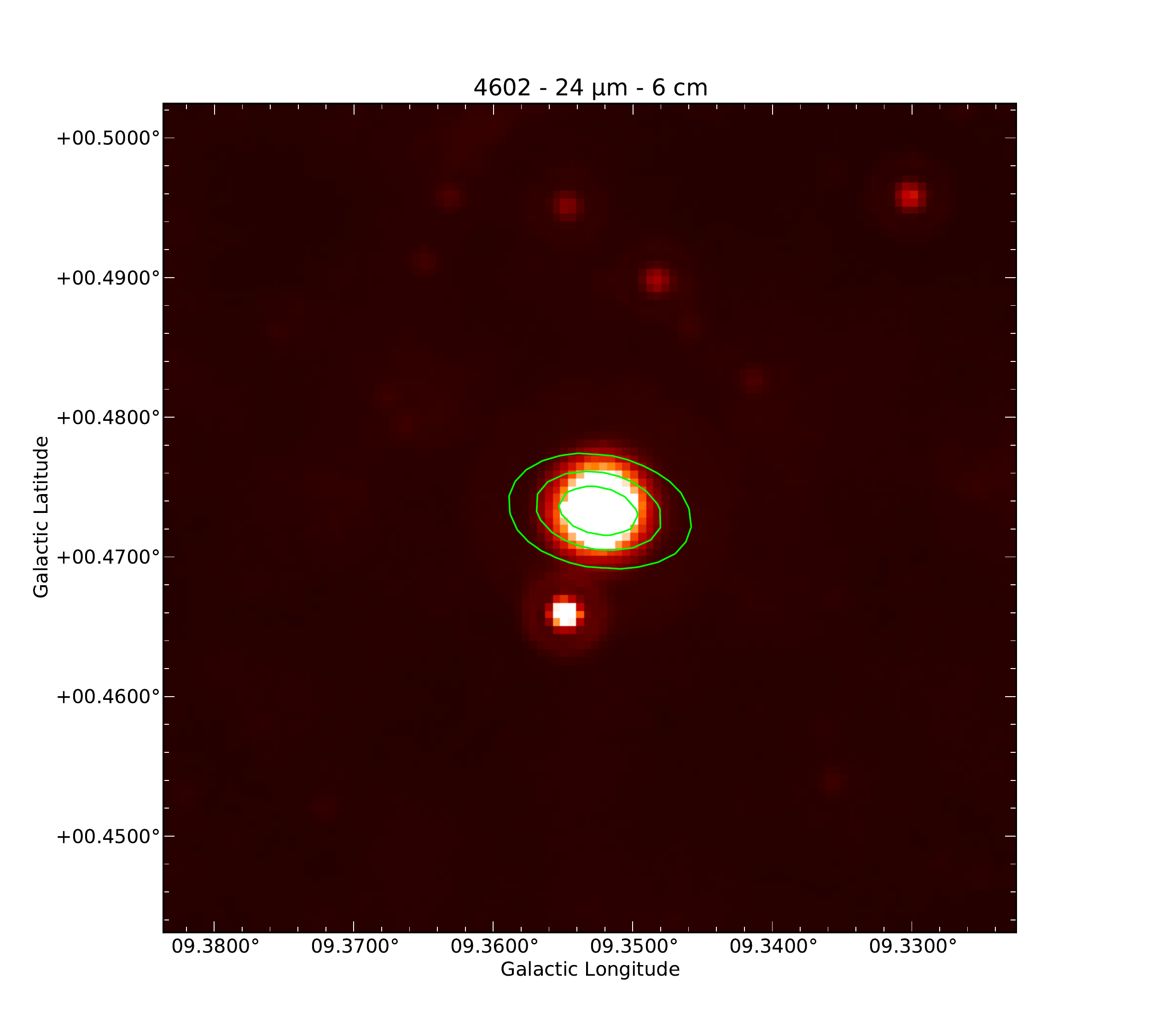}
\includegraphics[width=11cm]{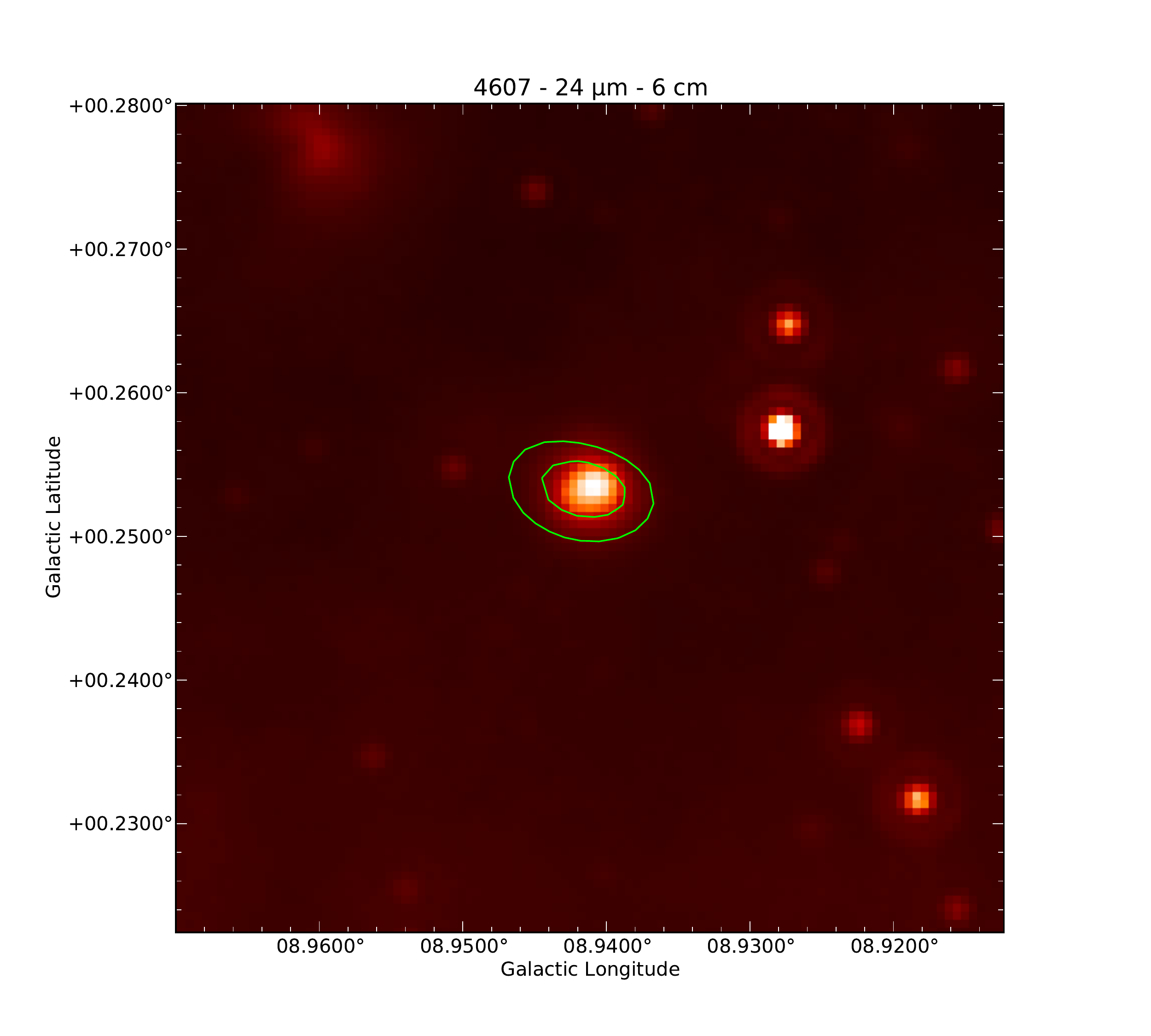}
\end{center}
\end{figure*}

\begin{figure*}

\begin{center}
\includegraphics[width=8cm]{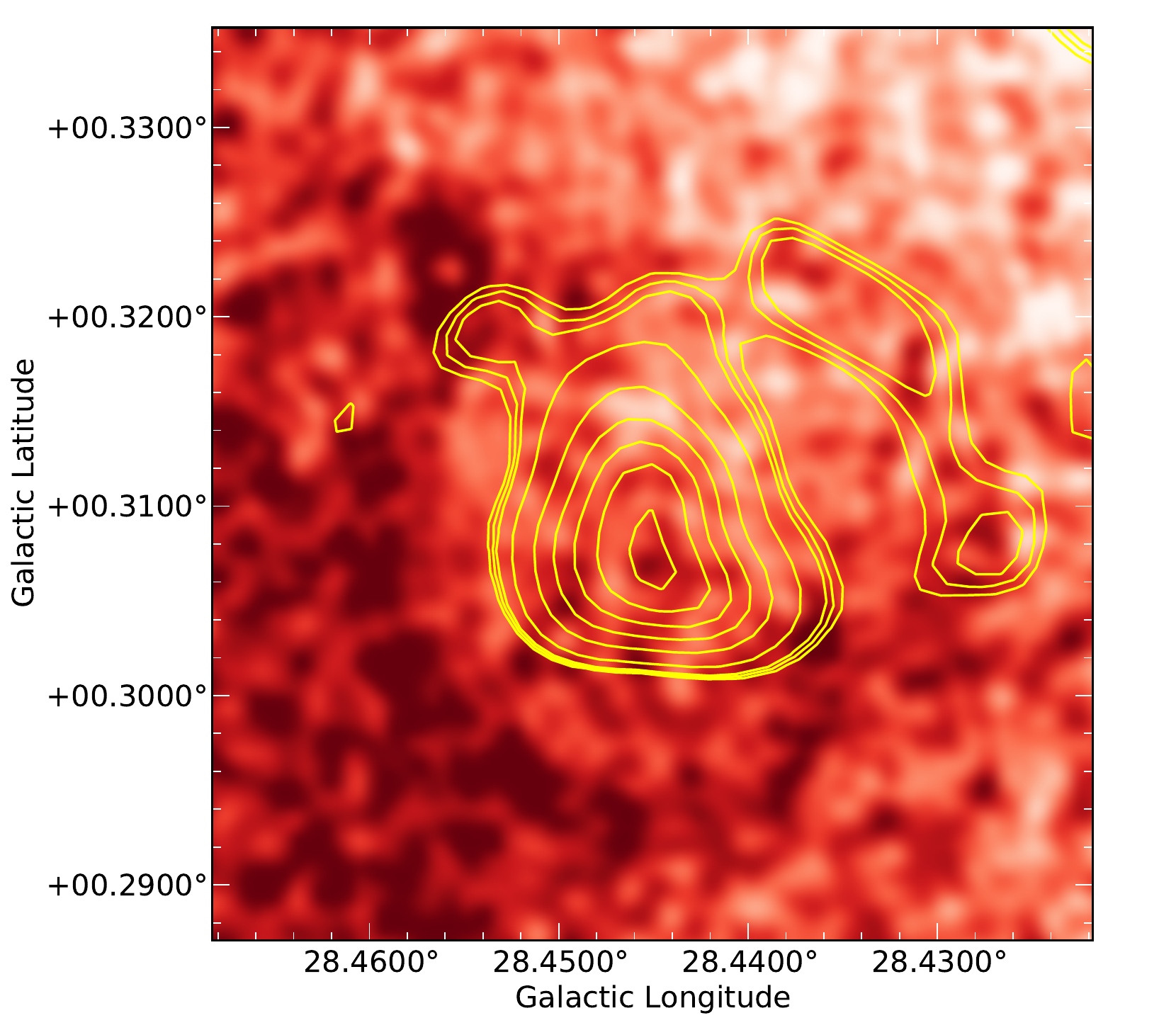}
\includegraphics[width=8cm]{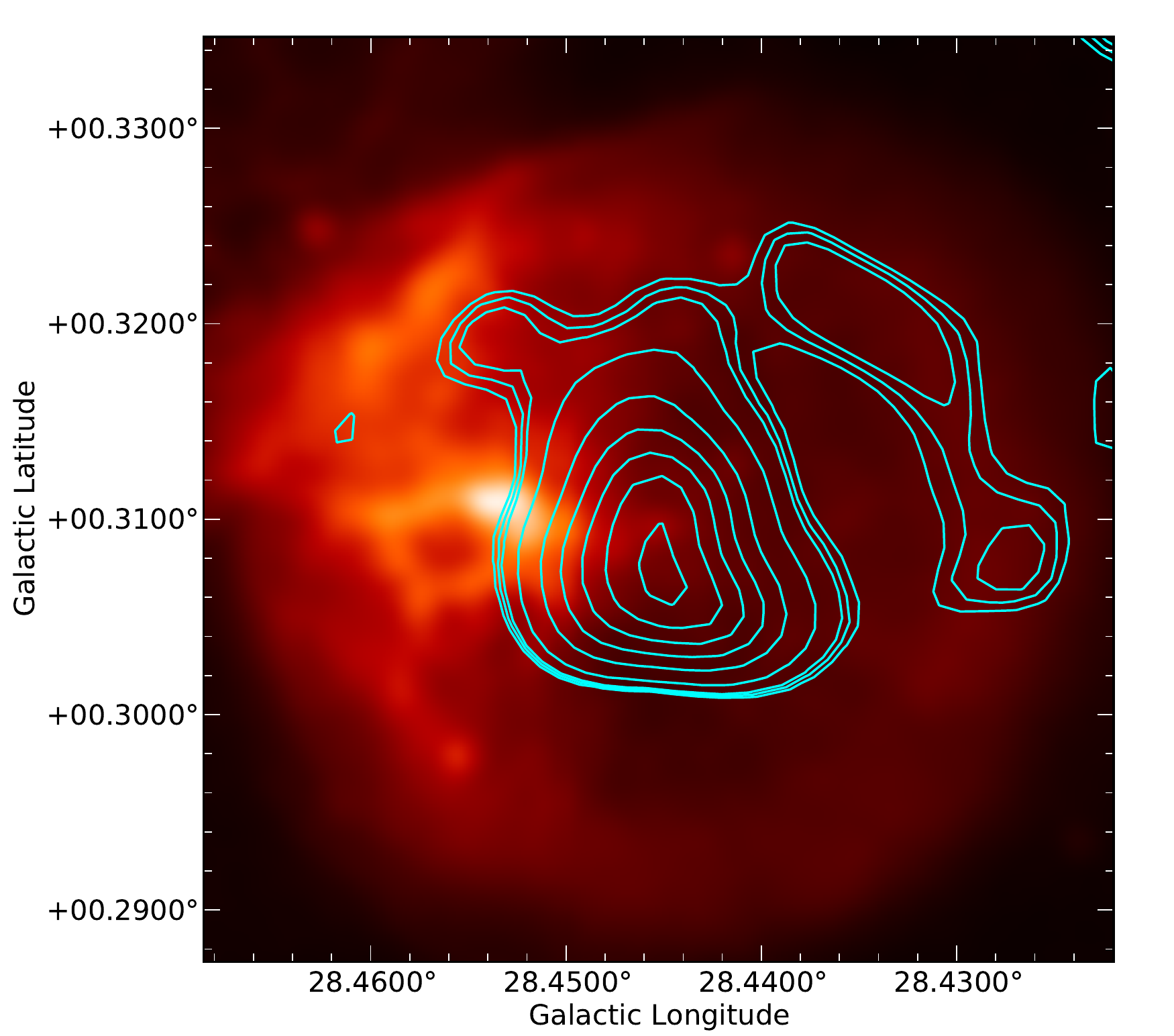}
\caption{Superposition of radio contours at $6\um{cm}$ for Bubble 3313 on MAGPIS 20-cm map (left, inverted colours) and MIPSGAL 24-$\umu$m image (right). Radio contour levels are (in both cases) $0.26$, $0.28$, $0.30$, $0.40$, $0.60$, $0.80$, $1.00$, $1.20$ and $1.40\um{mJy/beam}$. It is possible to notice how at $20\um{cm}$ part of the circular shell observed at $24\mic{m}$ is clearly detected. However, the 6-cm emission seems to come from the interior zone of the nebula, with a possible arc structure (top-right in the images) that traces the 24-$\umu$m emission.}
\end{center}
\end{figure*}

\end{document}